\pdfoutput=1

\documentclass[ amsmath, amssymb, aps, prd, twocolumn, lengthcheck, sort&compress, showpacs, nofootinbib, letterpaper ]{revtex4}

\usepackage{amsmath}
\usepackage{amssymb}
\usepackage{graphicx}
\usepackage{dsfont}
\usepackage{dcolumn}
\usepackage{units}
\usepackage{bm}
\usepackage{wasysym}
\usepackage{multirow}
\usepackage{slashed}
\usepackage[usenames]{color}
\usepackage[colorlinks=true,linkcolor=blue,citecolor=blue,urlcolor=blue]{hyperref}

      \newcommand{\conjg}[1]{\ensuremath{\hspace{1pt}\overline{\hspace{-1pt}#1\hspace{-1pt}}}\hspace{1pt}}
      \newcommand{\vect}[1]{\bm{#1}}

      \newcommand{\be}{\begin{equation}}
      \newcommand{\ee}{\end{equation}}

      \newcommand{\Slash}[1]{\slashed{#1}}

      \def\longlonglongrightarrow{
      \relbar\joinrel\relbar\joinrel\relbar\joinrel\relbar\joinrel\relbar\joinrel\relbar\joinrel\rightarrow}

\begin{document}

         \title{Nucleon Compton scattering in the Dyson-Schwinger approach  }

       \author{Gernot~Eichmann$^{1}$}
       \author{Christian~S.~Fischer$^2$}

\affiliation{$^1$Institut f\"ur Physik, Karl-Franzens-Universit\"at Graz, 8010 Graz, Austria  \\
             $^2$Institut f\"{u}r Theoretische Physik, Justus-Liebig-Universit\"at Giessen, D-35392 Giessen, Germany}

         \date{\today}

         \begin{abstract}
         We analyze the nucleon's Compton scattering amplitude in the Dyson-Schwinger/Faddeev approach.
         We calculate a subset of diagrams that implements the nonperturbative handbag contribution as well as all $t-$channel resonances.
         At the quark level, these ingredients are represented by the quark Compton vertex whose analytic properties we study in detail.
         We derive a general form for a fermion two-photon vertex that is consistent with its Ward-Takahashi identities and free of kinematic singularities,
         and we relate its transverse part to the onshell nucleon Compton amplitude.
         We solve an inhomogeneous Bethe-Salpeter equation for the quark Compton vertex in rainbow-ladder truncation and implement it in the nucleon Compton scattering
         amplitude. The remaining ingredients are the dressed quark propagator 
         and the nucleon's bound-state amplitude which are consistently solved from Dyson-Schwinger and covariant Faddeev equations.
         We verify numerically that the resulting quark Compton vertex and nucleon Compton amplitude both 
         reproduce the $\pi\gamma\gamma$ transition form factor when the pion pole in the $t-$channel
         is approached.
         \end{abstract}


         \maketitle

  \section{Introduction}

   \renewcommand{\arraystretch}{1.5}

     For several decades, Compton scattering has been an important tool to analyze the internal structure of the
     nucleon. At low energies, Compton scattering allows to probe the nucleon's ability to change its size and shape
     under the influence of an external electromagnetic field. These properties are encoded in the nucleon's
     polarizabilities for the case of real Compton scattering (RCS) and their generalized version for the
     case of a virtual incoming photon (VCS). Low-energy Compton scattering thus provides information on
     global as well as spatially resolved electromagnetic properties of the nucleon and, with increasing energy,
     allows to study the effects of its excitation spectrum. In contrast, Compton scattering at
     large energies probes the fundamental quark and gluon degrees of freedom. Here, real and virtual photons
     provide important information on the quark and gluon distributions and allow to extract the spatial
     distribution of the partons inside the nucleon. Consequently, Compton scattering is pursued at many
     experimental facilities around the world including MAMI (Mainz), JLab and HI$\gamma$S at Duke, see e.g.
     \cite{Belitsky:2001ns,Drechsel:2002ar,HydeWright:2004gh,Schumacher:2005an,Drechsel:2007sq,Downie:2011mm,Pasquini:2011zz,Griesshammer:2012we}
     for overviews.

     In addition, much can be learned from Compton-like processes where two virtual photons are exchanged between
     a charged source and the nucleon~\cite{Arrington:2011dn}. Such processes are particularly
     important in lepton-hadron scattering where they contribute to form-factor measurements and serve to explain
     the difference between polarization-transfer measurements and results using a Rosenbluth separation of the
     proton's electric to magnetic form-factor ratio. Furthermore, the Compton scattering amplitude in the
     forward limit contributes to the electromagnetic mass shifts in the nucleon system~\cite{WalkerLoud:2012bg} and encodes the nucleon structure functions.
     Its relevance for the proton radius puzzle has been under recent debate as well~\cite{Miller:2011yw,Carlson:2011dz,Birse:2012eb}.

     From a theoretical perspective, Compton scattering has been studied extensively in effective theories
     \cite{Bernard:1993bg,Hemmert:1997tj,Pascalutsa:2002pi,Beane:2004ra,Griesshammer:2012we,Gasparyan:2011yw} and from dispersion relations \cite{Drechsel:2002ar,Pasquini:2011zz},
     with many important results in both approaches. Effective theories are a natural tool in the low-energy
     region, where constraints due to chiral symmetry and its breaking pattern play a key role. Dispersion
     relations offer an alternative, reliable approach for the analysis of experimental data without a-priori
     limitations for the respective scales.

     While approaches built on explicit quark and gluon degrees of freedom are
     mandatory in the high-energy region, their use in the low-energy regime has been limited so far. First lattice calculations of
     polarizabilities have been performed, see e.g. \cite{Detmold:2010ts,Engelhardt:2011qq} and references therein, but they still
     suffer from large pion masses and the omission of disconnected diagrams. Model calculations for the
     nucleon polarizabilities are available (see e.g. \cite{Liebl:1994qc,Guichon:1995pu,Dong:2005kt} and references therein),
     but often lack a transparent relation to the underlying quantum field theory.

     \begin{figure*}[t]
     \center{
     \includegraphics[scale=0.107]{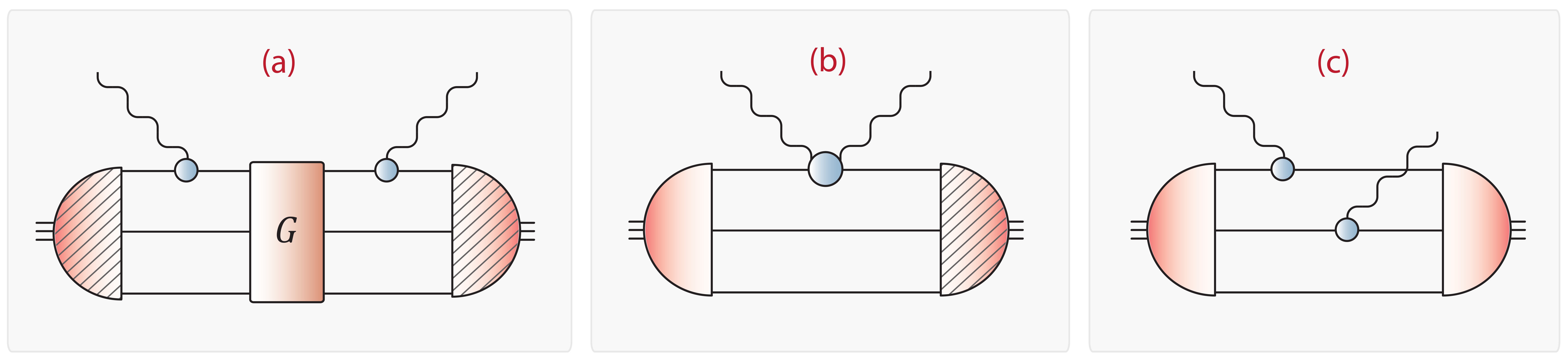}}
        \caption{Graphical representation of the nucleon Compton scattering amplitude in rainbow-ladder truncation, Eq.~\eqref{scattering-final} with Eq.~\eqref{Lambda-RL}.
                 All propagators are dressed, and permutations in the quark lines and symmetrizations of the photon legs are not displayed.
                 Diagram (a) corresponds to the term with $ \mathbf{\Lambda}^{\{\mu} G \,\mathbf{\Lambda}^{\nu\}}$ and diagrams (b)--(c) to $\mathbf{\Lambda}^{\mu\nu}$.
                 In terms of Eq.~\eqref{scattering-reduced}, diagram (a) illustrates the sum of the first and third lines, diagram (b) the second and diagram (c) the fourth line.
                 The kernels that connect the spectator quarks, i.e., the factors $(1-K_a)$ in Eq.~\eqref{scattering-reduced}, have been absorbed in the hatched amplitudes.
                 Diagram (b) contains the 1PI quark two-photon vertex and diagram (a) provides the Born parts, i.e., the pure handbag diagrams, to the full quark Compton vertex.
                 Note that the electromagnetic current of Eq.~\eqref{current-final} in rainbow-ladder truncation has the same form as diagram (b) if the two-photon vertex is replaced by the quark-photon vertex.}
        \label{fig:nca}
     \end{figure*}

     In this work we begin to explore another approach to low-energy Compton scattering. It is based on the
     Dyson-Schwinger equations (DSEs) of motion for QCD's Green functions at the quark-gluon level.
     In combination with Bethe-Salpeter (BSEs) and Faddeev equations, they
     provide a comprehensive framework for the study of non-perturbative properties of hadrons from QCD at all
     momenta and quark masses \cite{Alkofer:2000wg,Fischer:2006ub,Roberts:2007jh,Bashir:2012fs}. In this framework  the
     masses of baryons have been determined from the covariant three-body Faddeev equation
     \cite{Eichmann:2009qa,SanchisAlepuz:2011jn}, and electromagnetic as well as axial form factors have been
     calculated \cite{Eichmann:2011vu,Eichmann:2011pv,SanchisAlepuz:2012ej}; see~\cite{Eichmann:2011ej} for a brief overview.
     In a recent work, we generalized the framework to
     incorporate the coupling of two external currents to the nucleon~\cite{Eichmann:2011ec}, which opens
     the possibility to describe a variety of processes such as pion photo- and electroproduction, pion-nucleon
     scattering or Compton scattering in terms of their quark and gluon substructure. Here, the important
     constraint of electromagnetic gauge invariance dictates the appearance of well-defined classes of additional
     diagrams besides the Born terms, which naturally accommodates effects of resonances in $s-$, $u-$ as well as
     $t$-channel exchange diagrams.
     We now apply the general framework of Ref.~\cite{Eichmann:2011ec} to the specific case of Compton scattering.
     We work out the technical details of the representation of the Compton vertex on the quark level and consider
     first applications in the low-energy region.

   The paper is organized as follows.
   In Sec.~\ref{sec:nucleon-cs} we briefly recapitulate our derivation of the nucleon Compton scattering amplitude and
   discuss the relevant diagrams that appear in the description at the quark-gluon level.
   We isolate the handbag contribution and show that it is closely related to the quark Compton vertex.
   In Secs.~\ref{sec:kinematics},~\ref{sec:tensorbasis} and~\ref{sec:wti} we investigate the general properties
   of a fermion Compton vertex:
   in Sec.~\ref{sec:kinematics} we introduce our notation and discuss the Compton-scattering phase space;
   in Sec.~\ref{sec:tensorbasis} we present the orthogonal tensor basis that we will use in our calculations;
   and in Sec.~\ref{sec:wti} we derive
   a construction for the fermion Compton vertex that respects its Ward-Takahashi identity (WTI) and is free of kinematic singularities.
   In Sec.~\ref{sec:nucleon-compton-amplitude} we present first results for the nucleon Compton amplitude and its $\pi^0\gamma\gamma$
   pole contribution, and we conclude in Sec.~\ref{sec:conclusions}.
   The detailed techniques for solving the inhomogeneous BSE 
   for the quark Compton vertex and for computing the nucleon Compton amplitude are given in Appendices~\ref{sec:qcv-bse}
   and~\ref{sec:nucleon:handbag}, respectively.


\section{Nucleon Compton scattering amplitude} \label{sec:nucleon-cs}

  \subsection{Construction of the scattering amplitude} \label{sec:nucleon:ncv}

             In Ref.~\cite{Eichmann:2011ec} we have derived the following combined result for the electromagnetic current matrix element and the Compton scattering amplitude of a baryon:
             \begin{align}
                 J^\mu &= \conjg{\Psi}_f \,G_0 \,\mathbf{\Lambda}^\mu \,G_0 \Psi_i\,, \label{current-final}\\
                 \widetilde{J}^{\mu\nu} &= \conjg{\Psi}_f \,G_0 \left[  \mathbf{\Lambda}^{\{\mu} G \,\mathbf{\Lambda}^{\nu\}} - \mathbf{\Lambda}^{\mu\nu} \right] G_0 \Psi_i\,. \label{scattering-final}
             \end{align}
             We use a condensed notation where all Dirac, color and flavor indices are suppressed and loop-momentum integrations are implicit.
             The Lorentz indices $\mu$ and $\nu$ belong to the external currents, and the curly brackets denote a symmetrization in these indices.
             $\Psi_i$ and $\Psi_f$ are the incoming and outgoing bound-state amplitudes,
             where a bar denotes charge conjugation. $G_0 = S \otimes S \otimes S$ is the disconnected product of three dressed quark propagators $S$, and
             $G = G_0 + G_0\,T\,G_0$ is the full three-quark (i.e., six-point) Green function, with $T$ being the three-quark scattering matrix.
             The quantities $\mathbf{\Lambda}^\mu$ and $\mathbf{\Lambda}^{\mu\nu}$ are given by
             \begin{equation} \label{Lambda-mu}
             \begin{split}
                 \mathbf{\Lambda}^\mu &= \Big[ \,\Gamma^\mu \otimes S^{-1}\otimes S^{-1}  - \Gamma^\mu \otimes K_{(2)} \,\Big]_\text{perm}\\
                                      & - \Big[ \,S^{-1} \otimes K_{(2)}^\mu \, \Big]_\text{perm} -K_{(3)}^\mu\,,
             \end{split}
             \end{equation}
             \begin{equation} \label{Lambda-mu-nu}
             \begin{split}
                 \mathbf{\Lambda}^{\mu\nu} &=  \Big[ \,\Gamma^{\mu\nu}\otimes   S^{-1}\otimes S^{-1}  -  \Gamma^{\mu\nu} \otimes K_{(2)} \, \Big]_\text{perm}  \\
                                           & +  \Big[ \,\Gamma^{\{\mu}\otimes \Gamma^{\nu\}} \otimes S^{-1} \, \Big]_\text{perm} \\
                                           & -  \Big[ \,\Gamma^{\{\mu} \otimes K_{(2)}^{\nu\}} + S^{-1}\otimes K_{(2)}^{\mu\nu} \, \Big]_\text{perm}  -K_{(3)}^{\mu\nu}\,.
             \end{split}
             \end{equation}
             Here, $\Gamma^\mu$ and $\Gamma^{\mu\nu}$ are the dressed quark-photon and quark two-photon vertices,
             $K_{(2)}$ and $K_{(3)}$ are the two- and three-quark irreducible kernels, and $K_{(i)}^\mu$, $K_{(i)}^{\mu\nu}$
             are those kernels with one or two external photon lines attached. The subscript 'perm' indicates that each bracket has in total
             three permutations with respect to the quark lines.
             The Compton scattering amplitude is visualized in Fig.~\ref{fig:nca} for the special case of a rainbow-ladder truncation which is given in Eq.~\eqref{Lambda-RL} below.

             We note that Eqs.~(\ref{current-final}--\ref{scattering-final}) are completely general.
             Their derivation is based on the formalism of Refs.~\cite{Haberzettl:1997jg,Kvinikhidze:1998xn,Kvinikhidze:1999xp} which is
             neither restricted to baryons nor to electromagnetic currents and Compton scattering
             but holds for any type of current or hadron. In principle it can also be applied to pion electroproduction, nucleon-pion scattering, $\pi\pi$ scattering, etc.
             If the currents describe photons, Eqs.~(\ref{current-final}--\ref{scattering-final}) are consistent with
             the hadron's Bethe-Salpeter or Faddeev equation so that electromagnetic gauge invariance is satisfied.
             This is only valid as long as all underlying ingredients (the quark propagator, the
             two- and three-body kernels and the one- and two-photon vertices) are also consistent with each other, in the sense that they
             satisfy Dyson-Schwinger and Bethe-Salpeter equations and respect the appropriate Ward-Takahashi identities.

             We can extract several features of the scattering amplitude from analyzing the structure of Eq.~\eqref{scattering-final} as follows:
             (a) The scattering amplitude is invariant under $s-u$ crossing since the Lorentz indices are symmetrized.
             (b) The first term in~\eqref{scattering-final} with $\mathbf{\Lambda}^{\{\mu} G \,\mathbf{\Lambda}^{\nu\}}$
             generates the characteristic handbag diagrams via the three-quark disconnected term in $G=G_0 + G_0\,T\,G_0$.
             (c) The same term yields also cat's-ears diagram, where the photons couple to different quark lines; such diagrams are also generated by $\mathbf{\Lambda}^{\mu\nu}$
             via the second line in Eq.~\eqref{Lambda-mu-nu}.
             (d) The three-quark $T-$matrix in the interacting term of $G$ contains all baryon poles
             and therefore reproduces all $s-$ and $u-$channel nucleon resonances, starting with the nucleon's Born term.
             The scattering amplitude at the respective pole becomes:
             \begin{equation}
                 G \rightarrow \frac{G_0\,\Psi\,\conjg{\Psi}\,G_0}{P^2+M_i^2} \quad \Rightarrow \quad
                 \widetilde{J}^{\mu\nu} \rightarrow \frac{J_f^{\{\mu}J_i^{\nu\}}}{P^2+M_i^2}\,.
             \end{equation}
             (e) Finally, $t-$channel meson exchange is implemented as well via the quark two-photon vertex $\Gamma^{\mu\nu}$
             that appears in the first line of Eq.~\eqref{Lambda-mu-nu}; it  involves a quark-antiquark scattering matrix in the $t-$channel. We will elaborate
             this in the following subsection.

             For our practical calculations we will study Eq.~\eqref{scattering-final} by neglecting three-body irreducible interactions ($K_{(3)}=0$)
             and using a rainbow-ladder kernel $K_{(2)}=K_\text{RL}$. That kernel describes a dressed gluon exchange with a quark-gluon vertex ansatz
             that is proportional to $\gamma^\mu$ and only depends on the gluon momentum. Its main advantage despite its simplicity is that it
             satisfies the vector and axialvector Ward-Takahashi identities, which in turn guarantee electromagnetic current conservation for form factors
              and the Goldstone nature of the pion in the chiral limit. Furthermore,
             it has been shown to describe many properties of ground-state pseudoscalar and vector mesons as well as nucleon and $\Delta$ baryons reasonably
             well \cite{Bashir:2012fs,Maris:1997tm,Maris:1999nt,Alkofer:2004yf,Roberts:2007jh,Eichmann:2009qa,Eichmann:2011vu,Eichmann:2011pv,SanchisAlepuz:2011jn,Eichmann:2011aa,Mader:2011zf,SanchisAlepuz:2012ej,Eichmann:2011ej}.
             Nevertheless, the approximation is certainly too simplistic to capture all relevant details of the quark-gluon interaction. From a phenomenological
             point of view, the main missing structure comes from the pion cloud. Pion-cloud effects are known to be important for electromagnetic properties
             of hadrons such as form factors at small momenta or the magnitude of polarizabilities. First efforts to include these effects have been made in the
             meson sector~\cite{Fischer:2007ze,Fischer:2008wy} along with the discussion of other nonperturbative, gluonic corrections beyond rainbow-ladder
             \cite{Fischer:2005en,Fischer:2009jm,Chang:2009zb,Chang:2010hb}. In the baryon sector, such extensions have been computationally too demanding
             up to now. Consequently, in this exploratory work we also restrict ourselves to a rainbow-ladder kernel.

             In rainbow-ladder, the structure of Eqs.~(\ref{Lambda-mu}--\ref{Lambda-mu-nu}) becomes
             \begin{equation}\label{Lambda-RL}
             \begin{split}
                 \mathbf{\Lambda}^\mu &= \Big[ \,\Gamma^\mu \otimes S^{-1}\otimes S^{-1}  - \Gamma^\mu \otimes K_\text{RL} \,\Big]_\text{perm}\,, \\
                 \mathbf{\Lambda}^{\mu\nu} &=  \Big[ \,\Gamma^{\mu\nu}\otimes   S^{-1}\otimes S^{-1}  -  \Gamma^{\mu\nu} \otimes K_\text{RL} \, \Big]_\text{perm}  \\
                                           & +  \Big[ \,\Gamma^{\{\mu}\otimes \Gamma^{\nu\}} \otimes S^{-1} \, \Big]_\text{perm}\,.
             \end{split}
             \end{equation}
             To maintain all features described above, including electromagnetic gauge invariance, we must obtain the dressed quark propagator $S$ from its DSE, the
             quark-photon and two-photon vertices from their inhomogeneous BSEs, the nucleon's Faddeev amplitude $\Psi$
             from its covariant Faddeev equation, and the three-body scattering matrix $T$ from its Dyson equation; all within a rainbow-ladder truncation.

             Especially the last part, i.e., calculating the three-body scattering matrix $T$,
             is a huge endeavor that is not easily accomplished with current computational resources.
             In this paper we will focus exclusively on the handbag and $t-$channel structure, i.e., we will ignore the
             part involving the $T-$matrix as well as all cat's-ears-type diagrams.
             We are aware that electromagnetic gauge invariance cannot be realized with this subset of diagrams
             but only by taking into account the full structure of Eqs.~\eqref{scattering-final} and~\eqref{Lambda-RL}.
             Consequently, information on the low-energy behavior of the nucleon's scattering amplitude
             and the generalized polarizabilities it contains will be limited, at least in these initial studies.

  \subsection{Handbag and $t-$channel diagrams} \label{sec:handbag}

             Our goal in the following is to isolate the nonperturbative 'handbag' content in the scattering amplitude~\eqref{scattering-final}. It comprises
             all diagrams where the two photons couple to the same quark line which is, in addition, disconnected from the two remaining quarks.
             For the derivation we temporarily suppress all occurrences of the quark propagator in our notation.
             In turn, we include the quark-line permutations $a=1,2,3$ explicitly, so that
             the rainbow-ladder truncated current is  given by
             \begin{equation} \label{current-rl}
                J^{\mu}     = \conjg{\Psi}_f \Big[ \sum_{a}\,\Gamma_a^\mu \, (1-K_a)\Big] \Psi_i \,,
             \end{equation}
             and the combination of Eqs.~\eqref{scattering-final} and~\eqref{Lambda-RL} becomes
             \begin{equation}\label{scattering-reduced}
             \begin{split}
                \widetilde{J}^{\mu\nu} &= \conjg{\Psi}_f \Big[ \sum_{a} (1-K_a)\,\Gamma_a^{\{\mu} \,G \,\Gamma_a^{\nu\}}\, (1-K_a)\Big] \Psi_i \\
                           &- \conjg{\Psi}_f \Big[ \sum_{a} \,\Gamma_a^{\mu\nu}\,(1-K_a) \,\Big] \Psi_i \\
                           &+ \conjg{\Psi}_f \Big[ \sum_{a\neq a'} (1-K_a)\,\Gamma_a^{\{\mu} \,G \,\Gamma_{a'}^{\nu\}} \,(1-K_{a'})\Big] \Psi_i \\
                           &- \conjg{\Psi}_f \tfrac{1}{2} \Big[ \sum_{a\neq a'} \,\Gamma_a^{\{\mu}  \,\Gamma_{a'}^{\nu\}} \,\Big] \Psi_i\,.
             \end{split}
             \end{equation}
             $\Gamma^{\mu}_a$ and $\Gamma^{\mu\nu}_a$ are the one- and two-photon vertices that act on quark $a$, and $K_a$ is the
             kernel that connects the remaining two quarks. The handbag structure we are interested in can only emerge from the first two lines of Eq.~\eqref{scattering-reduced}
             whereas cat's-ears contributions ($a\neq a'$) appear in the third and fourth lines.
             The three-quark Green function becomes in this notation: $G=1+T$, and the $T-$matrix satisfies Dyson's equation:
             \begin{equation}\label{dyson-eq-2}
                 T = K\,(1+T)\,,  \qquad K=\sum_{a=1}^3 K_a\,.
             \end{equation}

     \begin{figure}[t]
     \center{
     \includegraphics[scale=0.082]{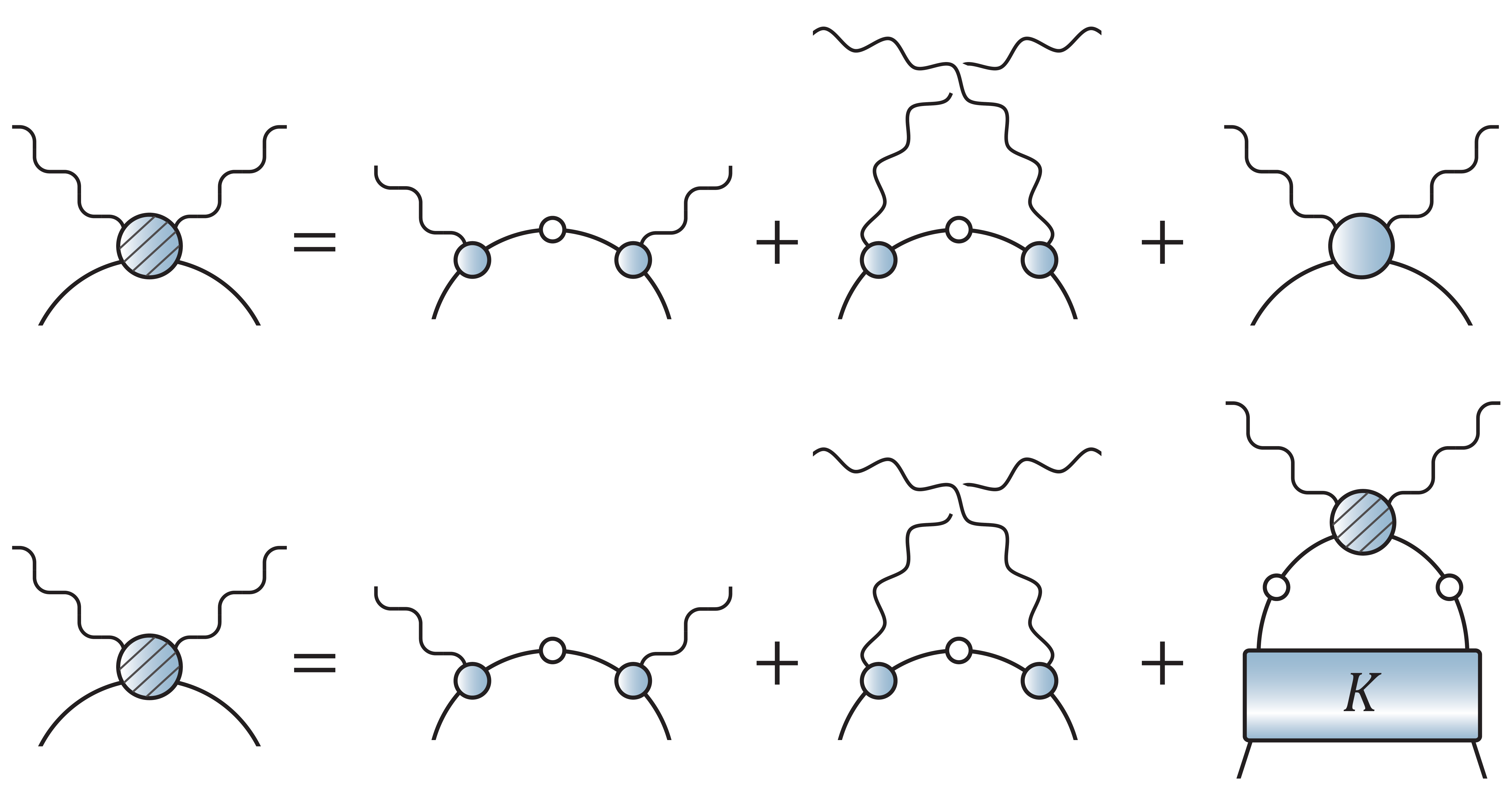}}
        \caption{\textit{Upper panel:} Separation of the fermion Compton vertex into Born terms and a 1PI part.
                         \textit{Lower panel:} inhomogeneous BSE for the quark Compton vertex in rainbow-ladder truncation, where the $q\bar{q}$ kernel $K$ additionally simplifies to a dressed gluon exchange. }
        \label{fig:qcv-born-small}
     \end{figure}

             To isolate the handbag structure, we extract the two-quark contribution with respect to the spectator quark $a$
             from the three-quark $T-$matrix:
             \begin{equation}
                 T =  T_a + T'\,, \qquad T_a = K_a\,(1+T_a)\,,
             \end{equation}
             where $T'$ is the remainder that connects quark $a$ with either of the remaining two quarks, or all three quarks via iteration of Eq.~\eqref{dyson-eq-2}.
             The last equation entails
             \begin{equation}
                 (1-K_a)\,(1+T_a) = 1\,,
             \end{equation}
             and inserting this together with $G=1+T_a + T'$ into Eq.~\eqref{scattering-reduced} yields
             \begin{equation}\label{scattering-handbag}
             \begin{split}
                \widetilde{J}^{\mu\nu} &= \conjg{\Psi}_f  \,\sum_{a} \Big[ \Gamma_a^{\{\mu} \,\Gamma_a^{\nu\}} -\Gamma_a^{\mu\nu}\Big]\, (1-K_a)\, \Psi_i \\
                           &=: -\conjg{\Psi}_f  \,\Big[\sum_{a} \widetilde{\Gamma}_a^{\mu\nu}\, (1-K_a)\Big]\, \Psi_i
             \end{split}
             \end{equation}
             plus further diagrams where photons couple to different quark lines, including
             those with $T-$matrix insertions that connect quark $a$ with its companions.
             Therefore, Eq.~\eqref{scattering-handbag} constitutes the handbag part of the Compton scattering amplitude.
             Comparison with the electromagnetic current~\eqref{current-rl} shows that both equations have an identical structure:
             both include only one factor $(1-K_a)$ and the vertices in both equations act on quark $a$ only.
             The quark-photon vertex $\Gamma^\mu$ in the first expression is replaced by a quark Compton vertex
             \begin{equation}\label{qcv-born+1pi}
                  \widetilde{\Gamma}^{\mu\nu} := \Gamma_\text{B}^{\mu\nu} +\Gamma ^{\mu\nu} = -\Gamma^{\{\mu} \, \Gamma ^{\nu\}} +\Gamma ^{\mu\nu}
             \end{equation}
             in the second, cf. Fig.~\ref{fig:qcv-born-small}. It is the sum of the one-particle-irreducible (1PI) part $\Gamma^{\mu\nu}$ plus a
             Born contribution $\Gamma_\text{B}^{\mu\nu}$ at the quark level.
             We will discuss this separation and its implications in more detail in Section~\ref{sec:wti}.
             We note that the nucleon Compton amplitude from Eq.~\eqref{scattering-handbag} is diagrammatically represented by Fig.~\ref{fig:nca}(b), with the 1PI vertex
             replaced by the full quark Compton vertex from Fig.~\ref{fig:qcv-born-small}.
             The respective diagrams including all kinematic dependencies are also illustrated in Fig.~\ref{fig:current-kinematics} in the appendix.

             At this point the question remains how the quark Compton vertex can be obtained in practice.
             A suitable starting point is the inhomogeneous BSE for the quark-photon vertex $\Gamma^\mu$:
             \begin{equation}\label{ibse-qpv}
                 \Gamma^\mu = \Gamma^\mu_0 + K\,G_0\,\Gamma^\mu\,,
             \end{equation}
             where $\Gamma^\mu_0 = Z_2 \,i\gamma^\mu$ is the tree-level vertex, $G_0=S\otimes S$ is now the disconnected product of a dressed quark and antiquark propagator,
             $K$ is the quark-antiquark kernel, and the second term involves a momentum-loop integral.
             'Gauging' the quark-photon vertex with an additional photon leg yields the 1PI part of the
             quark Compton vertex:
             \begin{equation}\label{qcv-der1}
             \begin{split}
                 \Gamma^{\mu\nu}  :\!\!&= (\Gamma^\mu)^\nu = (\Gamma^\mu_0 + K\,G_0\,\Gamma^\mu)^\nu \\
                                       &= \underbrace{K^\nu G_0\,\Gamma^\mu}_{=: \Gamma^{\mu\nu}_\text{R}} +
                                          \underbrace{K\,G_0^\nu\,\Gamma^\mu}_{=: K G_0\,\Gamma^{\mu\nu}_\text{B}}
                                          + K\,G_0\,\Gamma^{\mu\nu}\,.
             \end{split}
             \end{equation}
             The second term yields the Born contribution because
             \begin{equation}
             \begin{split}
                 G_0^\nu\,\Gamma^\mu &= S\,\Gamma^\mu S^\nu + S^\nu \Gamma^\mu S \\
                                     &= S \left( -\Gamma^\mu S\,\Gamma^\nu - \Gamma^\nu S\,\Gamma^\mu \right) S  \\
                                     &= S\,\Gamma^{\mu\nu}_\text{B} S = G_0\,\Gamma^{\mu\nu}_\text{B}\,.
             \end{split}
             \end{equation}
             The remainder  $\Gamma^{\mu\nu}_\text{R}$ involves $K^\nu$ and does not contribute in rainbow-ladder
             but we keep it for completeness.
             We have therefore
             \begin{equation}
                 \Gamma^{\mu\nu}  = \Gamma^{\mu\nu}_\text{R} + K\,G_0\,(\Gamma^{\mu\nu}_\text{B}+ \Gamma^{\mu\nu} ) \,.
             \end{equation}
             Adding the Born term on both sides of the equation and using Eq.~\eqref{qcv-born+1pi}, with
             $\Gamma^{\mu\nu}_0:=\Gamma^{\mu\nu}_\text{B} + \Gamma_\text{R}^{\mu\nu}$,  one finds an inhomogeneous BSE for the full vertex:
             \begin{equation}\label{ibse-qcv}
                 \widetilde{\Gamma}^{\mu\nu} = \Gamma^{\mu\nu}_0 + K\,G_0\,\widetilde{\Gamma}^{\mu\nu}\,.
             \end{equation}

             In rainbow-ladder, the driving term $\Gamma^{\mu\nu}_0$ simplifies to the Born contribution.
             The inhomogeneous BSE in that case is illustrated in Fig.~\ref{fig:qcv-born-small}.
             Eqs.~\eqref{ibse-qpv} and~\eqref{ibse-qcv} determine the one- and two-photon vertices
             for a given quark-antiquark kernel and quark propagator
             so that their respective Ward-Takahashi identities are automatically satisfied, cf.~Sec.~\ref{sec:wti}.
             Implementing the vertices in Eqs.~\eqref{current-rl} and~\eqref{scattering-handbag} yields
             the nucleon's electromagnetic current and the nonperturbative handbag part of the Compton scattering amplitude in a rainbow-ladder truncation.

             By virtue of the 1PI part in the quark Compton vertex, Eq.~\eqref{scattering-handbag} contains more than just the usual quark Born diagrams with two spectator quarks.
             Using Dyson's equation in the quark-antiquark channel, $T=K + K\,G_0\,T$, we can express the Compton vertex
             as the projection of the full Green function onto the driving term $\Gamma_0^{\mu\nu}$:
             \begin{equation}\label{qcv-T-matrix}
                 \widetilde{\Gamma}^{\mu\nu} =  \Gamma^{\mu\nu}_0 + T\,G_0\,\Gamma^{\mu\nu}_0 = G_0^{-1}\,G\,\Gamma^{\mu\nu}_0\,.
             \end{equation}
             The $q\bar{q}$ Green function $G$ (or, equivalently, the scattering matrix $T$) includes all meson bound-state poles.
             If the respective meson Bethe-Salpeter wave functions overlap with $\Gamma_0^{\mu\nu}$, these poles
             will appear in the quark Compton vertex, and consequently also in the Compton scattering amplitude of the nucleon.
             In fact, the quark Compton vertex is the sole origin of $t-$channel poles in the nucleon Compton amplitude.
             The remaining cat's-ears diagrams merely contain $\rho-$meson resonances
             in $Q^2$ and ${Q'}^2$ from each quark-photon vertex (which is the projection of $G$ onto the tree-level vertex $\Gamma^\mu_0$, cf.~Eq.~\eqref{ibse-qpv}),
             and the three-quark scattering matrix produces additional $s-$ and $u-$channel baryon poles.
             Consequently, the nonperturbative handbag content of Eq.~\eqref{scattering-handbag} contains also the full $t-$channel meson resonance structure.

             The explicit form of the inhomogeneous BSE~\eqref{ibse-qcv}
             for the quark Compton vertex and the handbag part of the nucleon Compton amplitude~\eqref{scattering-handbag}
             are given in in Appendices~\ref{sec:qcv-bse} and~\ref{sec:nucleon:handbag}, respectively.
             For the moment we note that the nucleon's Compton amplitude $\widetilde{J}^{\mu\nu}$
             and the quark Compton vertex $\widetilde{\Gamma}^{\mu\nu}$ have completely
             identical features. They are both fermion two-photon vertices with a rich Dirac-Lorentz and momentum structure.
             Both can be decomposed into a Born part and a 1PI remainder and
             are partially determined by a Ward-Takahashi identity. Therefore, before proceeding with the explicit numerical calculation,
             we will use the next three sections for studying the general properties of a fermion Compton vertex.


  \section{Kinematics and notation}  \label{sec:kinematics}

  \subsection{Kinematic variables} \label{sec:kinematics-variables}

\begin{figure}[t]
     \center{
     \includegraphics[scale=0.08]{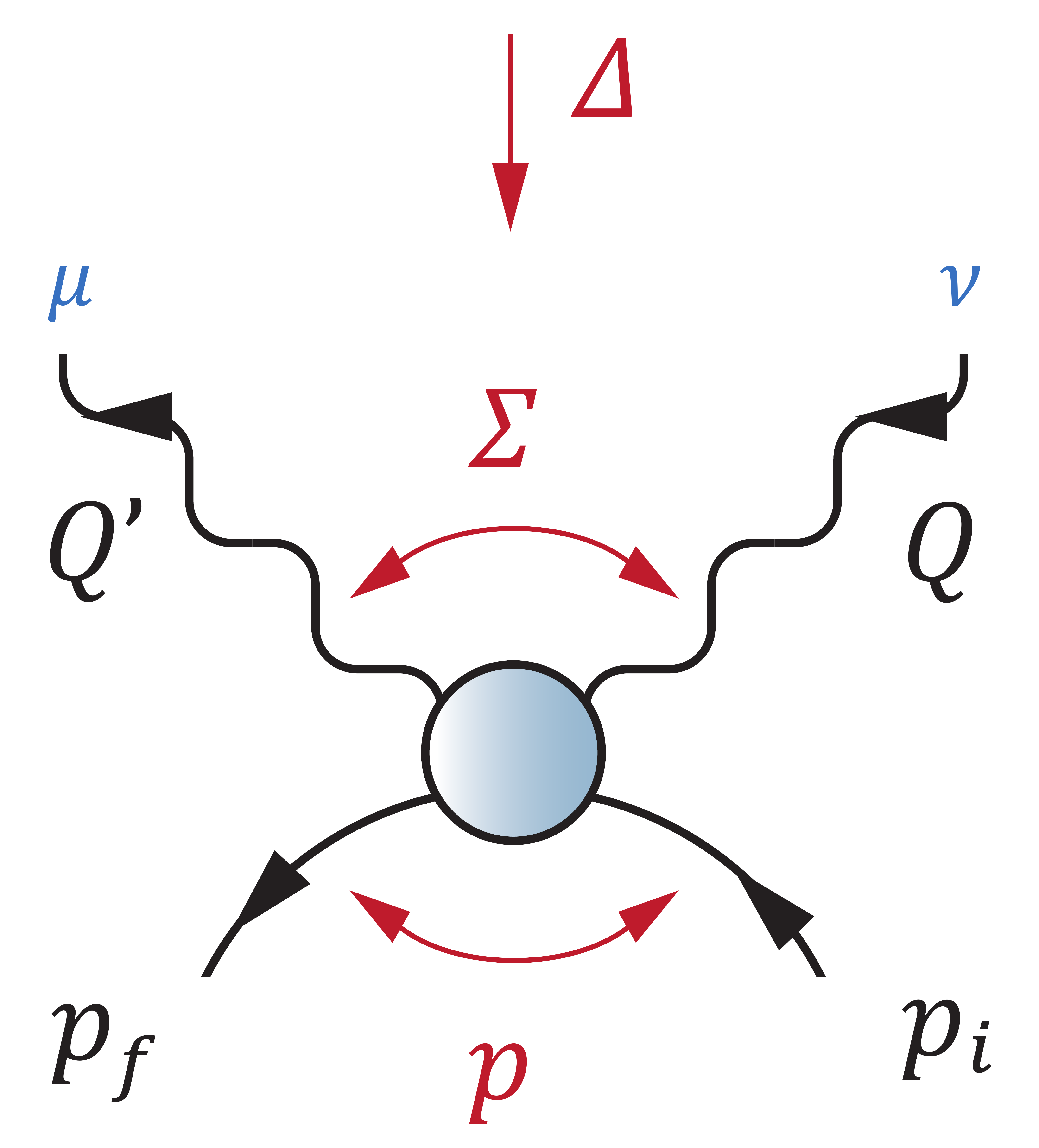}}
        \caption{Kinematics in the fermion Compton vertex.}
        \label{fig:qcv-kinematics}
     \end{figure}

             A fermion Compton vertex depends on three independent four-momenta which can be chosen as
             the four-momentum transfer $\Delta$, the average photon momentum $\Sigma$, and the
             average fermion momentum $p$, cf.~Fig.~\ref{fig:qcv-kinematics}. They are related to the incoming and outgoing fermion and photon momenta via
             \begin{equation}\label{kinematics-1}
                 \begin{array}{rl}
                     p &= \frac{1}{2} (p_i+p_f)\,, \\
                     \Sigma &= \frac{1}{2}(Q+Q')\,,
                 \end{array}\quad
                 \Delta=Q-Q'=p_f-p_i\,,
             \end{equation}
             with the inverse relations
             \begin{equation}
                 \begin{array}{rl}
                     p_i &= p-\frac{\Delta}{2}\,, \\
                     p_f &= p+\frac{\Delta}{2}\,,
                 \end{array}\qquad
                 \begin{array}{rl}
                     Q &= \Sigma+\frac{\Delta}{2}\,, \\
                     Q' &= \Sigma-\frac{\Delta}{2}\,.
                 \end{array}
             \end{equation}
             Depending on the features we want to study, we will alternately express the Compton vertex in terms of
             $\Gamma^{\mu\nu}(p,\Sigma,\Delta)$ or $\Gamma^{\mu\nu}(p,Q,Q')$.
             The first choice is more convenient for investigating the general symmetry properties of the vertex and
             for constructing orthogonal tensor bases, cf.~Sec.~\ref{sec:tensorbasis}; it is also highly advantageous for numerical calculations.
             It resembles the kinematics of the one-photon vertex and nucleon form factors, where $\Delta$ would be the photon momentum and $\Sigma=0$.
             The second choice is more natural for exploring the Ward-Takahashi identity for the vertex, and for constructing
             bases that are transverse with respect to both photon momenta and free of kinematic singularities; see Sec.~\ref{sec:wti}.
             From any set of three independent momenta one can form six Lorentz invariants, for example
             \begin{equation*}
             \begin{split}
                 &\left\{  \; p^2\,,  \; Q^2\,,  \; {Q'}^2\,,  \; p\cdot Q\,,  \; p\cdot Q'\,,  \; Q\cdot Q' \;\right\}   \\
                 \text{or} & \;\; \left\{  \; p^2\,,  \; \Sigma^2\,,  \; \Delta^2\,,  \; p\cdot\Sigma\,,  \; p\cdot\Delta\,,  \; \Sigma\cdot\Delta \;\right\}.
             \end{split}
             \end{equation*}

   \renewcommand{\arraystretch}{1.0}

             In the case of the nucleon Compton scattering amplitude, i.e., when the vertex is taken onshell,
             we rename the fermion momenta $p$, $p_i$, $p_f$ $\longrightarrow$ $P$, $P_i$, $P_f$; otherwise all relations remain unchanged.
             Here we have the additional constraint $P_i^2 = P_f^2 = -M^2$, where $M$ is the nucleon mass, which entails $P\cdot \Delta=0$ and $P^2 = -M^2-\Delta^2/4$.
             It is most convenient to work with the variables constructed from $\Delta$, $\Sigma$, and $P$:
             \begin{equation}\label{euclidean-variables-1}
                 \frac{\Delta^2}{4M^2} =: t\,, \quad
                 \frac{P^2}{M^2}  =-(1+t)\,, \quad
                 P\cdot\Delta =0\,,
             \end{equation}
             where $t$ is the (dimensionless) momentum transfer variable and the limit $t=0$ corresponds to forward scattering.
             The remaining variables involve the average photon momentum $\Sigma$ and we denote them by $X$, $Y$ and $Z$:
             \begin{equation}\label{XYZ-def}
                 \frac{\Sigma^2}{M^2} =: \sigma = tX\,, \quad
                 \widehat{P}\cdot\widehat{\Sigma_T}  =: Y\,,  \quad
                 \widehat{\Sigma}\cdot\widehat{\Delta}  =: Z\,.
             \end{equation}
             Here, a hat denotes a normalized four-momentum and the subscript $T$ refers to a transverse projection
             with respect to the total momentum $\Delta$, i.e.
             \begin{equation}
                 \Sigma_T^\mu = \Sigma^\mu - \frac{\Sigma\cdot\Delta}{\Delta^2}\,\Delta^\mu = (\delta^{\mu\nu}-\hat{\Delta}^\mu \hat{\Delta}^\nu)\,\Sigma^\nu\,.
             \end{equation}
             The Lorentz invariants $t$ and $X$ correspond to radial and $Y$, $Z$ to angular hyperspherical variables.
             $Y$ and $Z$ satisfy simple relations under photon crossing and charge conjugation, cf. Sec.~\ref{sec:vertex:tensorbasis}.
             $O(4)$ symmetry suggests that, kindred to other Euclidean bound-state calculations
             where hyperspherical angles can often be treated via the first few terms in a Chebyshev expansion,
             the dependencies of the Compton amplitude on the angular variables $Y$ and $Z$ might be small.
             In this work we consider only spacelike momenta to avoid difficulties associated with complex continuations in the nucleon Faddeev amplitude
             and singularity restrictions in the quark propagator; in that case we have
             \begin{equation}\label{accessible-phasespace}
                 t, \, X > 0\,, \qquad
                 Y, \, Z \in (-1,1)\,.
             \end{equation}
             This is illustrated by the colored regions in Fig.~\ref{fig:phasespace},
             which also makes clear that Eqs.~(\ref{euclidean-variables-1}--\ref{XYZ-def}) describe the 'natural' variables for Compton scattering in Euclidean kinematics.
             Specifically, we work in the frame where:
             \begin{align}
                 \frac{\Delta}{M} &= 2\sqrt{t}\left(\begin{array}{c} 0 \\ 0 \\ 0 \\ 1 \end{array}\right), \quad
                 \frac{\Sigma}{M}=\sqrt{\sigma}\left(\begin{array}{c} 0 \\ 0 \\ \sqrt{1-Z^2} \\ Z \end{array}\right), \label{simple-frame} \\
                 &  \frac{P}{M}=i \sqrt{1+t}\left(\begin{array}{c} 0 \\ \sqrt{1-Y^2} \\ Y \\ 0 \end{array}\right).  \label{simple-frame-nucleon}
             \end{align}
             Notwithstanding that, all subsequent equations will be written in a Lorentz-covariant (or, if applicable, Lorentz-invariant) manner.
             This is also mandatory as we will frequently switch reference frames when calculating  invariant dressing functions
             of the various ingredients that enter the scattering amplitude.

             When considering the quark Compton vertex instead of the nucleon Compton amplitude, 
             we must again lift the onshell constraints.
             We can arrange the six (now independent) Lorentz invariants of Eqs.~(\ref{euclidean-variables-1}--\ref{XYZ-def}) in two sets
             \begin{equation}\label{qcv-variables}
             \begin{split}
                 \omega &:= \left\{ \; p^2\,,  \;
                                   z=\widehat{p}\cdot\widehat{\Delta}\,, \;
                                   y=\widehat{p_T}\cdot\widehat{\Sigma_T} \; \right\}, \\
                 \lambda &:= \left\{ \; t = \frac{\Delta^2}{4M^2}\,,\;
                                      \sigma = \frac{\Sigma^2}{M^2}\,, \;
                                      Z = \widehat{\Sigma}\cdot\widehat{\Delta} \; \right\}.
             \end{split}
             \end{equation}
             This distinction will be practical in the context of the quark Compton-vertex BSE in App.~\ref{sec:qcv-bse},
             where the momentum variables in the set $\lambda$ appear only as external parameters
             whereas those in $\omega$ will be solved for dynamically.
             Since we replaced the nucleon momentum by an offshell fermion momentum, we have now
             \begin{equation}\label{simple-frame-p}
                 p=\sqrt{p^2}\left(\begin{array}{l} 0 \\ \sqrt{1-z^2}\,\sqrt{1-y^2} \\ \sqrt{1-z^2}\,y \\ z \end{array}\right). \\
             \end{equation}
             The definitions for $t$ and $\sigma$ in Eq.~\eqref{qcv-variables} anticipate know\-ledge of the nucleon's mass; however, that requirement can be easily removed.

  \subsection{Phase space in nucleon Compton scattering}\label{sec:phasespace}

   \renewcommand{\arraystretch}{1.0}

             Next, we want to
             investigate the phase space in nucleon Compton scattering
             and the applicability of our methods to the relevant kinematic limits.
             To this end we first relate the onshell variables $\{t,X,Y,Z\}$ of Eqs.~(\ref{euclidean-variables-1}--\ref{XYZ-def})
             to the incoming and outgoing photon virtualities
             \begin{equation}
                 \frac{Q^2}{4M^2} = \tau\,, \quad
                 \frac{{Q'}^2}{4M^2} = \tau'
             \end{equation}
             and the Mandelstam variables. The latter are obtained from the squares of the
             total momenta
             \begin{equation}
             \begin{split}
                 P+\Sigma &= P_i + Q = P_f + Q' \,, \\
                 P-\Sigma &= P_i - Q' = P_f - Q
             \end{split}
             \end{equation}
             in the $s$ and $u$ channels, see Fig.~\ref{fig:qcv-born}.
             In analogy to our definition of the momentum transfer $t$, it is
             convenient to define modified Mandelstam variables $s$ and $u$ that
             are dimensionless and positive in the physical region:
             \begin{equation}\label{su-def}
                 \frac{(P+\Sigma)^2}{M^2} = -(1+s)\,, \quad
                 \frac{(P-\Sigma)^2}{M^2} = -(1+u)\,.
             \end{equation}
             $s$, $t$ and $u$ are related to the conventional Mandelstam variables $\tilde{s}$, $\tilde{t}$, $\tilde{u}$ by
             \begin{equation}\label{Mandelstam}
                 \tilde{t} = -4M^2 \,t\,, \quad
                 \tilde{s} = M^2(1+s)\,, \quad
                 \tilde{u} = M^2(1+u)\,,
             \end{equation}
             where the extra minus signs in Eqs.~\eqref{su-def} and~\eqref{Mandelstam} originate
             from the Euclidean description.
             The limits $s=0$ and $u=0$ correspond to the
             location of the nucleon pole in each channel, and
             nucleon resonances in the Compton amplitude appear at $s>0$ or $u>0$.
             Photon crossing symmetry ($s\leftrightarrow u$) is expressed by
             the dimensionless crossing variable $\nu$, conventionally defined as
             \begin{equation}\label{crossing-variable}
                 \nu = -\frac{\Sigma\cdot P}{M^2} = \frac{\tilde{s}-\tilde{u}}{4M^2} = \frac{s-u}{4}\,.
             \end{equation}
             The two sets of variables $\{t,\tau,\tau',\nu\}$ and $\{t,X,Y,Z\}$ are then related via
             \begin{equation}\label{standard-from-XYZ}
             \begin{split}
                 \left\{\begin{array}{c} \tau \\ \tau' \end{array}\right\}  &= \frac{t}{4}\left( 1+X \pm 2\sqrt{X}\,Z\right), \\
                 \left\{ \begin{array}{c} s \\ u \end{array}\right\}  & =t\,(1-X) \pm 2\nu\,,
             \end{split}
             \end{equation}
             with $\nu = \sqrt{X}\sqrt{Z^2-1} \sqrt{t(1+t)}\,Y$, or vice versa:
             \begin{equation}\label{XYZ-from-standard}
             \begin{split}
                 X &= \frac{2(\tau+\tau')}{t}-1\,, \\
                 Y &= \sqrt{\frac{t}{1+t}}\,\frac{ \nu  }{\sqrt{(\tau+\tau'-t)^2-4\tau\tau'}}\,, \\
                 Z &= \frac{1}{\sqrt{t}}\frac{\tau-\tau'}{\sqrt{2(\tau+\tau')-t}}\,.
             \end{split}
             \end{equation}
             $Y$ plays the role of the crossing  variable since it is proportional to $\nu$, and
             $Z \sim \tau-\tau'$ is the skewness variable.
             As required, the Mandelstam sum $s+u=2t(1-X)$ leads to
             $Q^2+{Q'}^2+\tilde{s}+\tilde{t}+\tilde{u}= 2M^2$.

     \begin{figure}[t]
     \center{
     \includegraphics[scale=0.27]{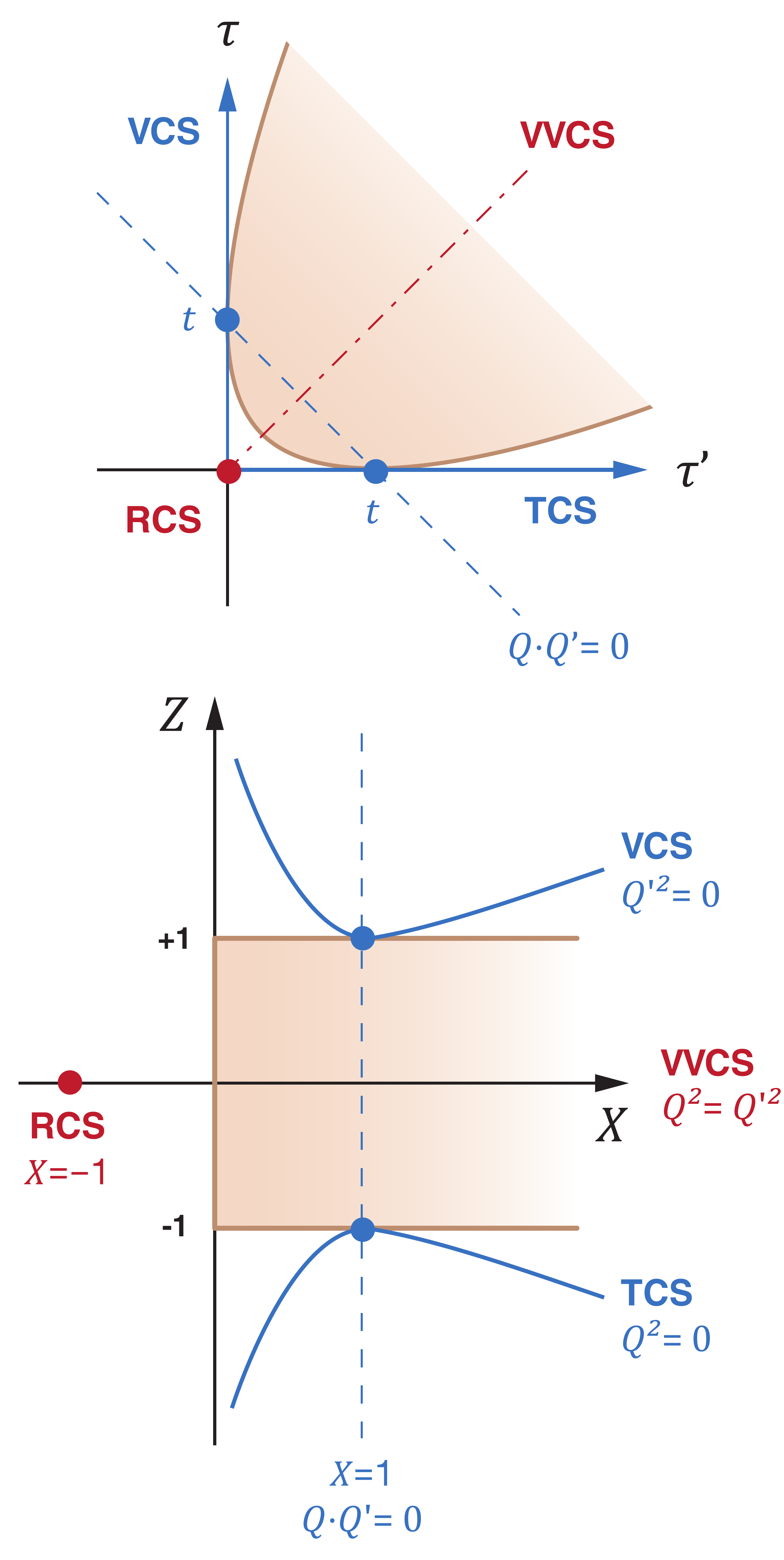}}
        \caption{Sketch of the Compton scattering phase space in the variables $\tau$ and $\tau'$ at fixed $t$ (\textit{upper panel})
        and in the variables $X$ and $Z$ for arbitrary $t\neq 0$ (\textit{lower panel}). The abbreviations RCS, VCS, VVCS and TCS
        denote real, virtual, doubly virtual and timelike Compton scattering. The colored area describes the spacelike region.}
        \label{fig:phasespace}
     \end{figure}

             We can now investigate various kinematic limits in the phase space defined by the variables $\{t,\tau,\tau',\nu\}$ or $\{t,X,Y,Z\}$,
             cf. Fig.~\ref{fig:phasespace}:
             \begin{itemize}

             \item \textbf{Real Compton scattering} ($\tau=\tau'=0$) implies $Z=0$ and $X=-1$.
                   The remaining independent variables $t$ and $Y$ constitute the Mandelstam plane, with
                   $\{s,u\} = 2t \pm 2\sqrt{t(1+t)}\,Y$.
                   The physical region corresponds to $t>0$ and $Y>1$, with $t=0$ forward scattering and $Y=1$ backward scattering.
                   The nucleon polarizabilities are defined in the real forward limit ($t=0$).

             \item \textbf{Virtual Compton scattering} ($\tau'=0$): here only the outgoing photon is real.
                    The skewness variable becomes a function of $X$:
                   \begin{equation}\label{Z-in-VCS}
                       Z = \frac{1+X}{2\sqrt{X}} \geq 1\,,
                   \end{equation}
                   and the remaining variables are $t$, $X$ and $Y$, with
                   \begin{equation}
                       \qquad \tau=\frac{t}{2}\,(1+X)\,, \quad
                       \nu = \pm\frac{1-X}{2}\, \sqrt{t(1+t)}\,Y\,.
                   \end{equation}

             \item \textbf{Timelike Compton scattering} ($\tau=0$) is the reverse process with an incoming real photon:
                   \begin{equation}
                       Z = -\frac{1+X}{2\sqrt{X}} \leq -1\,, \quad  \tau'=\frac{t}{2}\,(1+X)\,.
                   \end{equation}

             \item The \textbf{generalized polarizabilities} are defined
                   at the intersection of the VCS condition ${Q'}^2=0$ with the limit $Q\cdot Q'=0$, which implies $Z=X=1$.
                   In this case, the spacelike region reduces to $\nu=0$ since the dependence on $Y$ decouples, cf.~Eq.~\eqref{standard-from-XYZ},
                   and the only remaining variable is $t=\tau$.
                   This is the Born limit $s=u=0$ which exhibits the onshell nucleon pole in both $s$ and $u$ channels.

             \item \textbf{Forward VVCS} ($t=0$, $\tau=\tau'$): the VVCS limit corresponds to the line $\tau=\tau' \Leftrightarrow Z=0$ in Fig.~\ref{fig:phasespace}.
                   In the forward limit $t=0$ the spacelike region shrinks to the VVCS line since the dependence on $Z$ decouples.
                   For forward scattering
                   we must work with the variable $\sigma=tX$ from Eq.~\eqref{XYZ-def} since $X$ is no longer well-defined.
                   Because of $t=0$ and $Z=0$ the four-vector $\Delta^\mu$ vanishes, and hence $Q^\mu={Q'}^\mu=\Sigma^\mu$.
                   The remaining independent variables are $\sigma$ and $Y$ (or, equivalently, $\tau$ and $x_B$) with
                   \begin{equation}
                       \tau = \tau' = \frac{\sigma}{4}\,, \quad
                       \left\{ \begin{array}{c} s \\ u \end{array}\right\} = \sigma \left( -1 \pm \frac{1}{x_B}\right),
                   \end{equation}
                   where the Bjorken scaling variable is
                   \begin{equation}
                       x_B = -\frac{\Sigma^2}{2P\cdot \Sigma} = \frac{\sigma}{2\nu} = \frac{i\sqrt{\tau}}{Y}\,.
                   \end{equation}
                   In the physical region $Y$ must become imaginary; we have chosen sign conventions so that $x_B\rightarrow 0^+$ corresponds to $Y\rightarrow +i\infty$.
                   The forward VVCS amplitude encodes the nucleon structure functions $f_T$, $f_L$, $g_{TT}$ and $g_{LT}$.

             \end{itemize}

             It is clear from these considerations that many of the interesting physical applications with available experimental data
             correspond to phase space regions outside the 'spacelike' domain that is defined by Eq.~\eqref{accessible-phasespace}
             and illustrated in Fig.~\ref{fig:phasespace}. Directly accessible are:
             \begin{itemize}
             \item the generalized polarizabilities at $\nu=0$, including the usual polarizabilities in the real forward limit,
             \item forward VVCS for $\nu \lesssim |Q|/M \Leftrightarrow x_B \gtrsim \sqrt{\tau}$,
                   e.g. through a Chebyshev expansion to complex $Y$,
             \item and two-photon exchange processes with a charged source where the full spacelike region is integrated over.
             \end{itemize}
             Access to the remaining limits (real and deeply virtual Compton scattering, Bjorken regime)
             is in principle possible but requires
             higher numerical effort and sophisticated methods for dealing with complex singularities.
             For the time being, we will restrict ourselves to those applications which can be obtained from the spacelike calculation
             and, if necessary, perform extrapolations to the inaccessible phase space domains.


\section{Tensor basis for a fermion Compton vertex} \label{sec:tensorbasis}

      In this section we investigate the general properties of a fermion Compton vertex
      and construct a simple, orthonormal tensor basis that is useful for numerical calculations.
      The analysis can be applied to the nucleon's Compton amplitude
      as well as the quark Compton vertex. For the former,
      the incoming and outgoing nucleons are onshell which simplifies the structure
      considerably. We begin with the analysis of the general offshell case and return to the
      onshell simplification in Sec.~\ref{sec:tensorbasis-onshell}.

\subsection{General Compton vertex}\label{sec:vertex:tensorbasis}

    \renewcommand{\arraystretch}{0.8}

             We start by constructing the general Lorentz--Dirac basis for
             a fermion Compton vertex $\Gamma^{\mu\nu}(p,\Sigma,\Delta)$. The vertex
             is a matrix in Dirac space and depends on three four-vectors $p$, $\Sigma$ and $\Delta$, cf. Fig.~\ref{fig:qcv-kinematics}
             for the momentum routing.
             We work with the kinematic variables defined in Eq.~\eqref{qcv-variables}: the photon variables $t$, $\sigma$ and $Z$
             and the fermion variables $p^2$, $z$ and $y$.

             To facilitate the analysis we want to express the vertex in an orthonormal basis.
             We can achieve this by orthogonalizing the three momenta $\Delta$, $\Sigma$ and $p$, so that
             \begin{equation}\label{orthonormal-momenta}
                 d^\mu := \hat{\Delta}^\mu\,, \quad
                 s^\mu := \widehat{\Sigma_T}^\mu\,, \quad
                 r^\mu := \widehat{p_t}^\mu
             \end{equation}
             are three covariant unit vectors.
             The hats denote normalized four-vectors,
             and the subscript $T$ stands for a transverse projection with respect to the external momentum $\Delta$, e.g.:
             \begin{equation}
                 \Sigma_T^\mu = T^{\mu\nu}_\Delta\,\Sigma^\nu = (\delta^{\mu\nu}-\hat{\Delta}^\mu \hat{\Delta}^\nu)\,\Sigma^\nu\,.
             \end{equation}
             In addition,
             the subscript $t$ indicates a transverse projection with respect to both $\Delta$ and $\Sigma_T$, i.e.:
             \begin{equation}
                 p_t^\mu = p_T^\mu - \frac{p_T\cdot\Sigma_T}{\Sigma_T^2}\,\Sigma_T^\mu = p_T^\mu - (p_T\cdot s)\,s^\mu\,.
             \end{equation}
             The resulting vectors $r$, $s$ and $d$ satisfy $d^2=s^2=r^2=1$ and $d\cdot s = d\cdot r = s\cdot r=0$.

    \renewcommand{\arraystretch}{1.5}

             \begin{table*}[t]

                \begin{center}
                \begin{tabular}{   | @{\quad}  c @{\quad} | @{\quad} c @{\quad} |  c | c | } \cline{3-4}
                   \multicolumn{2}{c|}{}                                          &  \, $\mathsf{S}\,\mathsf{T}$ \, &  \, $\mathsf{B}\,\mathsf{C}$ \quad \\ \hline \rule{0mm}{0.5cm}
                 $\mathsf{X}_1^{\mu\nu}$      &  $v^\mu v^\nu + r^\mu r^\nu$    & $+\;\bullet$  &  $++$   \\
                 $\mathsf{X}_2^{\mu\nu}$      &  $v^\mu v^\nu - r^\mu r^\nu$    & $+\;\circ$  &  $++$  \\ [2mm]
                 $\mathsf{X}_3^{\mu\nu}$      &  $v^\mu r^\nu + r^\mu v^\nu $   & $+\;\circ$  &  $--$  \\
                 $\mathsf{X}_4^{\mu\nu}$      &  $v^\mu r^\nu - r^\mu v^\nu$    & $-\;\circ$  &  $++$  \\ [2mm]
                 $\mathsf{X}_5^{\mu\nu}$      &  $s^\mu s^\nu + d^\mu d^\nu$    & $+\;\bullet$  &  $++$  \\
                 $\mathsf{X}_6^{\mu\nu}$      &  $s^\mu s^\nu - d^\mu d^\nu$    & $+\;\circ$  &  $++$  \\[2mm]
                 $\mathsf{X}_7^{\mu\nu}$      &  $s^\mu d^\nu + d^\mu s^\nu$    & $+\;\circ$  &  $--$  \\
                 $\mathsf{X}_8^{\mu\nu}$      &  $s^\mu d^\nu - d^\mu s^\nu$    & $-\;\circ$  &  $++$  \\ [2mm]  \cline{1-4}

                \end{tabular} \hspace{2mm}
                \begin{tabular}{   | @{\quad} c @{\quad}|@{\quad} c @{\quad}| c | c | } \cline{3-4}
                   \multicolumn{2}{c|}{}                                      &  \, $\mathsf{S}\,\mathsf{T}$ \, &  \, $\mathsf{B}\,\mathsf{C}$ \quad \\ \hline \rule{0mm}{0.5cm}
                 $\mathsf{X}_9^{\mu\nu}$      &  $r^\mu s^\nu + s^\mu r^\nu$    & $+\;\circ$  &  $-+$  \\
                 $\mathsf{X}_{10}^{\mu\nu}$   &  $v^\mu s^\nu + s^\mu v^\nu$    & $+\;\circ$  &  $+-$  \\ [2mm]
                 $\mathsf{X}_{11}^{\mu\nu}$   &  $r^\mu s^\nu - s^\mu r^\nu$    & $-\;\circ$  &  $+-$  \\
                 $\mathsf{X}_{12}^{\mu\nu}$   &  $v^\mu s^\nu - s^\mu v^\nu$    & $-\;\circ$  &  $-+$  \\ [2mm]
                 $\mathsf{X}_{13}^{\mu\nu}$   &  $r^\mu d^\nu + d^\mu r^\nu$    & $+\;\circ$  &  $+-$  \\
                 $\mathsf{X}_{14}^{\mu\nu}$   &  $v^\mu d^\nu + d^\mu v^\nu$    & $+\;\circ$  &  $-+$  \\[2mm]
                 $\mathsf{X}_{15}^{\mu\nu}$   &  $r^\mu d^\nu - d^\mu r^\nu$    & $-\;\circ$  &  $-+$  \\
                 $\mathsf{X}_{16}^{\mu\nu}$   &  $v^\mu d^\nu - d^\mu v^\nu$    & $-\;\circ$  &  $+-$  \\ [2mm]  \hline

                \end{tabular} \hspace{15mm}
                \begin{tabular}{   | @{\quad} c @{\quad}|@{\quad} l @{\quad}| c | } \cline{3-3}
                   \multicolumn{2}{c|}{}                                      &  \, $\mathsf{B}\,\mathsf{C}$ \quad \\ \hline \rule{0mm}{0.5cm}
                 $\tau_1$   &  $\mathds{1}$                         &  $++$  \\
                 $\tau_2$   &  $\Slash{r}$                          &  $++$  \\
                 $\tau_3$   &  $\Slash{s}$                          &  $-+$  \\
                 $\tau_4$   &  $\Slash{d}$                          &  $+-$  \\ [2mm]
                 $\tau_5$   &  $\Slash{r}\,\Slash{s}$               &  $--$  \\
                 $\tau_6$   &  $\Slash{r}\,\Slash{d}$               &  $++$  \\
                 $\tau_7$   &  $\Slash{s}\,\Slash{d}$               &  $-+$  \\
                 $\tau_8$   &  $\Slash{r}\,\Slash{s}\,\Slash{d}$    &  $-+$  \\ [2mm]  \hline

                \end{tabular}
                \end{center}

               \caption{Tensor basis for a fermion Compton vertex.
                        The labels $\mathsf{S}$ and $\mathsf{T}$ indicate whether the respective basis element is symmetric or antisymmetric in the Lorentz indices
                        and whether it has a Lorentz trace or not.
                        $\mathsf{B}$ and $\mathsf{C}$ denote the Bose and charge-conjugation symmetries for each element.
                        The full basis is constructed according to Eq.~\eqref{basis-element}.
                        All basis elements must be additionally equipped with a factor $\tfrac{1}{\sqrt{2}}$ to ensure orthonormality in the sense of Eq.~\eqref{qcv-orthonormality}.}
               \label{qcv-basis-1}

        \end{table*}

         \renewcommand{\arraystretch}{1.0}

             Furthermore, instead of using $\gamma^\mu$ or some transverse combination thereof,
             the dependence of the vertex on \textit{three} independent momenta allows us to
             work with the axialvector
             \begin{equation}\label{def-v}
                 v^\mu = \varepsilon^{\mu\alpha\beta\gamma} r^\alpha s^\beta d^\gamma
             \end{equation}
             as the remaining independent four-vector.
             $v$ must appear in combination with $\gamma_5$ to transform like a vector under parity,
             which allows to construct a complete and orthogonal tensor basis from all possible combinations of
             \begin{equation}
                 \gamma_5\,v^\mu\,, \quad r^\mu\,, \quad s^\mu\,,\quad d^\mu\,, \quad
                 \mathds{1}\,,\quad \Slash{r}\,,\quad \Slash{s}\,,\quad \Slash{d}\,,
             \end{equation}
             with $\Slash{v}=\gamma_5 \,\Slash{r} \,\Slash{s} \,\Slash{d}$.
             For example, we can now write
             \begin{equation}\label{vrsd-gamma}
                 \gamma^\mu = v^\mu \Slash{v} + r^\mu \Slash{r} +s^\mu \Slash{s} +d^\mu \Slash{d} \,.
             \end{equation}
             The structure $\delta^{\mu\nu}$ no longer appears in the basis since it is linearly dependent:
             \begin{equation}\label{vrsd-delta}
                 \delta^{\mu\nu} = v^\mu v^\nu + r^\mu r^\nu + s^\mu s^\nu + d^\mu d^\nu\,.
             \end{equation}
             Similarly, the commutator $[ \gamma^\mu , \gamma^\nu ]$ can be expressed as
             \begin{equation}
             \begin{split}
                 \tfrac{1}{2} \left[ \gamma^\mu , \gamma^\nu \right] &=
                     \Slash{v}\,\Slash{r}\,(v^\mu r^\nu-r^\mu v^\nu)  + \Slash{r}\,\Slash{s}\,(r^\mu s^\nu-s^\mu r^\nu) \\
                  &+ \Slash{v}\,\Slash{s}\,(v^\mu s^\nu-s^\mu v^\nu)  + \Slash{r}\,\Slash{d}\,(r^\mu d^\nu-d^\mu r^\nu) \\
                  &+ \Slash{v}\,\Slash{d}\,(v^\mu d^\nu-d^\mu v^\nu)   + \Slash{s}\,\Slash{d}\,(s^\mu d^\nu-d^\mu s^\nu) \,, 
             \end{split}
             \end{equation}
             and one can write
             \begin{equation}\label{vrsd-epsilon}
                 \varepsilon^{\mu\nu\alpha\beta} s^\alpha d^\beta = v^\mu r^\nu - r^\mu v^\nu \,.
             \end{equation}
             In the Lorentz frame defined by Eqs.~\eqref{simple-frame} and~\eqref{simple-frame-p},
             $v$, $r$, $s$, and $d$ become the Euclidean unit vectors in $\mathds{R}^4$:
             \begin{equation}\label{vrsd}
                 v=\left(\begin{array}{c} 1 \\ 0 \\ 0 \\ 0 \end{array}\right), \;
                 r=\left(\begin{array}{c} 0 \\ 1 \\ 0 \\ 0 \end{array}\right), \;
                 s=\left(\begin{array}{c} 0 \\ 0 \\ 1 \\ 0 \end{array}\right), \;
                 d=\left(\begin{array}{c} 0 \\ 0 \\ 0 \\ 1 \end{array}\right),
             \end{equation}
             and the covariant relations~\eqref{vrsd-gamma}, \eqref{vrsd-delta} and \eqref{vrsd-epsilon} can be read off immediately.

             The vertex $\Gamma^{\mu\nu}(p,\Sigma,\Delta)$ consists of 128 independent Lorentz-Dirac basis elements.
             With the  relations above it is straightforward to construct a basis whose elements are
             \begin{itemize}
             \item orthonormal,
             \item symmetric or antisymmetric in the Lorentz indices,
             \item have a non-vanishing Lorentz trace or are traceless,
             \item have definite transformation properties under photon crossing and charge conjugation,
             \item fully factorize the Lorentz and Dirac structure.
             \end{itemize}
             The result is collected in Table~\ref{qcv-basis-1} and involves 16 Lorentz structures $\mathsf{X}_i^{\mu\nu}$ ($i=1\dots 16$)
             and eight Dirac structures $\tau_a$ ($a=1\dots 8$).
             The general basis element is constructed as:
             \begin{equation}\label{basis-element}
                 \tau_{i,a}^{\mu\nu}(r,s,d) = \mathsf{X}_i^{\mu\nu}(r,s,d)\left[\begin{array}{c} \mathds{1} \\ \gamma_5 \end{array}\right]_i \tau_a(r,s,d)\,,
             \end{equation}
             where the bracket means that the negative-parity
             elements which involve only one instance of the axialvector $v^\mu$ (i.e., those with $i=3,4,10,12,14,16$) must be equipped with a factor $\gamma_5$ to restore positive parity.
             The decomposition of the fermion Compton vertex can then be written as
             \begin{equation}\label{qcv-basis-decomposition}
                 \Gamma^{\mu\nu}(p,\Sigma,\Delta) = \sum_{i=1}^{16} \sum_{a=1}^8 f_{i,a}(\omega,\lambda)\,\tau_{i,a}^{\mu\nu}(r,s,d)\,,
             \end{equation}
             where the 128 dressing functions $f_{i,a}$ depend on the six Lorentz-invariants that appear in the sets $\omega$ and $\lambda$,
             and the basis elements depend on the three orthonormal momenta $r$, $s$, $d$.
             The advantage of this construction is that Lorentz-invariant relations between the basis elements become very simple:
             all inherent scalar products are either zero or one, and Lorentz traces can be conveniently separated from Dirac traces.

             The column labels $\mathsf{S}$ and $\mathsf{T}$ in Table~\ref{qcv-basis-1} define orthogonal subspaces with respect to the Lorentz indices:
             the elements with $\mathsf{S}=\pm$ are symmetric or antisymmetric in the Lorentz indices,
             those with $\mathsf{T}=\bullet$ have a nonvanishing Lorentz trace, and the ones with $\mathsf{T}=\circ$ are traceless.
             We show in App.~\ref{sec:ibse-kernel} that the three possible $\{ \mathsf{S}, \mathsf{T} \}$
             combinations define different subspaces which decouple in the inhomogeneous BSE for the quark Compton vertex.
             In fact, all operations on the Lorentz indices and/or external momenta $s^\mu$ and $d^\mu$ leave the relative-momentum integration unchanged,
             and the basis in Table~\ref{qcv-basis-1} leads to the maximum number of decoupled equations, namely 11.

             The remaining columns labeled $\mathsf{B}$ and $\mathsf{C}$ correspond to the Bose and charge-conjugation symmetry properties
             of the basis elements.
             The full vertex must be Bose-symmetric (i.e., invariant under photon crossing) and invariant under charge conjugation:
             \begin{align}
                 \Gamma^{\mu\nu}(p,\Sigma,\Delta) &\stackrel{!}{=} \Gamma^{\nu\mu}(p,-\Sigma,\Delta)\,, \label{Bose-symmetry}\\
                 \conjg{\Gamma}^{\mu\nu}(p,\Sigma,\Delta) &= C\,\Gamma^{\nu\mu}(-p,-\Sigma,-\Delta)^T C^T \nonumber \\
                                                          &\stackrel{!}{=} \Gamma^{\mu\nu}(p,\Sigma,-\Delta)\,. \label{CC-symmetry}
             \end{align}
             The basis elements in Table~\ref{qcv-basis-1} are either symmetric or antisymmetric under either of these operations.
             (For example: the element $\mathsf{X}_{9}^{\mu\nu}$ is Bose-antisymmetric but symmetric under charge conjugation.)
             The symmetry of the full basis element $\tau_{i,a}^{\mu\nu}$ of Eq.~\eqref{basis-element} is the combination of the individual symmetries for $\mathsf{X}_i^{\mu\nu}$
             and $\tau_a$ (e.g., $\tau_{9,3}^{\mu\nu}$ is both Bose- and charge-conjugation symmetric).
             An exception are the opposite-parity elements which carry an additional factor $\gamma_5$: when combined with any of the spin structures $a=2,3,4,8$, they switch
             $C-$parity because $\gamma_5$ has to be permuted back through an odd number of $\gamma$ matrices.

    \renewcommand{\arraystretch}{1.0}

             Furthermore, the momentum variables $y$ and $Z$ from Eq.~\eqref{qcv-variables} flip their sign under photon crossing,
             and $z$ and $Z$ switch sign under charge conjugation. The variables $p^2$, $t$ and $\sigma$ are invariant under both.
             The invariance of the full Compton vertex under both operations implies that each basis element
             in Table~\ref{qcv-basis-1} with symmetry $\mathsf{B} \mathsf{C}$ must be matched
             by an appropriate dressing function $f_{i,a}$ with the same symmetry.
             This yields the following symmetry properties for the dressing functions $f_{i,a}$
             for given $\mathsf{B},\mathsf{C}\in \{\pm 1\}$:
             \begin{equation}\label{BC-symmetry-dressings}
                 f(z,y,-Z) = \mathsf{B}\,f(z,-y,Z) = \mathsf{C}\,f(-z,y,Z)\,,
             \end{equation}
             where we omitted the labels $i,a$ and the momentum dependencies in $p^2$, $t$ and $\sigma$.
             Eq.~\eqref{BC-symmetry-dressings} entails that the dressing functions can be factorized into angular prefactors which carry the symmetry
             and remainders which are completely symmetric in all variables. For each $\mathsf{B} \mathsf{C}$ combination
             there are two possible choices for these prefactors:
             \begin{equation}\label{BC-symmetry-prefactors}
             \begin{split}
                  \mathsf{B} \mathsf{C} = ++ \quad & \Leftrightarrow \quad 1\quad\text{or}\quad yzZ\,, \\
                  \mathsf{B} \mathsf{C} = -+ \quad & \Leftrightarrow \quad y\quad\text{or}\quad zZ\,, \\
                  \mathsf{B} \mathsf{C} = +- \quad & \Leftrightarrow \quad z\quad\text{or}\quad yZ\,, \\
                  \mathsf{B} \mathsf{C} = -- \quad & \Leftrightarrow \quad Z\quad\text{or}\quad yz\,.
             \end{split}
             \end{equation}

             Finally, the basis has the advantage of being orthonormal. The orthonormality relation depends on the charge-conjugated basis elements, defined by
             \begin{equation}
             \begin{split}
                 &\conjg{\tau}_{i,a}^{\nu\mu}(r,s,d) = C\,\tau_{i,a}^{\mu\nu}(-r,-s,-d)^T C^T \\
                 & = \mathsf{X}_i^{\mu\nu}(r,s,d)\,  C\,\tau_a(-r,-s,-d)^T C^T \left[\begin{array}{c} \mathds{1} \\ \gamma_5 \end{array}\right]_i \\
                 & = \mathsf{X}_i^{\mu\nu}(r,s,d)\, \conjg{\tau}_a(r,s,d) \left[\begin{array}{c} \mathds{1} \\ \gamma_5 \end{array}\right]_i\,.
             \end{split}
             \end{equation}
             It is given by
             \begin{equation}\label{qcv-orthonormality}
             \begin{split}
                  \tfrac{1}{4} \text{Tr} \left\{ \conjg{\tau}_{i,a}^{\nu\mu}\, \tau_{j,b}^{\mu\nu}\right\}
                 &=  \mathsf{X}_i^{\mu\nu} \mathsf{X}_j^{\mu\nu} \,\tfrac{1}{4}\, \text{Tr} \left\{ \conjg{\tau}_a
                      \left[\begin{array}{c} \mathds{1} \\ \gamma_5 \end{array}\right]_i
                      \left[\begin{array}{c} \mathds{1} \\ \gamma_5 \end{array}\right]_j
                      \tau_b \right\} \\
                 &= \delta_{ij}\,\tfrac{1}{4}\, \text{Tr} \left\{ \conjg{\tau}_a \,\tau_b \right\} = \delta_{ij}\,\delta_{ab}\,.
             \end{split}
             \end{equation}
             Since the $\mathsf{X}_i$ are orthogonal by themselves, $\mathsf{X}_i^{\mu\nu} \mathsf{X}_j^{\mu\nu}=\delta_{ij}$,
             the product of the square brackets is the unit matrix.

             As the basis vectors are normalized to unity and contain transverse projectors with respect to $\Delta$ and $\Sigma_T$,
             one will encounter kinematic constraints in the limits $\Delta^2=0$, $\Sigma_T^2=0$ and $p_t^2=0$. This is just the boundary of the spacelike region
             where any of the variables $t$, $\sigma$ or $p^2$ vanishes or $Z$, $z$ or $y$ become $\pm 1$.
             Hence, the dressing functions in those limits are interrelated and/or can have kinematical zeros.
             This is not a problem in the practical numerical calculation as long as we 
             avoid to work in these exact kinematical limits but rather approach them from the interior of the spacelike region.
             Nevertheless, it may affect the numerical accuracy in their vicinity.

            So far we have not worked out the transversality of the photon at the level of the basis elements of the Compton vertex.
            This is an important issue since electromagnetic gauge invariance entails that the onshell nucleon Compton amplitude is fully transverse,
            i.e., transverse with respect to $Q'$ in the index $\mu$ and $Q$ in the index $\nu$.
            Moreover, also the quark Compton vertex can be split into a piece that satisfies its Ward-Takahashi identity (containing longitudinal and transverse parts) 
            plus a remainder that is purely transverse, cf.~Sec.~\ref{sec:wti}.
            Since the complete dynamics of the system is encoded in the transverse pieces, it is
            in principle sufficient to work with a transverse basis only.
            The distinction between transverse and longitudinal remains intact when the quark Compton vertex is implemented in the nucleon Compton diagrams,
            and electromagnetic gauge invariance requires that the sum of all longitudinal diagrams must cancel at the nucleon level.

            In Sec.~\ref{sec:fully-transverse} we will construct a fully transverse basis that is free of kinematic singularities; unfortunately
            that basis is no longer orthogonal and therefore cumbersome in the numerical calculation.
            Here we will pursue a different goal and construct a transverse basis from Table~\ref{qcv-basis-1}.
             Since those basis elements 
             factorize the Dirac from the Lorentz structure,
             it is sufficient to analyze the transversality relations at the level of the 16 Lorentz structures.
             The transverse part satisfies the conditions ${Q'}^\mu\,\Gamma^{\mu\nu} = 0$ and $Q^\nu\,\Gamma^{\mu\nu} = 0$
             which can be worked out to construct a fully transverse, orthonormal basis that consists of 9 elements, see App.~\ref{qcv-basis-LT}:
             \begin{equation}\label{qcv-transverse-basis-Y}
             \begin{split}
                   \mathsf{Y}_1 &= \mathsf{X}_1 \,, \\
                   \mathsf{Y}_2 &= \mathsf{X}_2 \,, \\
                   \mathsf{Y}_3 &= \frac{(1-X)(\mathsf{X}_5+\mathsf{X}_6) -2b\,(\mathsf{X}_8 -a \,\mathsf{X}_7- b\,\mathsf{X}_6)}{\sqrt{2\,n_1 n_2}} \,, \\
                   \mathsf{Y}_4 &= \mathsf{X}_4 \,, \\
                   \mathsf{Y}_5 &= \mathsf{X}_3 \,, \\[1mm]
                   \mathsf{Y}_6 &=  \frac{1}{\sqrt{n_1}} \left(\mathsf{X}_{11}+a \,\mathsf{X}_9-b \,\mathsf{X}_{13}\right)  \\
                   \mathsf{Y}_7 &=  \frac{1}{\sqrt{n_2}} \left(\mathsf{X}_{9}+a  \,\mathsf{X}_{11}-b \,\mathsf{X}_{15}\right) - \frac{2a\,\mathsf{Y}_6}{\sqrt{n_1 n_2}}\,,\\[1mm]
                   \mathsf{Y}_8 &= \frac{1}{\sqrt{n_1}} \left(\mathsf{X}_{12}+a  \,\mathsf{X}_{10}-b \,\mathsf{X}_{14}\right),\\
                   \mathsf{Y}_9 &= \frac{1}{\sqrt{n_2}} \left(\mathsf{X}_{10}+a  \,\mathsf{X}_{12}-b \,\mathsf{X}_{16}\right)- \frac{2a\,\mathsf{Y}_8}{\sqrt{n_1 n_2}}\,. \\[2mm]
             \end{split}
             \end{equation}
             We have abbreviated
             \begin{equation}
                 a = \sqrt{X}\,Z\,, \quad b = \sqrt{X}\,\sqrt{1-Z^2}
             \end{equation}
             and
             \[ n_1 = 1 + X\,, \qquad n_2 = n_1 - \frac{4a^2 }{n_1}\,.\]
             The new basis elements $\mathsf{Y}_i$ can now have kinematic singularities if $n_1=0$ or $n_2=0$.
             From Eq.~\eqref{Z-in-VCS} we see that $n_2=0$ in virtual and timelike Compton scattering while $n_1$ vanishes for real Compton scattering, cf. Fig.~\ref{fig:phasespace}.
             The respective dressing functions must have nodes in these limits so that the singular basis elements decouple.
             For example, the elements $\mathsf{Y}_{3,7,9}$ would decouple in the VCS limit, so that only six independent (Lorentz) basis elements survive.

             The $\mathsf{BC}$ symmetries for the $\mathsf{Y}_i$ are inherited from the $\mathsf{X}_i$;
             note that the variable $a$ switches sign both under charge conjugation and photon crossing.
             The full transverse basis is then constructed in complete analogy
             to Eqs.~\eqref{basis-element}--\eqref{qcv-basis-decomposition}, namely by multiplying the $\mathsf{Y}_i$ with the Dirac structures $\tau_a$
             from Table~\ref{qcv-basis-1} and with appropriate $\gamma_5$ insertions for the elements $i=4,5,8,9$.
             Therefore, we arrive in total
             at $9\times 8 =72$ independent transverse Lorentz-Dirac elements.
             For completeness, the remaining $7\times 8=56$ longitudinal and transverse-longitudinal basis elements $\mathsf{Y}_{10 \dots 16}^{\mu\nu}$ are worked out in App.~\ref{qcv-basis-LT}.
             The full vertex is expressed as
             \begin{equation}\label{qcv-basis-decomposition-Y}
                 \widetilde{\Gamma}^{\mu\nu}(p,\Sigma,\Delta) = \sum_{i=1}^{16} \sum_{a=1}^8 \tilde{f}_{i,a}(\omega,\lambda)\,   \mathsf{Y}_i^{\mu\nu}\left[\begin{array}{c} \mathds{1} \\ \gamma_5 \end{array}\right]_i \tau_a(r,s,d)\,,
             \end{equation}
             with $\gamma_5$ insertions for $i=4,5,8,9,15,16$.

   \renewcommand{\arraystretch}{1.5}

             \begin{table}[t]

                \begin{center}
                \begin{tabular}{   | @{\quad} c @{\quad} | c | } \cline{2-2}
                   \multicolumn{1}{c|}{}     &  \, $\mathsf{B}\,\mathsf{C}$ \quad \\ \hline \rule{0mm}{0.5cm}
                 $\mathsf{Y}_1$                       &  $++$  \\
                 $\mathsf{Y}_2$                       &  $++$  \\
                 $\mathsf{Y}_3$                       &  $++$  \\ [2mm]
                 $\mathsf{Y}_7\,\Slash{s}$            &  $++$  \\
                 $\mathsf{Y}_4\,\gamma_5$             &  $++$  \\
                 $\mathsf{Y}_5\,\gamma_5\,\Slash{s}$  &  $++$  \\ [2mm] \hline

                \end{tabular} \hspace{2mm}
                \begin{tabular}{   | @{\quad} c @{\quad}| c | } \cline{2-2}
                   \multicolumn{1}{c|}{}     &  \, $\mathsf{B}\,\mathsf{C}$ \quad \\ \hline \rule{0mm}{0.5cm}
                 $\mathsf{Y}_1\,\Slash{s}$            &  $-+$  \\
                 $\mathsf{Y}_2\,\Slash{s}$            &  $-+$  \\
                 $\mathsf{Y}_3\,\Slash{s}$            &  $-+$  \\ [2mm]
                 $\mathsf{Y}_7$                       &  $-+$  \\
                 $\mathsf{Y}_8\,\gamma_5$             &  $-+$  \\
                 $\mathsf{Y}_9\,\gamma_5\,\Slash{s}$  &  $-+$  \\ [2mm] \hline

                \end{tabular} \hspace{2mm}
                \begin{tabular}{   | @{\quad} c @{\quad}| c | } \cline{2-2}
                   \multicolumn{1}{c|}{}    &  \, $\mathsf{B}\,\mathsf{C}$ \quad \\ \hline \rule{0mm}{0.5cm}
                 $\mathsf{Y}_6$                      &  $+-$  \\
                 $\mathsf{Y}_8\,\gamma_5\,\Slash{s}$   &  $+-$  \\
                 $\mathsf{Y}_9\,\gamma_5$            &  $+-$  \\ [2mm]
                 $\mathsf{Y}_6\,\Slash{s}$           &  $--$  \\
                 $\mathsf{Y}_4\,\gamma_5\,\Slash{s}$   &  $--$  \\
                 $\mathsf{Y}_5\,\gamma_5$            &  $--$  \\ [2mm] \hline
                \end{tabular}
                \end{center}

               \caption{Transverse, orthonormal tensor basis for the onshell nucleon Compton amplitude, together with the photon-crossing and charge-conjugation symmetries for each element.
                        The tensor structures $\mathsf{Y}_i$ correspond to the dressing functions $F_i$ in Eq.~\eqref{qcv-nucleon-transverse}  and the structures $\mathsf{Y}_i\,\slashed{s}$ to the functions $G_i$.}
                        \label{qcv-basis-2}

        \end{table}

    \renewcommand{\arraystretch}{1.0}

  \subsection{Nucleon Compton amplitude} \label{sec:tensorbasis-onshell}

             Next, we apply our general considerations to the special case of the nucleon's Compton amplitude.
             To this end we replace the relative momentum $p$ again by the average nucleon momentum $P$ and the variable $y$ by $Y$.
             The onshell constraints $P_i^2=P_f^2=-M^2$ entail
             \begin{equation}\label{onshell-1}
                 P^2=-M^2(1+t)\,, \qquad
                 z=\hat{P}\cdot\hat{\Delta}=0\,,
             \end{equation}
             so that there are four instead of six
             independent Lorentz-invariants. $P$ is now already transverse to $\Delta$, i.e., $P_T=P$.
             Sandwiching the vertex between onshell spinors is equivalent to applying
             the positive-energy projectors
             \begin{equation}\label{positive-energy-projectors}
                 \Lambda_+^f := \frac{\mathds{1}+\widehat{\Slash{P}}_f}{2}\,, \qquad
                 \Lambda_+^i := \frac{\mathds{1}+\widehat{\Slash{P}}_i}{2}
             \end{equation}
             from left and right, so that the matrix-valued nucleon Compton amplitude has the structure
             \begin{equation}\label{qcv-nucleon}
                 \widetilde{J}^{\mu\nu}(P,\Sigma,\Delta) = \Lambda_+^f \,\Gamma^{\mu\nu}(P,\Sigma,\Delta)\,\Lambda_+^i\Big|_\text{Eq.~\eqref{onshell-1}}\,.
             \end{equation}

             Acting with the projectors on the Dirac basis elements $\tau_a(r,s,d)$ leaves only two independent structures,
             namely $\mathds{1}$ and $\Slash{s}$ (or, equivalently, $\mathds{1}$ and $\Slash{\Sigma}$), whereas
             the remaining ones are either zero or become linearly dependent via the Dirac equations
             \begin{equation}
                 \Lambda_+^f \Slash{P}_f = iM \Lambda_+^f\,, \qquad \Slash{P}_i \,\Lambda_+^i = iM\Lambda_+^i\,.
             \end{equation}
             From Table~\ref{qcv-basis-1}, this leaves $16\times 2=32$ independent tensor structures.
             In contrast to a general fermion Compton vertex, the onshell nucleon Compton amplitude is fully transverse, cf.~Eq.~\eqref{ncv-wti} below.
             The transversality condition reduces this number to $9\times 2=18$, and it is sufficient to work with the $\mathsf{Y}_i$ given in Eq.~\eqref{qcv-transverse-basis-Y}
             and Table~\ref{qcv-basis-2}.
             The resulting Compton amplitude assumes the form
             \begin{equation}\label{qcv-nucleon-transverse}
                 \widetilde{J}^{\mu\nu}(P,\Sigma,\Delta) = \sum_{i=1}^{9} \mathsf{Y}_i^{\mu\nu} \Lambda_+^f \left[\begin{array}{c} \mathds{1} \\ \gamma_5 \end{array}\right]_i
                 \left( F_i + G_i \, \Slash{s}\right)\Lambda_+^i\,,
             \end{equation}
             where the $F_i$ and $G_i$ are functions of $t$, $X$, $Y$ and $Z$. All other symmetry relations
             are maintained, and Eq.~\eqref{BC-symmetry-dressings} simplifies to
             \begin{equation}
                 f(Y,-Z) = \mathsf{B}\,f(-Y,Z) = \mathsf{C}\,f(Y,Z)
             \end{equation}
             for $f \in \{ F_i, G_i\}$. The possible symmetry prefactors of Eq.~\eqref{BC-symmetry-prefactors} become:
             \begin{equation}
             \begin{split}
                  \mathsf{B} \mathsf{C} = ++ \quad & \Leftrightarrow \quad 1 \,, \\
                  \mathsf{B} \mathsf{C} = -+ \quad & \Leftrightarrow \quad Y \,, \\
                  \mathsf{B} \mathsf{C} = +- \quad & \Leftrightarrow \quad YZ \,, \\
                  \mathsf{B} \mathsf{C} = -- \quad & \Leftrightarrow \quad Z\ \,.
             \end{split}
             \end{equation}
             These relations have also been studied in the context of dispersion relations to determine the low-energy behavior of the individual dressing functions of the
             Compton amplitude~\cite{Drechsel:1997xv,Pasquini:2001yy,Drechsel:2002ar,Gorchtein:2009wz}.
             (Note that from Eqs.~\eqref{crossing-variable} and \eqref{XYZ-from-standard}: $Z\sim Q^2-{Q'}^2$ and $Y\sim P\cdot\Sigma=P\cdot Q = P\cdot Q'$).


\section{Ward-Takahashi identity and transversality}  \label{sec:wti}

      The basis construction of the previous section is convenient for studying the general properties of the Compton vertex
      and also for the numerical implementation. However, it is less useful for working out the properties related to electromagnetic gauge invariance,
      i.e., the Ward-Takahashi identity, the transversality conditions and related analyticity constraints.
      We will deal with these issues in the following and decompose the Compton vertex into a 'gauge part', i.e.,
      the contribution that is fixed by its Ward-Takahashi identity, and further transverse pieces that are free of kinematic singularities and constraints.

  \subsection{Fermion-photon vertex}

             We start with a discussion of the fermion-photon vertex as it provides the template for the two-photon case.
             It satisfies the Ward-Takahashi identity
                  \begin{equation}\label{qpv-wti}
                      Q^\mu \,\Gamma^\mu (k,Q) = S^{-1}(k_+)-S^{-1}(k_-)\,,
                  \end{equation}
             where $Q$ is the photon momentum, $k$ is the relative momentum of the quark, and $k_\pm = k \pm Q/2$ are the quark momenta.
             The inverse dressed quark propagator reads
             \begin{equation}\label{quark-propagator}
                  S^{-1}(k) =   i\Slash{k}\,A(k^2) + B(k^2) \,,
             \end{equation}
             and the renormalization-point independent mass function of the fermion is given by $M(k^2)=B(k^2)/A(k^2)$.
             Eq.~\eqref{qpv-wti} is solved by the Ball-Chiu vertex~\cite{Ball:1980ay}
                  \begin{equation}\label{vertex:BC}
                      \Gamma_\text{BC}^\mu(k,Q) =   i\gamma^\mu\,\Sigma_A + 2 k^\mu (i\Slash{k}\, \Delta_A  + \Delta_B),
                  \end{equation}
             where the functions
                 \begin{equation}\label{QPV:sigma,delta}
                 \begin{split}
                     \Sigma_A(k,Q) &:= \frac{A(k_+^2)+A(k_-^2)}{2} ,\\
                     \Delta_A(k,Q) &:= \frac{A(k_+^2)-A(k_-^2)}{k_+^2-k_-^2}, \\
                     \Delta_B(k,Q) &:= \frac{B(k_+^2)-B(k_-^2)}{k_+^2-k_-^2}
                 \end{split}
                 \end{equation}
             are completely determined by the dressed fermion propagator and free of kinematic singularities.

             The full vertex is then the sum of the Ball-Chiu part and a transverse piece that is not constrained by the WTI:
                  \begin{equation}\label{vertex:qpv}
                      \Gamma^\mu(k,Q) =  \Gamma_\text{BC}^\mu(k,Q) +  \Gamma^{\mu}_\text{T}(k,Q)\,.
                  \end{equation}
            $\Gamma^\mu_\text{T}$ consists of eight independent tensor structures.
            Analyticity at vanishing photon momentum requires $\Gamma^{\mu}_T$ to vanish in the limit $Q^\mu=0$, either
            via appropriate momentum dependencies of the basis elements, vanishing dressing functions, or kinematic relations between the dressing functions in that limit.
            In order to find eight kinematically independent dressing functions,
            we want to express $\Gamma^{\mu}_T$ in a basis that is
            free of kinematic singularities and 'minimal' with respect to its powers in the photon momentum.
            Since the construction of the two-photon vertex is closely related to the one-photon case, we illustrate the problem here in detail.

    \renewcommand{\arraystretch}{1.2}

            The general fermion-photon vertex with quantum numbers $J^{PC}=1^{--}$ vertex consists of 12 tensor structures
            which can be chosen as
            \begin{equation}\label{qpv-general-basis}
            \begin{array}{ll}
               (+) & \gamma^\mu    \\
               (-) & \left[ \gamma^\mu, \slashed{k} \right]   \\
               (+) & \left[ \gamma^\mu, \slashed{Q} \right] \\
               (+) & \left[ \gamma^\mu, \slashed{k}, \slashed{Q} \right]
            \end{array}\quad
            \begin{array}{ll}
               (+) & k^\mu \\
               (+) & k^\mu \slashed{k}\\
               (-) & k^\mu \slashed{Q}\\
               (+) & k^\mu [ \slashed{k},\slashed{Q}]
            \end{array}\quad
            \begin{array}{ll}
               (-) & Q^\mu\\
               (-) & Q^\mu \slashed{k}  \\
               (+) & Q^\mu \slashed{Q}  \\
               (-) & Q^\mu [ \slashed{k},\slashed{Q}] .
            \end{array}
            \end{equation}
            To ensure definite charge-conjugation symmetry (indicated by the signs in the brackets)
            we have used the commutator for the product of two $\gamma$ matrices and the totally antisymmetric combination
             \begin{equation}\label{generalized-commutator}
                 \left[A,B,C\right] := \left[A,B\right] C + \left[B,C\right]A+\left[C,A\right]B
             \end{equation}
            for three $\gamma$ matrices.
            If the odd basis tensors are multiplied with a factor $k\cdot Q$, the full vertex satisfies
            \begin{equation}
                \conjg{\Gamma}^\mu(k,Q) = C\,\Gamma^\mu(-k,-Q)^T C^T = - \Gamma^\mu(k,-Q)
            \end{equation}
            with scalar dressing functions that are even in $k\cdot Q$.

            The transverse part of the vertex consists of eight tensor structures that are constructed from Eq.~\eqref{qpv-general-basis}.
            The two elements $[\gamma^\mu, \slashed{Q}]$ and $[ \gamma^\mu, \slashed{k}, \slashed{Q} ]$ are transverse by themselves.
            In principle one could apply the transverse projector
            \begin{equation}\label{transverse-projector}
               T_Q^{\mu\nu} = \delta^{\mu\nu} - \frac{Q^\mu Q^\nu}{Q^2}
            \end{equation}
            to the remaining elements from the first two columns of Eq.~\eqref{qpv-general-basis} to obtain the basis decomposition
             \begin{equation}\label{transverse-vertex}
             \begin{split}
                    -i \Gamma^\mu_T &=  g_1 \gamma^\mu_T + g_2\,k\!\cdot\! Q\,\tfrac{i}{2}\,[\gamma^\mu_T, \slashed{k}] \\
                                  & + g_3\,\tfrac{i}{2}\,[\gamma^\mu,\slashed{Q}] + g_4\,\tfrac{1}{6}\,[\gamma^\mu, \slashed{k}, \slashed{Q}] \\
                                  & + k^\mu_T \,\big( ig_5 + g_6\,\slashed{k}  + g_7\,k\! \cdot\!  Q\,\slashed{Q}   + g_8 \,\tfrac{i}{2}\,[\slashed{k}, \slashed{Q}] \big) \,,
             \end{split}
             \end{equation}
             where
             \begin{equation}
                 \gamma^\mu_T =  T_Q^{\mu\nu} \gamma^\nu\,, \qquad
                 k^\mu_T = T_Q^{\mu\nu} k^\nu \,.
             \end{equation}
             We have attached prefactors so that the scalar dressing functions $g_i(k^2, \,k\cdot Q, \,Q^2)$ are
             even in $k\cdot Q$ and real for $k^2 > 0$, $Q^2 \in \mathds{R}$.
            However, since the projector~\eqref{transverse-projector} contains a kinematic singularity at $Q^2 \rightarrow 0$, the resulting dressing functions
            are kinematically dependent: the four combinations
            \begin{equation}\label{qpv:kinematic-relations}
                g_1 + (k\cdot Q)^2 g_7\,, \quad
                g_2 - g_8 \,, \quad
                g_5\,, \quad
                g_6
            \end{equation}
            must vanish with $Q^2$ for $Q^2 \rightarrow 0$.
            Instead of the projector~\eqref{transverse-projector} one could equally apply $Q^2 \,T^{\mu\nu}_Q$ which has no kinematic singularity;
            unfortunately this overcompensates the problem since $g_1$, $g_2$, $g_7$, $g_8$ do not need to vanish individually when $Q^2$ goes to zero.

            A basis decomposition where all dressing functions are truly kinematically independent is given by~\cite{Skullerud:2002ge,Williams:2007zzh,Alkofer:2008tt}
             \begin{equation}\label{transverse-vertex-no-kin-sing}
             \begin{split}
                   -i\Gamma^\mu_T &=  f_1\,Q^2\,\gamma^\mu_T + f_2\,k\!\cdot\! Q\,Q^2\,\tfrac{i}{2}\,[\gamma^\mu_T, \slashed{k}] \\
                                  & + f_3\,\tfrac{i}{2}\,[\gamma^\mu,\slashed{Q}] + f_4\,\tfrac{1}{6}\,[\gamma^\mu, \slashed{k}, \slashed{Q}] \\
                                  & + if_5\,Q^2\,k^\mu_T  + f_6\,Q^2\,k^\mu_T\,\slashed{k}   \\
                                  & + f_7\,k\! \cdot\!  Q\,(k\!\cdot\!Q\,\gamma^\mu-k^\mu \slashed{Q})  \\
                                  & + f_8 \,\tfrac{i}{2}\,[k\!\cdot\!Q\,\gamma^\mu - k^\mu \slashed{Q}, \slashed{k}] .
             \end{split}
             \end{equation}
             It satisfies the requirements of Eq.~\eqref{qpv:kinematic-relations} since
             \begin{equation}
                 \begin{array}{rl}
                     f_1\,Q^2 &=  g_1 + (k\cdot Q)^2 g_7\,,  \\
                     f_2\,Q^2 &=  g_2-g_8\,,  \\
                     f_3 &= g_3\,, \\
                     f_4 &= g_4\,,
                 \end{array}\qquad
                 \begin{array}{rl}
                     f_5\,Q^2 &=  g_5\,, \\
                     f_6\,Q^2 &=  g_6\,,  \\
                    -f_7 &= g_7\,, \\
                     f_8 &= g_8\,.
                 \end{array}
             \end{equation}
             Apart from global factors $k\cdot Q$, the four tensor structures corresponding to $f_{3,4,7,8}$ are linear
             and the remaining four are quadratic in the photon momentum.

             The question remains whether Eq.~\eqref{transverse-vertex-no-kin-sing} can be obtained from a systematic construction principle.
             To this end we define the quantities
             \begin{equation}\label{new-transverse-projectors}
             \begin{split}
                   t_{ab}^{\mu\nu} &:= a\cdot b\,\delta^{\mu\nu} - b^\mu a^\nu\,,  \\
                   \varepsilon^{\mu\nu}_{ab} &:= \gamma_5\,\varepsilon^{\mu\nu\alpha\beta}a^\alpha b^\beta\,,
             \end{split}
             \end{equation}
             with $a^\mu, b^\mu \in \{ k^\mu, \, Q^\mu \}$. They are both regular in the limits $a\rightarrow 0$ or $b\rightarrow 0$.
             $t_{ab}^{\mu\nu}$ is transverse to $a^\mu$ and $b^\nu$,
             \begin{equation}
                 a^\mu \, t_{ab}^{\mu\nu} = 0\,, \qquad t_{ab}^{\mu\nu}\,b^\nu = 0\,,
             \end{equation}
             whereas $\varepsilon^{\mu\nu}_{ab}$ is transverse to $a$ and $b$ in both Lorentz indices.
             The usual transverse projectors can thus be written as $T^{\mu\nu}_Q=t^{\mu\nu}_{QQ}/Q^2$ and
             $T^{\mu\nu}_{Q'}=t^{\mu\nu}_{Q'Q'}/{Q'}^2$.

             With the help of these definitions one can generate the basis~\eqref{transverse-vertex-no-kin-sing} as follows.
             Take the four tensor structures that are independent of the photon momentum:
             \begin{equation}\label{qpv-momentum-independent-4}
                \gamma^\nu, \quad
                [\gamma^\nu, \Slash{k}]\,, \quad
                k^\nu, \quad
                k^\nu \Slash{k}\,.
             \end{equation}
             Contract them with $t^{\mu\nu}_{QQ}$, $t^{\mu\nu}_{Qk}$ and $\varepsilon^{\mu\nu}_{Qk}$ to generate
             eight transverse basis elements that are kinematically independent and linear or quadratic in the four-momentum $Q^\mu$:
             \begin{equation}
             \begin{split}
                    t_{QQ}^{\mu\nu}\left\{ \begin{array}{c}
                                           \gamma^\nu   \\ \left[\gamma^\nu,\Slash{k}\right]   \\  k^\nu   \\   k^\nu \Slash{k}
                                           \end{array}\right\} &=
                                Q^2 \left\{ \begin{array}{c}
                                           \gamma^\mu_T \\ \left[\gamma^\mu_T,\Slash{k}\right] \\  k^\mu_T \\   k^\mu_T \,\slashed{k}
                                           \end{array}\right\}, \\
                    t_{Qk}^{\mu\nu}\left\{ \begin{array}{c}
                                           \gamma^\nu   \\ \left[\gamma^\nu,\Slash{k}\right]
                                           \end{array}\right\} &=
                                    \left\{ \begin{array}{c}
                                           k\!\cdot\!Q\,\gamma^\mu - k^\mu \slashed{Q}  \\ \left[ k\!\cdot\!Q\,\gamma^\mu - k^\mu \slashed{Q}, \slashed{k}\right]
                                           \end{array}\right\}, \\
                    \varepsilon_{Qk}^{\mu\nu} \left\{ \begin{array}{c}
                                           \gamma^\nu   \\ \left[\gamma^\nu,\Slash{k}\right]
                                           \end{array}\right\} &=
                                    \left\{ \begin{array}{c}
                                           \tfrac{1}{6} \left[\gamma^\mu, \Slash{k}, \Slash{Q} \right]  \\ t_{Qk}^{\mu\nu}\left[\gamma^\nu,\Slash{k}\right] -k^2 \left[ \gamma^\mu, \Slash{Q}\right]
                                           \end{array}\right\}.
             \end{split}
             \end{equation}
             Instead of using $t_{Qk}^{\mu\nu}$ and $\varepsilon_{Qk}^{\mu\nu}$, one could contract the four elements in Eq.~\eqref{qpv-momentum-independent-4} also with
             $t^{\mu\nu}_{Q\gamma}=\slashed{Q}\,\delta^{\mu\nu}-\gamma^\mu Q^\nu$ and use commutators where necessary.
             However, this does not generate any new elements:
             \begin{equation}
             \begin{split}
                    \tfrac{1}{2}\big[ t_{Q\gamma}^{\mu\nu},\gamma^\nu \big]             &= -\left[ \gamma^\mu, \Slash{Q}\right], \\
                    \tfrac{1}{2}\big[ t_{Q\gamma}^{\mu\nu},\gamma^\nu,\Slash{k} \big]    &= \left[\gamma^\mu, \Slash{k}, \Slash{Q} \right], \\
                    t_{Q\gamma}^{\mu\nu}\,k^\nu                            &= -4 \,t_{Qk}^{\mu\nu}\,\gamma^\nu\,, \\
                    \big[ t_{Q\gamma}^{\mu\nu} \,k^\nu, \Slash{k}\big] &= -t_{Qk}^{\mu\nu}\left[ \gamma^\nu, \Slash{k} \right].
             \end{split}
             \end{equation}
             Finally, attach appropriate factors $k \cdot  Q$ to ensure charge-conjugation invariance of the dressing functions.

     \begin{figure*}[t]
     \center{
     \includegraphics[scale=0.11]{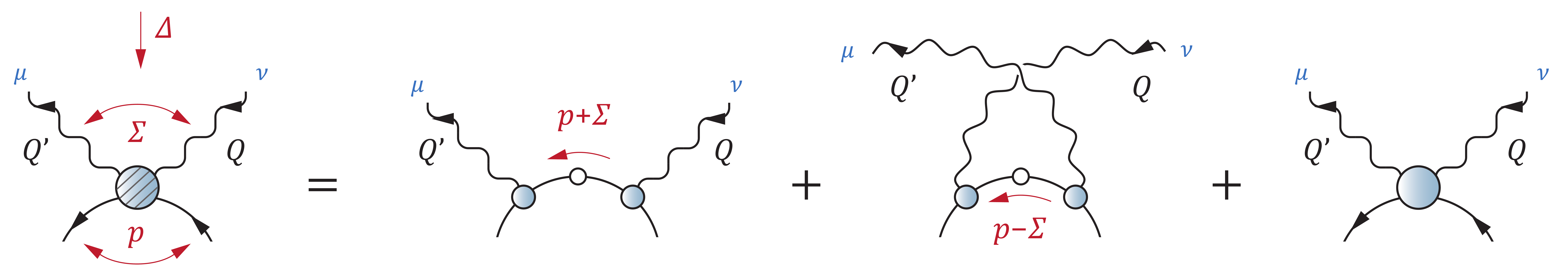}}
        \caption{Separation of the fermion Compton vertex into Born terms and a 1PI part. }
        \label{fig:qcv-born}
     \end{figure*}

    \renewcommand{\arraystretch}{1.4}

             We will henceforth use Eq.~\eqref{transverse-vertex-no-kin-sing} as our reference basis for the
             transverse part of the fermion-photon vertex.
             We write it in a compact way:
             \begin{equation}\label{qpv-newbasis}
             \begin{split}
                 \begin{array}{rl}
                 \tau_1^\mu &= t^{\mu\nu}_{QQ}\,\gamma^\nu\,, \\
                 \tau_2^\mu &= t^{\mu\nu}_{QQ}\,k\!\cdot\! Q\,  \tfrac{i}{2} [\gamma^\nu,\Slash{k}]\,, \\
                 \tau_3^\mu &= \tfrac{i}{2}\,[\gamma^\mu,\slashed{Q}]\,, \\
                 \tau_4^\mu &= \tfrac{1}{6}\,[\gamma^\mu, \slashed{k}, \slashed{Q}]\,,
                 \end{array}\quad
                 \begin{array}{rl}
                 \tau_5^\mu &= t^{\mu\nu}_{QQ}\,ik^\nu\,, \\
                 \tau_6^\mu &= t^{\mu\nu}_{QQ}\,k^\nu \Slash{k}\,, \\
                 \tau_7^\mu &= t^{\mu\nu}_{Qk}\,k\!\cdot\! Q\,\gamma^\nu\,, \\
                 \tau_8^\mu &= t^{\mu\nu}_{Qk}\,\tfrac{i}{2}\,[\gamma^\nu,\Slash{k}]\,.
                 \end{array}
             \end{split}
             \end{equation}
             The full vertex is thus given by
             Eq.~\eqref{vertex:qpv}, with the transverse part
             \begin{equation}\label{qpv-transverse-new}
                 -i\Gamma^\mu_T(k,Q) = \sum_{i=1}^8 f_i(k^2, k\cdot Q, Q^2)\, \tau_i^\mu(k,Q)\,.
             \end{equation}
             The dimensionful dressing functions $f_i(k^2, \,k\cdot Q, \,Q^2)$ are again
             even in $k\cdot Q$.
             They are kinematically independent
             and can remain constant at vanishing photon momentum.
    \renewcommand{\arraystretch}{1.2}
             The basis~\eqref{qpv-newbasis} is essentially identical to Eq.~(A.8) in Ref.~\cite{Skullerud:2002ge}
             and Eq.~(A2) in Ref.~\cite{Alkofer:2008tt}. The relations between our $\tau_i^\mu$ and the
             transverse tensor structures $T_i^\mu$ in those papers are
             \begin{equation}
                 \begin{array}{rl}
                    \tau_1 &= -T_3\,, \\
                    \tau_2 &= -\tfrac{1}{2}\,k\!\cdot\! Q\,T_4\,, \\
                    \tau_3 &= T_5\,, \\
                    \tau_4 &= T_8\,,
                 \end{array}\qquad
                 \begin{array}{rl}
                    \tau_5 &= T_1\,, \\
                    \tau_6 &= \tfrac{1}{2}\,T_2\,, \\
                    \tau_7 &= -\tfrac{1}{2}\,k\!\cdot\! Q\,T_6\,,\\
                    \tau_8 &= \tfrac{1}{2}\,T_7\,.
                 \end{array}
             \end{equation}
             The dressing functions associated with $\tau_3$ and $\tau_4$ contribute to the onshell anomalous magnetic moment, cf.~Ref.~\cite{Chang:2010hb} and Eq.~\eqref{qpv-onshell-dressing-functions} below,
             and $\tau_7$ constitutes the transverse part of the Curtis-Pennington vertex~\cite{Curtis:1990zs}.

             Finally, to obtain a connection with the nucleon's onshell current,
             we investigate the limit where the incoming and outgoing fermion lines are taken on the mass shell, i.e.,
             $k_\pm^2 = -m^2$ or
             \begin{equation}\label{onshell-limit}
                 k^2 = -m^2 -Q^2/4\,, \qquad k\cdot Q = 0\,.
             \end{equation}
             The onshell vertex
    \renewcommand{\arraystretch}{0.7}
             \begin{equation} \label{qpv-temp-onshell}
                 J^\mu(k,Q) =  \Lambda_+^f\,\Gamma^\mu(k,Q)\,\Lambda_+^i \Big|_\text{Eq.~\eqref{onshell-limit}}
             \end{equation}
             is sandwiched between Dirac spinors that are eigenvectors of the positive-energy projectors
    \renewcommand{\arraystretch}{1.2}
             \begin{equation}
                 \begin{array}{l}
                   \Lambda_+^f = \Lambda_+(k_+),  \\
                   \Lambda_+^i = \Lambda_+(k_-),
                 \end{array}\qquad
                 \Lambda_+(p)=\frac{\mathds{1}+\hat{\Slash{p}}}{2}\,.
             \end{equation}
             By virtue of the projectors, only two of the basis elements in Eq.~\eqref{qpv-newbasis} remain independent, and the vertex can be written in the standard form
             \begin{equation}\label{current-onshell-standard}
                 J^\mu(k,Q) =  i\Lambda_+^f\left( F_1\,\gamma^\mu + \frac{iF_2}{4m}\,[\gamma^\mu,\slashed{Q}]\right)\Lambda_+^i\,,
             \end{equation}
             where $F_1$, $F_2$ are dimensionless functions of $Q^2$ only. Via Eq.~\eqref{vertex:qpv} they
             consist of Ball-Chiu parts and transverse components
             which are related to the functions $\Sigma_A$, $\Delta_A$, $\Delta_B$ and $f_j$ in the onshell limit:
             \begin{equation}\label{qpv-onshell-dressing-functions}
             \begin{split}
                 F_1(Q^2) &= A(-m^2) + 2m \left[B'(-m^2)-m A'(-m^2)\right] \\
                          &+ Q^2 \left[ f_1 - m \,(f_5 + m f_6) -\frac{f_4-m f_8}{2} \right]\Bigg|_\text{Eq.~\eqref{onshell-limit}}, \\
                 \frac{F_2(Q^2)}{2m} &= f_3-m f_4-\left[B'(-m^2)-m A'(-m^2)\right] \\
                          &+ \frac{Q^2}{2} \left[ f_5 + m f_6 - \frac{f_8}{2}\right]\Bigg|_\text{Eq.~\eqref{onshell-limit}}  \,.
             \end{split}
             \end{equation}
             Here we have exploited the fact that $\Sigma_A$, $\Delta_A$, $\Delta_B$ simplify on the mass shell
             to $A(-m^2)$, $A'(-m^2)$ and $B'(-m^2)$, respectively.
             Since $m$ is defined via Eq.~\eqref{onshell-limit}, i.e., $m^2=-k_+^2=-k_-^2$, all relations hold even in the case where the fermion propagator
             has no timelike pole (although in that case Eq.~\eqref{qpv-temp-onshell} cannot be derived from a pole condition but is merely a definition).
             On the other hand, if a mass shell exists where the propagator behaves like a free particle,
             we have in addition $M(-m^2) = m$, $M'(-m^2)=0$ and $A(-m^2)=1$, and the square brackets $ B'-mA'$ in Eq.~\eqref{qpv-onshell-dressing-functions} vanish.
             The dressing functions $f_3$ and $f_4$ contribute to the fermion's anomalous magnetic moment.

    \renewcommand{\arraystretch}{1.1}

  \subsection{Compton vertex, Born terms and WTI}

             We proceed by generalizing our findings to the two-photon case.
             The 1PI part of the fermion Compton vertex satisfies the following Ward-Takahashi identity, see~\cite{Scherer:1996ux} and references therein:
             \begin{equation}\label{qcv-wti-r}
             \begin{split}
                 & {Q'}^\mu \,\Gamma^{\mu\nu}(p,\Sigma,\Delta) =  \Gamma^\nu(p_f^+,Q) - \Gamma^\nu(p_f^-,Q)\,,  \\
                 & Q^\nu \,\Gamma^{\mu\nu}(p,\Sigma,\Delta) =  \Gamma^\mu(p_i^+,-Q')- \Gamma^\mu(p_i^-,-Q')\,.
             \end{split}
             \end{equation}
             Here, $\Gamma^\mu(k,Q)$ is the dressed fermion-photon vertex that depends
             on a relative momentum $k$ and a total photon momentum $Q$. The relative momenta
             that appear in the fermion-photon vertices are
             \begin{equation}
                 p_f^\pm = p \pm \frac{Q'}{2}\,, \qquad
                 p_i^\pm = p \pm \frac{Q}{2}\,,
             \end{equation}
             and $p$ is the relative momentum in the Compton vertex.
             The second equation in Eq.~\eqref{qcv-wti-r} follows from the first one via Bose symmetry (replace $\mu \leftrightarrow \nu$, $Q \leftrightarrow -Q'$ and thus $\Sigma \leftrightarrow -\Sigma$,
             and use Eq.~\eqref{Bose-symmetry}).
             The full Compton vertex $\widetilde{\Gamma}^{\mu\nu} := \Gamma_\text{B}^{\mu\nu} + \Gamma^{\mu\nu}$ is the sum of the 1PI part and a one-particle reducible, Bose-symmetric and charge-conjugation invariant Born piece, cf.~Fig.~\ref{fig:qcv-born}:
             \begin{equation}\label{qcv-born}
             \begin{split}
                 \Gamma_\text{B}^{\mu\nu}  =& -\Gamma^\mu(p_i^+,-Q')\,S(p+\Sigma)\,\Gamma^\nu(p_f^+,Q) \\
                                                    & -\Gamma^\nu(p_f^-,Q)\,S(p-\Sigma)\,\Gamma^\mu(p_i^-,-Q')\,,
             \end{split}
             \end{equation}
             where $S(p)$ is the dressed fermion propagator.

             To obtain the WTI for the full vertex $\widetilde{\Gamma}^{\mu\nu}$, we insert the WTI for the fermion-photon vertex,
                  \begin{equation} 
                      Q^\mu \,\Gamma^\mu (k,Q) = S^{-1}(k_+)-S^{-1}(k_-)
                  \end{equation}
              with $k_\pm = k \pm Q/2$, in Eq.~\eqref{qcv-born}:
             \begin{equation*}\label{qcv-wti4}
             \begin{split}
                 & {Q'}^\mu \,\Gamma^{\mu\nu}_\text{B}(p,\Sigma,\Delta) = \\[2mm]
                 & = \left[S^{-1}(p+\tfrac{\Delta}{2})-S^{-1}(p+\Sigma)\right] S(p+\Sigma)\,\Gamma^\nu(p_f^+,Q) \\
                 & \quad + \Gamma^\nu(p_f^-,Q) \,S(p- \Sigma) \left[S^{-1}(p-\Sigma)-S^{-1}(p-\tfrac{\Delta}{2})\right].
             \end{split}
             \end{equation*}
             We expressed the photon momenta $Q$ and $Q'$ by the average photon momentum $\Sigma$ and the momentum transfer $\Delta$.
             Adding Eq.~\eqref{qcv-wti-r} yields the result
             \begin{equation}\label{qcv-wti-full}
             \begin{split}
                 & {Q'}^\mu \,\widetilde{\Gamma}^{\mu\nu}(p,\Sigma,\Delta) = \\
                 & \quad + S^{-1}(p+\tfrac{\Delta}{2})\,S(p+\Sigma)\,\Gamma^\nu(p_f^+,Q) \\
                 & \quad - \Gamma^\nu(p_f^-,Q) \,S(p- \Sigma)\,S^{-1}(p-\tfrac{\Delta}{2})\,.
             \end{split}
             \end{equation}
             We can furthermore attach quark propagators from the left and right to obtain
             \begin{equation}
             \begin{split}
                 & {Q'}^\mu\left[ S(p+\tfrac{\Delta}{2})\,\widetilde{\Gamma}^{\mu\nu}(p,\Sigma,\Delta)\,S(p-\tfrac{\Delta}{2})\right] = \\
                 & \qquad = S(p+\Sigma)\,\Gamma^\nu(p_f^+,Q)\,S(p-\tfrac{\Delta}{2}) \\
                 & \qquad - S(p+\tfrac{\Delta}{2})\,\Gamma^\nu(p_f^-,Q)\,S(p-\Sigma)\,,
             \end{split}
             \end{equation}
             which can be written as a WTI for the vertices with external (dressed) fermion lines attached:
             \begin{equation}\label{qcv-wti-wf}
                  {Q'}^\mu \widetilde{\chi}^{\mu\nu}(p,\Sigma,\Delta) = \chi^\nu(p_f^+,Q) - \chi^\nu(p_f^-,Q) \,,
             \end{equation}
             where
             \begin{equation}\label{qcv-wf-def}
             \begin{split}
                 \widetilde{\chi}^{\mu\nu}(p,\Sigma,\Delta) &:= S(p+\tfrac{\Delta}{2})\,\widetilde{\Gamma}^{\mu\nu}(p,\Sigma,\Delta)\,S(p-\tfrac{\Delta}{2})\,,\\
                 \chi^\mu(p,Q) &:= S(p+\tfrac{Q}{2})\,\Gamma^\mu(p,Q)\,S(p-\tfrac{Q}{2})\,.
             \end{split}
             \end{equation}
             We see that the WTI for the 1PI part $\Gamma^{\mu\nu}$ has the same form as the one for the full vertex $\widetilde{\chi}^{\mu\nu}$ with external fermion legs.
             The remaining equation for $Q^\nu \widetilde{\chi}^{\mu\nu}$ is again obtained from photon-crossing symmetry.

             Note that the right-hand side of Eq.~\eqref{qcv-wti-full} vanishes in the case of the nucleon's Compton amplitude $\widetilde{J}^{\mu\nu}$:
             the inverse nucleon propagators for the incoming and outgoing nucleon are proportional to the negative-energy projectors $\Lambda_-^i$, $\Lambda_-^f$ and vanish
             on the nucleon's mass shell, i.e., when sandwiched between the positive-energy projectors of Eq.~\eqref{positive-energy-projectors}.
             This is just the statement that the Compton amplitude
             $\widetilde{J}^{\mu\nu} =  J^{\mu\nu}_\text{B} + J^{\mu\nu}$ is transverse:
             \begin{equation}\label{ncv-wti}
                   {Q'}^\mu   \widetilde{J}^{\mu\nu} = {Q'}^\mu   J^{\mu\nu}_\text{B} + {Q'}^\mu   J^{\mu\nu} = 0\,,
             \end{equation}
             where $J^{\mu\nu}_\text{B}$ is the analogue of the Born part~\eqref{qcv-born} for the nucleon.
             The two terms in Eq.~\eqref{ncv-wti} vanish individually in the Born limit where both intermediate nucleons in the Born term are onshell: $(P\pm\Sigma)^2=-M^2$.
             In that case both nucleon-photon vertices reduce to the onshell currents, Eq.~\eqref{current-onshell-standard}, and become transverse by themselves.
             However, ${Q'}^\mu J^{\mu\nu}_\text{B}$ also vanishes in general kinematics as long as the nucleon-photon vertices in
             the Born part have the specific onshell form of Eq.~\eqref{current-onshell-standard},
             without positive-energy projectors for the intermediate nucleon.
             The Born term constructed in that way is fixed by the nucleon electromagnetic form factors and usually subtracted from the Compton amplitude.
             Since the corresponding 1PI part must be transverse as well, one has a gauge-invariant decomposition into 'Born' and 'residual' parts.
             We will return to this point in Sec.~\ref{sec:bornterms}.

  \subsection{WTI-preserving Compton vertex}

             In the following we want to construct the analogue of Eq.~\eqref{vertex:BC} for the two-photon case, i.e., the most general fermion Compton vertex
             that is compatible with its Ward-Takahashi identity~\eqref{qcv-wti-r} and free of kinematic singularities.
             Constructions for similar seagull vertices with external photon and meson/diquark legs have been devised in Refs.~\cite{Wang:1996zu,Oettel:1999gc,Eichmann:2009zx}; however,
             the requirement of Bose symmetry for the fermion Compton vertex does not permit a direct generalization of these studies to the present case.
             The structure of the scalar Compton vertex has also been analyzed in Ref.~\cite{Bashir:2009xx}, although without a discussion of kinematic singularities.

             In analogy to the Ball-Chiu construction, the WTI determines the fermion Compton vertex up to transverse parts
            with respect to both photons. From inserting~\eqref{vertex:qpv} in~\eqref{qcv-wti-r} we see that it must have the schematic structure
            \begin{equation}\label{qcv-wti-splitting}
                \Gamma^{\mu\nu} =  \Gamma^{\mu\nu}_\text{BC} + \Gamma^{\mu\nu}_\text{T} + \Gamma^{\mu\nu}_\text{TT}\,,
            \end{equation}
            where each term is Bose-symmetric on its own. The first two pieces are necessary to satisfy the WTI,
            whereas $\Gamma^{\mu\nu}_\text{TT}$
            is transverse with respect to both photons:
            \begin{equation}
                {Q'}^\mu \,\Gamma_\text{TT}^{\mu\nu} = 0\,, \quad
                \Gamma_\text{TT}^{\mu\nu}\,Q^\nu  = 0\,.
            \end{equation}
            The Ball-Chiu part $\Gamma^{\mu\nu}_\text{BC}$ is
            obtained from~\eqref{vertex:BC} via
             \begin{equation}
             \begin{split}
                  {Q'}^\mu \,\Gamma^{\mu\nu}_\text{BC} &=  \Gamma_\text{BC}^\nu(p_f^+,Q) - \Gamma_\text{BC}^\nu(p_f^-,Q)\,,  \\
                  Q^\nu \,\Gamma^{\mu\nu}_\text{BC} &=  \Gamma_\text{BC}^\mu(p_i^+,-Q')- \Gamma_\text{BC}^\mu(p_i^-,-Q')
             \end{split}
             \end{equation}
            and fixed by the dressed fermion propagator:
            \begin{equation}
            \begin{split}
                {Q'}^\mu \,\Gamma_\text{BC}^{\mu\nu} \,Q^\nu &= S^{-1}(p+\Sigma)+S^{-1}(p-\Sigma) \\
                                                                             &- S^{-1}(p+\tfrac{\Delta}{2}) - S^{-1}(p-\tfrac{\Delta}{2}).
            \end{split}
            \end{equation}
            The remaining contribution $\Gamma^{\mu\nu}_\text{T}$ is generated by the transverse part of the fermion-photon vertex.
            It is not fully transverse with respect to both photon legs but merely satisfies
            \begin{equation}
                {Q'}^\mu \,\Gamma^{\mu\nu}_\text{T} \,Q^\nu = 0\,.
            \end{equation}
            In the following we will derive the Ball-Chiu part $\Gamma^{\mu\nu}_\text{BC}$ and
            return to the remaining pieces $\Gamma^{\mu\nu}_\text{T}$ and $\Gamma^{\mu\nu}_\text{TT}$
            in Secs.~\ref{sec:partially-transverse} and~\ref{sec:fully-transverse}, respectively.

    \renewcommand{\arraystretch}{1.1}

            We exemplify the derivation for a scalar vertex
            \begin{equation}
                \Gamma^\mu(k,Q) =   2 k^\mu  \Delta_B \,,
            \end{equation}
            where we have set $A(k^2)=0$; the generalization to the vertex~\eqref{vertex:BC}
            for a spin-$\nicefrac{1}{2}$ fermion will be straightforward.
            Evaluating the WTIs in that case yields
            \begin{equation}\label{diff-quot-1}
            \begin{split}
                  {Q'}^\mu \,\Gamma^{\mu\nu}_\text{BC} =  {Q'}^\nu& \Big[\Delta_B(p_f^+,Q) + \Delta_B(p_f^-,Q)\Big]\\
                                                         + 2p^\nu &  \Big[\Delta_B(p_f^+,Q) - \Delta_B(p_f^-,Q)\Big] \,,  \\
                  Q^\nu \,\Gamma^{\mu\nu}_\text{BC}  =  Q^\mu&   \Big[\Delta_B(p_i^+,-Q') + \Delta_B(p_i^-,-Q')\Big]  \\
                                                         + 2p^\mu &  \Big[\Delta_B(p_i^+,-Q') - \Delta_B(p_i^-,-Q')\Big]\,.
            \end{split}
            \end{equation}
            The difference quotients that appear here were defined in Eq.~\eqref{QPV:sigma,delta}. Their differences can have zeros
            in several variables, and our goal is to disentangle those zeros and express the remainders in terms of quantities that are free of kinematic singularities.
            The arguments of the function $B$ in the last equation are the momenta $p\pm \tfrac{Q}{2} \pm \tfrac{Q'}{2}$,
            and we must study the following momentum dependencies:
            \begin{equation}
                B\left(p^2+\tfrac{1}{4} (Q^2+{Q'}^2) \pm p\cdot Q \pm p\cdot Q' \pm \tfrac{1}{2}\,Q\cdot Q'\right).
            \end{equation}
            We simplify the notation by abbreviating\footnote{We will use these abbreviations only in the present and the following subsection; $u$ should not be confused
            with the Mandelstam variable that we defined in Section~\ref{sec:phasespace}.}
            \begin{equation}
               u = p\cdot Q\,,  \qquad u' = p\cdot Q'\,, \qquad w = \tfrac{1}{2}\,Q\cdot Q'
            \end{equation}
            and $x_0 = p^2+\tfrac{1}{4} (Q^2+{Q'}^2)$.
            Generalizing from the one-dimensional case, we
            define appropriate difference quotients via the linear terms (with respect to each variable) in a Taylor expansion:
            \begin{equation}\label{diff-quot-def}
            \begin{split}
                & B(x_0 + u + u' + w) = B_0 + u \,\Delta_u + u' \,\Delta_{u'} + w \,\Delta_w \\
                & \;   + uu'\,\Delta_{uu'} + uw\,\Delta_{uw}  + u'w\,\Delta_{u'w} + uu'w\,\Delta_{uu'w}\,.
            \end{split}
            \end{equation}
            Including the signs $\alpha$, $\beta$, $\gamma \in \{ \pm 1\}$, we can write
            \begin{equation}
                  B_{\alpha\beta\gamma}  := B(x_0 + \alpha u + \beta u' + \gamma w)\,,
            \end{equation}
            so that these difference quotients are obtained from the inverse relations
            \begin{equation*}\label{diff-quot-3}
               \left[ \begin{array}{@{\!\;}c@{\!\;}}  B_0 \\ u\,\Delta_u \\ u'\,\Delta_{u'} \\ w \,\Delta_w \\ uu'\,\Delta_{uu'} \\ uw\,\Delta_{uw} \\ u'w\,\Delta_{u'w} \\   uu'w\,\Delta_{uu'w}   \end{array} \right] =
               \frac{1}{8}
               \left[ \begin{array}{@{\!\;}c@{\!\;\,}c@{\!\;\,}c@{\!\;\,}c@{\!\;\,}c@{\!\;\,}c@{\!\;\,}c@{\!\;\,}c@{\!\;}}
                                    + & + & + & + & + & + & + & + \\
                                    + & + & + & + & - & - & - & - \\
                                    + & + & - & - & + & + & - & - \\
                                    + & - & + & - & + & - & + & - \\
                                    + & + & - & - & - & - & + & + \\
                                    + & - & + & - & - & + & - & + \\
                                    + & - & - & + & + & - & - & + \\
                                    + & - & - & + & - & + & + & -
                                  \end{array}\right] \!
               \left[ \begin{array}{@{\!\;}c@{\!\;}}    B_{+++}   \\ B_{++-} \\ B_{+-+} \\ B_{+--} \\ B_{-++} \\ B_{-+-} \\ B_{--+} \\ B_{---}  \end{array}\right].
            \end{equation*}

            Inserting Eq.~\eqref{diff-quot-def} for each instance of the function $B$ in Eq.~\eqref{diff-quot-1} yields
            \begin{equation}\label{diff-quot-4}
            \begin{split}
                  {Q'}^\mu \,\Gamma^{\mu\nu}_\text{BC} &=  2\,{Q'}^\nu  F_1  + 4p^\nu  (u' F_2 - wu\,F_3)    \,,  \\
                  Q^\nu \,\Gamma^{\mu\nu}_\text{BC}  &=  2\,Q^\mu  F_1'      + 4p^\mu (u\,F'_2 - wu' F'_3)   \,,
            \end{split}
            \end{equation}
            where $F_1$, $F_2$ and $F_3$ are defined by
            \begin{equation}\label{F123}
            \begin{split}
                F_1 &= \frac{u^2 \Delta_{u} - w^2 \Delta_{w}}{u^2-w^2} - uu'w\,\frac{\Delta_{uu'}-\Delta_{u'w}}{u^2-w^2}, \\
                F_2 &= \frac{u^2 \Delta_{uu'} - w^2 \Delta_{u'w}}{u^2-w^2} \,, \\
                F_3 &= \frac{\Delta_u-\Delta_w}{u^2-w^2}
            \end{split}
            \end{equation}
            and their Bose conjugates ($u\leftrightarrow -u'$) are given by
            \begin{equation}\label{F123'}
            \begin{split}
                F'_1 &= \frac{{u'}^2 \Delta_{u'} - w^2 \Delta_{w}}{{u'}^2-w^2} - uu'w\,\frac{\Delta_{uu'}-\Delta_{uw}}{{u'}^2-w^2}, \\
                F'_2 &= \frac{{u'}^2 \Delta_{uu'} - w^2 \Delta_{uw}}{{u'}^2-w^2} \,, \\
                F'_3 &= \frac{\Delta_{u'}-\Delta_w}{{u'}^2-w^2}\,.
            \end{split}
            \end{equation}
            We have achieved an intermediate goal: the difference quotients defined by Eq.~\eqref{diff-quot-def} are regular in the limits $u\rightarrow 0$, $u'\rightarrow 0$ and $w\rightarrow 0$;
            the $F_i$ and $F_i'$ are also regular in the limits $u^2 \rightarrow w^2$ and ${u'}^2\rightarrow w^2$; and
            their prefactors in~\eqref{diff-quot-4} are linear
            in the respective photon momentum $Q$ or $Q'$.

            To proceed, it will be convenient to work with Bose-symmetric combinations of $F_i$, $F_i'$: 
            \begin{equation}
                \conjg{F}_i := \frac{F_i+F'_i}{2}\,, \qquad
                \delta F_i  := \frac{F_i-F'_i}{u^2-{u'}^2}\,.
            \end{equation}
            The $\delta F_i$ are also regular in the limit $u^2 \rightarrow {u'}^2$.
            They are not all independent:
            contracting the first equation in~\eqref{diff-quot-4} with $Q^\nu$,
            the second with ${Q'}^\mu$ and equating both yields the relation
            \begin{equation}\label{deltaF3}
               \delta F_1 = \conjg{F}_3 + \frac{u^2+{u'}^2}{2}\,\delta F_3 - uu' \,\frac{\delta F_2}{w}\,.
            \end{equation}
            Note that from Eqs.~\eqref{F123}--\eqref{F123'}, $\delta F_2/w$ is also regular in the limit $w\rightarrow 0$.
            If we now define
            \begin{equation}\label{BC-Gi}
            \begin{split}
                G_1 &= 2\conjg{F}_1 + (u-u')^2\,\delta F_1 \,, \\
                G_2 &= 4\conjg{F}_2 - 4w\left[\conjg{F}_3 + \frac{(u-u')^2}{2}\,\left[ \frac{\delta F_2}{w} + \delta F_3\right]\right], \\
                G_3 &= 2(u-u')\,\delta F_1 \,,
            \end{split}
            \end{equation}
            we can express Eq.~\eqref{diff-quot-4} in terms of $G_1$, $G_2$ and $G_3$:
            \begin{equation}
            \begin{split}
                  {Q'}^\mu \,\Gamma^{\mu\nu}_\text{BC} =  {Q'}^\nu & (G_1+u'G_3) + p^\nu  (u'G_2 -2w\,G_3 ) \,,  \\
                  Q^\nu \,\Gamma^{\mu\nu}_\text{BC}  =  Q^\mu &  (G_1-u\,G_3) + p^\mu  (u\,G_2+2w\,G_3 )\,.
            \end{split}
            \end{equation}
            Up to transverse terms in ${Q'}^\mu$ and $Q^\nu$, this equation  has the solution
            \begin{equation}
                 \Gamma^{\mu\nu}_\text{BC} = G_1\,  \delta^{\mu\nu}  + G_2\,p^\mu p^\nu  + G_3 \left( p^\mu {Q'}^\nu - Q^\mu p^\nu\right).
            \end{equation}
            This is the resulting vertex for a scalar particle; it is Bose-symmetric and free of kinematic singularities.
            If the vertex is taken onshell, one has $u=u'$ (because $p\cdot\Delta=0$, cf.~Eq.~\eqref{euclidean-variables-1}) and hence $\conjg{F}_i=F_i=F'_i$, and the $G_i$ simplify to
            \begin{equation}
                G_1 = 2F_1 \,, \quad  G_2 = 4\,(F_2 - wF_3)\, , \quad G_3 = 0\,.
            \end{equation}

    \renewcommand{\arraystretch}{1.2}

            The generalization to the full Ball-Chiu vertex~\eqref{vertex:BC}, which includes also the vector dressing functions $\Sigma_A$ and $\Delta_A$,
            proceeds along the same lines. After repeating the steps~\eqref{diff-quot-1} and (\ref{diff-quot-def}--\ref{deltaF3}) for the fermion dressing function $A(k^2)$,
            we obtain the final result
            \begin{equation}\label{qpv-bc-final}
                 \Gamma_\text{BC}^{\mu\nu} = \sum_{j=1}^3 \left( i\Slash{p} \,G_j^A + G_j^B\right) \tau_j^{\mu\nu}
                                              +\sum_{j=5}^8 iG_j^A\,\tau_j^{\mu\nu}\,,
            \end{equation}
            where we temporarily define the Bose-(anti-)symmetric tensor structures
            \begin{equation}
            \begin{array}{r@{\!\;}l}
               \tau_1^{\mu\nu} &= \delta^{\mu\nu} , \\
               \tau_2^{\mu\nu} &= p^\mu p^\nu , \\
               \tau_3^{\mu\nu} &= p^\mu {Q'}^\nu - Q^\mu p^\nu ,
            \end{array}
            \quad
            \begin{array}{r@{\!\;}l}
               \tau_5^{\mu\nu} &= p^\mu \gamma^\nu +\gamma^\mu p^\nu , \\
               \tau_6^{\mu\nu} &= p^\mu \gamma^\nu -\gamma^\mu p^\nu , \\
               \tau_7^{\mu\nu} &= \gamma^\mu {Q'}^\nu + Q^\mu \gamma^\nu , \\
               \tau_8^{\mu\nu} &= \gamma^\mu {Q'}^\nu - Q^\mu \gamma^\nu .
            \end{array}
            \end{equation}
            The functions $G_{1\dots 3}$ are defined in Eq.~\eqref{BC-Gi}; we have attached
            the superscripts $A$ and $B$ to distinguish the two fermion dressing functions they are based upon.
            The vector dressing function $A$ entails four additional functions $G_{5\dots 8}$ which depend on the previous ones:
            \begin{equation}\label{G5-G8}
            \begin{split}
                G_5 &= 2\conjg{F}_1 \,, \\
               -G_6 &= (u^2-{u'}^2)\,\delta F_1\,, \\
            8\, G_7 &= (u+u') \,G_2\,,  \\
            -8\,G_8 &= (u-u')(G_2+8w\,\delta F_1)\,.
            \end{split}
            \end{equation}
            Altogether there are six independent tensor structures associated with $G_{1,2,3}^A$ and $G_{1,2,3}^B$ which contribute to $\Gamma_\text{BC}^{\mu\nu}$.
            Note that the kinematic prefactors in Eqs.~\eqref{BC-Gi} and~\eqref{G5-G8},
            \begin{equation}
            \begin{split}
                u+u' &= 2p\cdot\Sigma \sim zZ+y\sqrt{1-z^2}\sqrt{1-Z^2}   \,,\\
                u-u' &= p\cdot\Delta \sim z\,,
            \end{split}
            \end{equation}
            compensate the Bose- and charge-conjugation symmetries of the basis elements, cf.~Eq.~\eqref{BC-symmetry-prefactors}:
            $\tau_1$, $\tau_2$ and $\tau_5$ are symmetric ($\mathsf{B} \mathsf{C} = ++$),
            whereas appropriate prefactors must be attached to $\tau_6$ ($\mathsf{B} \mathsf{C} =--$), $\tau_7$ ($\mathsf{B} \mathsf{C} =-+$),
            and $\tau_3$, $\tau_8$ ($\mathsf{B} \mathsf{C} =+-$)
            to ensure invariance of the full vertex, Eqs.~(\ref{Bose-symmetry}--\ref{CC-symmetry}).

    \renewcommand{\arraystretch}{1.2}

            To conclude this section, we note that the resulting structure of the vertex $\widetilde{\chi}^{\mu\nu}$ from Eq.~\eqref{qcv-wf-def}, where the Born terms are included
            and external propagator legs attached, is completely identical to Eq.~\eqref{qpv-bc-final}.
            The WTI for the fermion-photon vertex with external propagator legs is given by
            \begin{equation}
                Q^\mu \chi^\mu(k,Q) = S(k_-)-S(k_+)\,,
            \end{equation}
            which, compared to~\eqref{qpv-wti}, merely amounts to a replacement $A\rightarrow \sigma_v$ and $B\rightarrow -\sigma_s$,
            i.e., the dressing functions of the inverse fermion propagator are replaced by those of the propagator itself:
            \begin{equation}
            \begin{split}
                \sigma_v(k^2) &= \frac{1}{A(k^2)}\,\frac{1}{k^2+M(k^2)^2}\,, \\
                \sigma_s(k^2) &= \frac{M(k^2)}{A(k^2)}\,\frac{1}{k^2+M(k^2)^2}\,.
            \end{split}
            \end{equation}
            The complete analysis carries then through to the Compton vertex $\widetilde{\chi}_\text{BC}^{\mu\nu}$.

  \subsection{Partially transverse terms}\label{sec:partially-transverse}

    \renewcommand{\arraystretch}{1.4}

            We still need to work out the contribution $\Gamma^{\mu\nu}_\text{T}$ in Eq.~\eqref{qcv-wti-splitting}
            that is needed to satisfy the Compton WTI and
            depends on the transverse part $\Gamma^\mu_\text{T}$ of the fermion-photon vertex.
            With the basis decomposition of Eq.~\eqref{qpv-transverse-new}
            one can construct the contributions $\Gamma^{\mu\nu}_{\text{T},i}$ to the Compton vertex
            \begin{equation}\label{qcv-partialtrans}
                  \Gamma^{\mu\nu}_\text{T}(p,\Sigma,\Delta) = \sum_{i=1}^8 \Gamma^{\mu\nu}_{\text{T},i} (p,\Sigma,\Delta)
            \end{equation}
            for each basis element in Eq.~\eqref{qpv-newbasis}
            separately. The right-hand side of the Compton WTI will contain differences of basis elements $\tau_i^\mu$ as well as sums and differences of the
            dressing functions $f_i$. For the former, we isolate the relative momentum $k^\mu$ in each basis element:
            \begin{equation}
            \begin{split}
                \tau_i^\mu(k,Q) = \left\{ \begin{array}{l@{\qquad}l}
                                           \tau_i^\mu(Q) & i=1,3 \\
                                           k^\alpha \tau_i^{\mu\alpha}(Q) & i=4,5 \\
                                           k^\alpha k^\beta \tau_i^{\mu\alpha\beta}(Q)  & i=2,6,7,8.
                                           \end{array}\right.
            \end{split}
            \end{equation}
            For the elements that are linear in $k^\alpha$, the $k-$independent remainders are
            \begin{equation}
                \tau_4^{\mu\alpha} = \tfrac{1}{6}\,\left[ \gamma^\mu, \gamma^\alpha, \Slash{Q} \right], \qquad
                \tau_5^{\mu\alpha} = i t^{\mu\alpha}_{QQ}\,,
            \end{equation}
            where we exploited the definitions in Eqs.~\eqref{generalized-commutator} and~\eqref{new-transverse-projectors}.
            For the $k-$quadratic basis elements one obtains
            \begin{equation}
            \begin{split}
                \tau_2^{\mu\alpha\beta} &= t^{\mu\nu}_{QQ}\,\tfrac{i}{4}[ \gamma^\nu, Q^\alpha \gamma^\beta + Q^\beta \gamma^\alpha ], \\
                \tau_6^{\mu\alpha\beta} &= \tfrac{1}{2}\left( t^{\mu\alpha}_{QQ}\,\gamma^\beta + t^{\mu\beta}_{QQ}\,\gamma^\alpha\right), \\
                \tau_7^{\mu\alpha\beta} &= Q^\alpha Q^\beta \gamma^\mu - \tfrac{1}{2}\,(\delta^{\mu\alpha}Q^\beta + \delta^{\mu\beta} Q^\alpha)\,\Slash{Q},  \\
                \tau_8^{\mu\alpha\beta} &= \tfrac{i}{4} \left[ \gamma^\mu, Q^\alpha \gamma^\beta + Q^\beta \gamma^\alpha \right] \\
                                        &- \tfrac{i}{4} \left[ \Slash{Q}, \delta^{\mu\alpha}\gamma^\beta + \delta^{\mu\beta} \gamma^\alpha \right],
            \end{split}
            \end{equation}
            where we symmetrized in the indices $\alpha$ and $\beta$.

    \renewcommand{\arraystretch}{1.1}

            The dressing functions
            \[
                f_i(k,Q) = f_i(k^2, k\cdot Q, Q^2)
            \]
            depend on three Lorentz-invariants, so that the following combinations appear in the WTIs:
            \begin{equation}
            \begin{split}
                f_i(p_f^\pm,Q) &= f_i\big( p^2 + \tfrac{1}{4}{Q'}^2 \pm u', u \pm w, Q^2 \big), \\
                f_i(p_i^\mp,-Q') &= f_i'(p_f^\pm,Q) \\
                                 &= f_i\big( p^2 + \tfrac{1}{4}{Q}^2 \mp u, -u' \pm w, {Q'}^2 \big).
            \end{split}
            \end{equation}
            They are related by Bose exchange $Q\leftrightarrow -Q'$, $u\leftrightarrow -u'$.
            In analogy to Eq.~\eqref{diff-quot-def} we define difference quotients by
            \begin{equation}\label{diff-quot-t}
            \begin{split}
                f^i_{\alpha\beta} :&= f_i\big( p^2 + \tfrac{1}{4}{Q'}^2 + \alpha u', u + \beta w, Q^2 \big) \\
                                   &= \big[ f_0 + \alpha u' \widetilde{\Delta}_{u'} + \beta w \,\widetilde{\Delta}_w + \alpha\beta u'w \,\widetilde{\Delta}_{u'w} \big]_i\,,\\
             (f^i)'_{\alpha\beta}  &= f_i\big( p^2 + \tfrac{1}{4}{Q}^2 - \alpha u, -u' + \beta w, {Q'}^2 \big) \\
                                   &= \big[ f_0' - \alpha u \,\widetilde{\Delta}_{u} + \beta w \,\widetilde{\Delta}_w - \alpha\beta uw \,\widetilde{\Delta}_{uw} \big]_i\,,
             \end{split}
             \end{equation}
             where $\alpha$, $\beta \in \{\pm\}$ and $f_0 = f(p^2+\tfrac{1}{4}{Q'}^2, u, Q^2)$.
            The tilde is here merely a reminder that the functions differ from those in Eq.~\eqref{diff-quot-def}.
             The difference quotients are obtained from
             the $f^i_{\alpha\beta}$ via
            \begin{equation}
               \left[ \begin{array}{@{\!\;}c@{\!\;}}  f_0 \\ u'\,\widetilde{\Delta}_{u'} \\ w \,\widetilde{\Delta}_w \\ u'w\,\widetilde{\Delta}_{u'w}  \end{array} \right]_i =
               \frac{1}{4}\!
               \left[ \begin{array}{@{\!\;}c@{\!\;\,}c@{\!\;\,}c@{\!\;\,}c@{\!\;}}
                                    + & + & + & +  \\
                                    + & + & - & -  \\
                                    + & - & + & -  \\
                                    + & - & - & +
                                  \end{array}\right]\!
               \left[ \begin{array}{@{\!\;}c@{\!\;}} f_{++} \\ f_{+-} \\ f_{-+} \\ f_{--}  \end{array}\right]_i.
            \end{equation}
            The following combinations appear in the WTIs:
            \begin{equation*}
            \begin{split}
                f_i(p_f^+,Q)-f_i(p_f^-,Q) &= f^i_{++}-f^i_{--} = {Q'}^\mu H_i^\mu,\\
                f_i(p_f^+,Q)+f_i(p_f^-,Q) &= f^i_{++}+f^i_{--} = 2 H_i, \\
                f_i(p_i^+,-Q')-f_i(p_i^-,-Q') &= (f^i_{--}-f^i_{++})' = Q^\nu {H_i'}^\nu,\\
                f_i(p_i^+,-Q')+f_i(p_i^-,-Q') &= (f^i_{--}+f^i_{++})' = 2 H_i',
            \end{split}
            \end{equation*}
            where we can exploit Eq.~\eqref{diff-quot-t} to obtain
            \begin{equation}
            \begin{split}
                H_i^\mu &=     ( 2p^\mu \widetilde{\Delta}_{u'} + Q^\mu \widetilde{\Delta}_w)_i , \\
                {H_i'}^\nu &=  ( 2p^\nu \widetilde{\Delta}_{u} - {Q'}^\nu \widetilde{\Delta}_w)_i, \\
                H_i &=  (f_0 + u'w \,\widetilde{\Delta}_{u'w}  )_i , \\
                H_i' &=  (f_0' - uw \,\widetilde{\Delta}_{uw}  )_i .
            \end{split}
            \end{equation}

            Evaluating the WTIs is now straightforward. For the two basis elements $i=1,3$
            that are independent of the relative momentum we arrive at:
            \begin{equation}
            \begin{split}
                  {Q'}^\mu \,\Gamma_{\text{T},i}^{\mu\nu} =  {Q'}^\mu H_i^\mu\, i\tau_i^\nu(Q)\,,  \\
                  Q^\nu \,\Gamma_{\text{T},i}^{\mu\nu}  = Q^\nu {H_i'}^\nu\,i\tau_i^\mu(-Q')\,.
            \end{split}
            \end{equation}
            Since the $\tau_i^\mu$ are already transverse in their arguments, we can simply add both contributions to obtain
            a Bose-symmetric solution of both equations (up to fully transverse terms):
            \begin{equation}
            \begin{split}
                -i\Gamma_{\text{T},i}^{\mu\nu} &= H_i^\mu \tau_i^\nu(Q) + {H_i'}^\nu \tau_i^\mu(-Q').
            \end{split}
            \end{equation}
            For the elements $i=4,5$ we obtain similarly
            \begin{equation}
            \begin{split}
                -i\Gamma_{\text{T},i}^{\mu\nu} &= H_i^\mu \tau_i^\nu(p,Q) + {H_i'}^\nu \tau_i^\mu(p,-Q')  \\
                                              & + H_i\, \tau_i^{\nu\mu}(Q) + H'_i\, \tau_i^{\mu\nu}(-Q'),
            \end{split}
            \end{equation}
            and for the elements $i=2,6,7,8$ we have
            \begin{equation}
            \begin{split}
                -i\Gamma_{\text{T},i}^{\mu\nu} &= H_i^\mu \tau_i^\nu(p,Q) + {H_i'}^\nu \tau_i^\mu(p,-Q')  \\
                                              &-\tfrac{1}{4}\left[  H_i^\mu \tau_i^\nu(-Q',Q) + {H_i'}^\nu \tau_i^\mu(Q,-Q')\right]  \\
                                              & + 2k^\alpha\left[ H_i\, \tau_i^{\nu\mu\alpha}(Q) + H'_i\, \tau_i^{\mu\nu\alpha}(-Q')\right].
            \end{split}
            \end{equation}
            Note that for a scalar Compton vertex only $\Gamma_{\text{T},5}^{\mu\nu}$ contributes.

            The final result for the WTI-preserving 1PI fermion Compton vertex is the sum of Eq.~\eqref{qpv-bc-final},
            which is completely fixed by the dressed fermion propagator, plus the eight $\Gamma_{\text{T},i}^{\mu\nu}$ that constitute $\Gamma^{\mu\nu}_\text{T}$ in Eq.~\eqref{qcv-partialtrans}.
            The latter are fixed by the transverse part of the fermion-photon vertex. The full Compton vertex including the Born term is then given by
            \begin{equation}\label{qcv-wti-splitting-full}
                \widetilde{\Gamma}^{\mu\nu} =  \Gamma^{\mu\nu}_\text{B} + \Gamma^{\mu\nu}_\text{BC} + \Gamma^{\mu\nu}_\text{T} + \Gamma^{\mu\nu}_\text{TT}\,.
            \end{equation}
            The first three terms on the r.h.s. are necessary to satisfy the WTI of Eq.~\eqref{qcv-wti-full}. The above construction
            can also be useful in view of the nucleon Compton amplitude $\widetilde{J}^{\mu\nu}$: since the r.h.s. of the WTI vanishes onshell,
            Eq.~\eqref{qcv-wti-splitting-full} provides a gauge-invariant decomposition into a fully transverse 'generalized Born part'
            $J^{\mu\nu}_\text{B} + J^{\mu\nu}_\text{BC} + J^{\mu\nu}_\text{T}$ that depends on the (offshell) nucleon propagator and nucleon-photon vertex,
            and another fully transverse remainder $J^{\mu\nu}_\text{TT}$ which must be determined dynamically.

    \renewcommand{\arraystretch}{1.0}

  \subsection{Fully transverse part of the Compton vertex}\label{sec:fully-transverse}

             The remaining part $\Gamma^{\mu\nu}_\text{TT}$ of the fermion Compton vertex in the representation~\eqref{qcv-wti-splitting-full} is
             transverse with respect to both ${Q'}^\mu$ and $Q^\nu$. Its structure is of particular importance
             for the onshell nucleon Compton amplitude:
             since the r.h.s. of the onshell Compton WTI~\eqref{ncv-wti} vanishes, both the full amplitude  $\widetilde{J}^{\mu\nu}$ and the residual part $J^{\mu\nu}$
             (if a suitable gauge-invariant Born term is subtracted)  are fully transverse and therefore subject to possible kinematic constraints.

             We recall from the analysis in Sec.~\eqref{sec:tensorbasis} that there are $9 \times 8 = 72$ independent transverse basis elements in the general offshell case
             and $9 \times 2 = 18$ transverse basis elements on the mass shell.
             In order to study real or virtual Compton scattering,
             we must find the analogue to Eqs.~(\ref{qpv-newbasis}--\ref{qpv-transverse-new}).
             The tensor structures in the resulting basis
             should be free of kinematic singularities in the relevant kinematic limits
             so that the respective dressing functions are kinematically independent.
             For the onshell Compton scattering amplitude, such a basis has been constructed by Tarrach~\cite{Tarrach:1975tu},
             with a later modification in Refs.~\cite{Drechsel:1997xv,Pasquini:2001yy}.
             In the following we attempt to build such a basis from a systematic point of view.

    \renewcommand{\arraystretch}{1.4}

             \bigskip

             \textbf{Construction of the basis.}
             One can generalize the construction of Eqs.~(\ref{new-transverse-projectors}--\ref{qpv-newbasis}) to the Compton case in the following way. We
             start with the ten basis elements that are independent of the photon momenta:
             \begin{equation}\label{EFG-rho}
             \begin{array}{rl}
                  \rho_1^{\alpha\beta} &= \delta^{\alpha\beta}, \\
                  \rho_3^{\alpha\beta} &= \tfrac{1}{2}\left[ \gamma^{\alpha}, \gamma^\beta \right], \\
                  \rho_5^{\alpha\beta} &= p^\alpha \gamma^\beta + \gamma^\alpha p^\beta, \\
                  \rho_7^{\alpha\beta} &= p^\alpha \gamma^\beta - \gamma^\alpha p^\beta, \\
                  \rho_9^{\alpha\beta} &= p^\alpha p^\beta,
             \end{array}\quad
             \begin{array}{rl}
                  \rho_2^{\alpha\beta} &= \delta^{\alpha\beta}\,\slashed{p}\,, \\
                  \rho_4^{\alpha\beta} &= \tfrac{1}{6}\left[ \gamma^\alpha, \gamma^\beta, \slashed{p} \right], \\
                  \rho_6^{\alpha\beta} &= \tfrac{1}{2}\left[ p^\alpha \gamma^\beta + \gamma^\alpha p^\beta, \slashed{p} \right], \\
                  \rho_8^{\alpha\beta} &= \tfrac{1}{2}\left[ p^\alpha \gamma^\beta - \gamma^\alpha p^\beta, \slashed{p} \right], \\
                  \rho_{10}^{\alpha\beta}  &= p^\alpha p^\beta\,\slashed{p}\,,
             \end{array}
             \end{equation}
             where the generalized commutator $[\,\cdot\,,\,\cdot\,,\,\cdot\,]$ has been defined in Eq.~\eqref{generalized-commutator}.
             Second, we want to contract them with all possible tensor structures that are transverse with respect to both photons and free of kinematic singularities.
             To this end, we use the definitions from Eq.~\eqref{new-transverse-projectors},
             \begin{equation} \label{new-transverse-projectors-2}
             \begin{split}
                   t_{ab}^{\mu\nu} &= a\cdot b\,\delta^{\mu\nu} - b^\mu a^\nu\,,  \\
                   \varepsilon^{\mu\nu}_{ab} &= \gamma_5\,\varepsilon^{\mu\nu\alpha\beta}a^\alpha b^\beta\,,
             \end{split}
             \end{equation}
             to define the quantities:
             \begin{equation}\label{EFG}
             \begin{split}
                 \mathsf{E}_\pm^{\mu\alpha,\beta\nu}(a,b) &:= \tfrac{1}{2}\left(\varepsilon^{\mu\alpha}_{Q'a'} \, \varepsilon^{\beta\nu}_{bQ} \pm \varepsilon^{\mu\alpha}_{Q'b'} \, \varepsilon^{\beta\nu}_{aQ} \right), \\
                 \mathsf{F}_\pm^{\mu\alpha,\beta\nu}(a,b) &:= \tfrac{1}{2}\left( t^{\mu\alpha}_{Q'a'} \, t^{\beta\nu}_{bQ} \pm t^{\mu\alpha}_{Q'b'} \, t^{\beta\nu}_{aQ} \right), \\
                 \mathsf{G}_\pm^{\mu\alpha,\beta\nu}(a,b) &:= \tfrac{1}{2}\left(\varepsilon^{\mu\alpha}_{Q'a'} \, t^{\beta\nu}_{bQ} \pm t^{\mu\alpha}_{Q'b'} \, \varepsilon^{\beta\nu}_{aQ} \right),
             \end{split}
             \end{equation}
             where $a^\mu,b^\mu \in \{p^\mu, Q^\mu, {Q'}^\mu\}$, and primed quantities are obtained from
             \begin{equation}
                 a = p,\, Q,\, Q' \quad \leftrightarrow \quad a' = p,\, Q',\, Q\,.
             \end{equation}
             These tensors are symmetric or antisymmetric under Bose and charge conjugation,
             transverse with respect to ${Q'}^\mu$ and $Q^\nu$,
             linear in all momenta $Q$, $Q'$, $a$ and $b$, and free of kinematic singularities.
             They are not all linearly independent;
             for example, $\mathsf{E}_-(a,b)$ and $\mathsf{F}_-(a,b)$ vanish if $a=b$.

    \renewcommand{\arraystretch}{1.0}

             In the last step we contract Eq.~\eqref{EFG} with the $\rho_i^{\alpha\beta}$ and
             collect the resulting elements in subsets with increasing powers in the photon momenta, in order to avoid kinematic relations between the dressing functions.
             Since there are only $72$ independent transverse elements, one will encounter linear dependencies
             between the possible combinations
             \begin{equation}\label{fcv-overcounting}
             \begin{split}
                 &\mathsf{E}_\pm^{\mu\alpha,\beta\nu}(a,b)\, \rho_i^{\alpha\beta}(p) \,, \\
                 &\mathsf{F}_\pm^{\mu\alpha,\beta\nu}(a,b)\, \rho_i^{\alpha\beta}(p) \,, \\
                 &\mathsf{G}_\pm^{\mu\alpha,\beta\nu}(a,b)\, \rho_i^{\alpha\beta}(p) \,.
             \end{split}
             \end{equation}
             The detailed analysis yields a maximum set of 16 independent elements which carry two powers of the photon momenta;
             they can be chosen as
             \begin{equation}\label{qcv-offshell-basis-1}
             \left\{ \begin{array}{c}
                     \mathsf{E}_+^{\mu\alpha,\beta\nu}(p,p) \\
                     \mathsf{F}_+^{\mu\alpha,\beta\nu}(p,p) \\
                     \mathsf{G}_\pm^{\mu\alpha,\beta\nu}(p,p)
                     \end{array}\right\} \times \rho_{1\dots 4}^{\alpha\beta}(p)\,.
             \end{equation}
             There are 40 further cubic elements
             \begin{equation}
             \begin{split}
             \left\{ \begin{array}{c}
                     \mathsf{F}_\pm^{\mu\alpha,\beta\nu}(p,Q) \\
                     \mathsf{G}_\pm^{\mu\alpha,\beta\nu}(p,Q)
                     \end{array}\right\} &\times \rho_{1\dots 6}^{\alpha\beta}(p), \\
             \left\{ \begin{array}{c}
                     \mathsf{F}_\pm^{\mu\alpha,\beta\nu}(p,Q') \\
                     \mathsf{G}_\pm^{\mu\alpha,\beta\nu}(p,Q')
                     \end{array}\right\} &\times \rho_{3\dots 6}^{\alpha\beta}(p)
             \end{split}
             \end{equation}
             and 16 that come with four powers of $Q$, $Q'$:
             \begin{equation}\label{qcv-offshell-basis-3}
             \begin{split}
                     \mathsf{F}_+^{\mu\alpha,\beta\nu}(Q,Q) & \,\rho_{1\dots 8}^{\alpha\beta}(p), \\
                     \mathsf{G}_\pm^{\mu\alpha,\beta\nu}(Q',Q)& \,\rho_{3\dots 6}^{\alpha\beta}(p).
             \end{split}
             \end{equation}
             This forms a complete 72-dimensional basis for the fully transverse part of the vertex
             which is free of kinematic singularities and constraints.

    \renewcommand{\arraystretch}{1.5}

             \begin{table*}[t]

             \begin{equation*}\label{drechsel-tarrach}
             \begin{array}{rl}
                -T_1^{\mu\nu} &= t_{Q'Q}^{\mu\nu} \\ 
                 T_2^{\mu\nu} &= \mathsf{F}_+^{\mu\nu}(p,p) \\
                -T_3^{\mu\nu} &= \mathsf{F}_+^{\mu\nu}(Q,Q) \\
                 T_4^{\mu\nu} &= 2\,\mathsf{F}_+^{\mu\nu}(p,Q) \\[1mm]
                 T_6^{\mu\nu} &= \tfrac{1}{2}\,\mathsf{F}_+^{\mu\alpha,\beta\nu}(Q,Q)\,\rho_3^{\alpha\beta} \\[1mm]
                 T_7^{\mu\nu} &= 4\,\mathsf{G}_-^{\mu\nu}(\gamma,p) \\
                 T_8^{\mu\nu} &= \mathsf{G}_-^{\mu\nu}(\gamma,Q) \\
                 T_9^{\mu\nu} &= \mathsf{G}_+^{\mu\nu}(\gamma,Q) \\[1mm]
                 T_{10}^{\mu\nu} &= 4iM \left(2\,\mathsf{F}_+^{\mu\nu}(\gamma,p) + \mathsf{G}_-^{\mu\nu}(\gamma,Q') \right) \\
                                    & \quad + 2\,p\cdot\Sigma\,\mathsf{F}_+^{\mu\nu}(\gamma,\gamma) - 8\,\mathsf{F}_+^{\mu\nu}(p,p)
             \end{array} \qquad\qquad
             \begin{array}{rl}
                 T_{11}^{\mu\nu} &= \mathsf{G}_+^{\mu\nu}(\gamma,Q') \\[1mm]
                 T_{12}^{\mu\nu} &= 2iM\,\mathsf{F}_+^{\mu\nu}(\gamma,Q)  - 2\,\mathsf{F}_+^{\mu\nu}(p,Q)  +M^2\omega_+\,\mathsf{F}_+^{\mu\nu}(\gamma,\gamma)    \\[0.5mm]
                 T_{13}^{\mu\nu} &= 2iM\,\mathsf{F}_-^{\mu\nu}(\gamma,Q)  - 2\,\mathsf{F}_-^{\mu\nu}(p,Q)  +M^2\omega_-\,\mathsf{F}_+^{\mu\nu}(\gamma,\gamma)    \\
                 T_{14}^{\mu\nu} &= \tfrac{2}{3}\left(2iM\,\mathsf{G}_+^{\mu\nu}(\gamma,Q')-\left[ \mathsf{F}_+^{\mu\alpha,\beta\nu}(\gamma,Q'),\rho_5^{\alpha\beta}\right]\right)\\
                -T_{17}^{\mu\nu} &=  \mathsf{F}_+^{\mu\nu}(\gamma,\gamma) \\[1mm]
                 T_{18}^{\mu\nu} &= 4 \,\mathsf{G}_-^{\mu\nu}(\gamma,Q')\\
                 T_{19}^{\mu\nu} &= 2\,\mathsf{F}_+^{\mu\alpha,\beta\nu}(Q,Q)\,\rho_9^{\alpha\beta} \\
                 T_{20}^{\mu\nu} &= 4\,\mathsf{F}_-^{\mu\alpha,\beta\nu}(p,Q)\,\rho_3^{\alpha\beta} + 4iM\,\mathsf{G}_+^{\mu\nu}(\gamma,Q)\\
                                    & \quad+ 2iM \omega_- \left(\mathsf{G}_-^{\mu\nu}(\gamma,Q)-\mathsf{G}_-^{\mu\nu}(\gamma,Q')\right) \\
                 T_{21}^{\mu\nu} &= 4\,\mathsf{F}_+^{\mu\alpha,\beta\nu}(p,Q)\,\rho_3^{\alpha\beta} + 4iM\,\mathsf{G}_-^{\mu\nu}(\gamma,Q) \\
                                    & \quad+ 2iM \omega_+ \left(\mathsf{G}_-^{\mu\nu}(\gamma,Q)-\mathsf{G}_-^{\mu\nu}(\gamma,Q')\right).
             \end{array}
             \end{equation*}

               \caption{Representation of the basis of Refs.~\cite{Drechsel:1997xv} with our definitions from Eqs.~\eqref{EFG-rho}, \eqref{EFG}, and~\eqref{EFG-gamma}.
                        We have abbreviated $\omega_\pm = (Q^2\pm{Q'}^2)/(4M^2)$. }
               \label{drechsel-tarrach}

             \end{table*}

    \renewcommand{\arraystretch}{1.0}

             On the mass shell, only $72/4=18$ basis elements remain by virtue of the positive-energy projectors
             defined in Eq.~\eqref{positive-energy-projectors}.
             A possible choice for a linearly independent onshell basis is
             \begin{equation}\label{qcv-onshell-basis}
             \Lambda_+^f \times \left\{ \begin{array}{l}
                     \mathsf{E}_+^{\mu\alpha,\beta\nu}(p,p) \\
                     \mathsf{F}_+^{\mu\alpha,\beta\nu}(p,p) \\
                     \mathsf{G}_\pm^{\mu\alpha,\beta\nu}(p,p) \\
                     \mathsf{F}_\pm^{\mu\alpha,\beta\nu}(p,Q) \\
                     \mathsf{G}_\pm^{\mu\alpha,\beta\nu}(p,Q) \\
                     \mathsf{F}_+^{\mu\alpha,\beta\nu}(Q,Q)
                     \end{array}\right\}  \times
             \left\{ \begin{array}{c}
                      \rho_1^{\alpha\beta}(p)  \\
                     \rho_4^{\alpha\beta}(p)
                     \end{array}\right\}\times\Lambda_+^i\,.
             \end{equation}
             This subset is particularly convenient since,
             because of the identities
             \begin{equation}\label{qcv-rho-4}
             \begin{split}
                 \Lambda_+^f \,\rho_4^{\alpha\beta}(p)\, \Lambda_+^i &= -\varepsilon^{\alpha\beta\rho\sigma}\,p^\rho\,\Lambda_+^f \gamma_5 \gamma^\sigma \Lambda_+^i\,, \\
                 \Lambda_+^f \gamma_5 \,\rho_4^{\alpha\beta}(p)\, \Lambda_+^i &= -\varepsilon^{\alpha\beta\rho\sigma}p^\rho\,\Lambda_+^f \gamma^\sigma \Lambda_+^i\,,
             \end{split}
             \end{equation}
             one only deals with the Dirac structures
             \begin{equation}
                 \Lambda_+^f\,\Lambda_+^i\,, \quad
                 \Lambda_+^f\gamma_5\,\Lambda_+^i\,, \quad
                 \Lambda_+^f\gamma^\sigma \Lambda_+^i\,, \quad
                 \Lambda_+^f\gamma_5\gamma^\sigma \Lambda_+^i\,.
             \end{equation}
             Except for potentially nontrivial relations induced by the Levi-Civita symbol in Eq.~\eqref{qcv-rho-4},
             sandwiching with the onshell projectors should therefore not alter the powers in the photon momenta.
             Consequently, one would have eight quadratic, eight cubic and two quartic basis elements.

              \bigskip

             \textbf{Scalar Compton vertex.}
             Returning to our special case of a scalar Compton vertex, we note that five spin-independent combinations
             are encapsulated both in Eqs.~(\ref{qcv-offshell-basis-1}--\ref{qcv-offshell-basis-3}) and Eq.~\eqref{qcv-onshell-basis},
             namely those that are proportional to $\rho_1^{\alpha\beta}=\delta^{\alpha\beta}$ and do not
             depend on $\mathsf{G}_\pm^{\mu\alpha,\beta\nu}$:
             \begin{equation} \label{transverse-scalar}
             \begin{split}
                 \mathsf{E}_+^{\mu\nu}(p,p)  &= \varepsilon_{Q'p}^{\mu\alpha}\,\varepsilon_{pQ}^{\alpha\nu} = p^2\,t^{\mu\nu}_{Q'Q} - t_{Q'p}^{\mu\alpha}\,t_{pQ}^{\alpha\nu}\,, \\
                 \mathsf{F}_+^{\mu\nu}(p,p)  &= t_{Q'p}^{\mu\alpha}\,t_{pQ}^{\alpha\nu} \,,\\
                 \mathsf{F}_\pm^{\mu\nu}(p,Q)&= \tfrac{1}{2}\left( t_{Q'p}^{\mu\alpha}\,t_{QQ}^{\alpha\nu} \pm t_{Q'Q'}^{\mu\alpha}\,t_{pQ}^{\alpha\nu}\right), \\
                 \mathsf{F}_+^{\mu\nu}(Q,Q)  &= t_{Q'Q'}^{\mu\alpha}\,t_{QQ}^{\alpha\nu}\,,
             \end{split}
             \end{equation}
             where we abbreviated the contraction of the indices $\alpha$ and $\beta$ by
             \begin{equation}
                 \mathsf{F}_\pm^{\mu\nu}(a,b) := \mathsf{F}_\pm^{\mu\alpha,\alpha\nu}(a,b) = \mathsf{F}_\pm^{\mu\alpha,\beta\nu}(a,b)\,\rho_1^{\alpha\beta}(p) \,.
             \end{equation}
             They describe Compton scattering on a scalar particle and are identical to the $T_{1\dots 5}^{\mu\nu}$ given in Eq.~(15) in Ref.~\cite{Drechsel:1996ag}:
             \begin{equation}
             \begin{split}
                -T_1^{\mu\nu} &= t_{Q'Q}^{\mu\nu}, \\
                 T_2^{\mu\nu} &= \mathsf{F}_+^{\mu\nu}(p,p)\,, \\
                -T_3^{\mu\nu} &= \mathsf{F}_+^{\mu\nu}(Q,Q)\,, \\
                 T_4^{\mu\nu} &= 2\,\mathsf{F}_+^{\mu\nu}(p,Q)\,, \\
                 T_5^{\mu\nu} &= 2\,\mathsf{F}_-^{\mu\nu}(p,Q)\,.
             \end{split}
             \end{equation}
             For comparison with Ref.~\cite{Drechsel:1996ag}, note that
              on the mass shell one has $p\cdot Q = p\cdot Q'$, and in Euclidean conventions
              all scalar products pick up a minus sign whereas $g^{\mu\nu}$ is replaced by $-\delta^{\mu\nu}$.
              We also swapped the indices $\mu\leftrightarrow\nu$.

             \bigskip
             \textbf{Relation with other bases.}
             Next, we would like to establish a connection with Tarrach's basis for the onshell nucleon Compton amplitude~\cite{Tarrach:1975tu}
             which has been frequently used in the context of dispersion relations; see e.g.~\cite{Drechsel:1997xv,Drechsel:1998zm,Drechsel:2002ar}.
             We consider the modified version by Drechsel \textit{et al.} who write the onshell scattering amplitude as (Ref.~\cite{Drechsel:1997xv}, Eq.~(A5)):
             \begin{equation}
             \begin{split}
                 \widetilde{J}^{\mu\nu} &= \sum_{i \in j} B_i(Q^2,{Q'}^2, Q\cdot Q', p\cdot \Sigma)\,\Lambda_+^f \, T_i^{\mu\nu} \Lambda_+^i\,, \\
                       j&=\{ 1 \dots 21\} \setminus \{ 5, 15, 16\}.
             \end{split}
             \end{equation}
             In comparison to Ref.~\cite{Drechsel:1997xv}, our scattering amplitude $\widetilde{J}^{\mu\nu}$ is matrix-valued since we contract with positive-energy projectors
             instead of nucleon spinors ($\Lambda_+^i u_i = u_i$, $\conjg{u}_f \,\Lambda_+^f = \conjg{u}_f$).
             The $T_i^{\mu\nu}$ in the above equation correspond to those in Ref.~\cite{Drechsel:1997xv}:
             they are essentially Tarrach's basis elements except for the first four, which have been replaced by the scalar structures of Ref.~\cite{Drechsel:1996ag};
             and the elements $T_{5, 15, 16}$ were swapped by $T_{19,20,21}$ which can also be found in Tarrach's paper~\cite{Tarrach:1975tu}.

             In principle, the $T_i^{\mu\nu}$ will be linear combinations of our basis elements in Eq.~\eqref{qcv-onshell-basis}; however,
             one might ask if our construction from above can also be used to generate them directly.
             To that end we extend Eq.~\eqref{new-transverse-projectors-2} to accommodate the $\gamma-$matrix:
             \begin{equation}\label{new-transverse-projectors-3}
             \begin{split}
                   t_{a\gamma}^{\mu\nu} &:= \slashed{a}\,\delta^{\mu\nu} - \gamma^\mu a^\nu\,,  \\
                   \varepsilon^{\mu\nu}_{a\gamma} &:= \gamma_5\,\varepsilon^{\mu\nu\alpha\beta}a^\alpha \gamma^\beta\,,
             \end{split}
             \end{equation}
             and define similarly as before
             \begin{equation}\label{EFG-gamma}
             \begin{split}
                 \mathsf{F}_+^{\mu\alpha,\beta\nu}(\gamma,\gamma) &:= \left[ t^{\mu\alpha}_{Q'\gamma} \,,\, t^{\beta\nu}_{\gamma Q} \right], \\
                 \mathsf{F}_\pm^{\mu\alpha,\beta\nu}(\gamma,b) &:= \tfrac{1}{2}\left( t^{\mu\alpha}_{Q'\gamma} \, t^{\beta\nu}_{bQ} \pm t^{\mu\alpha}_{Q'b'} \, t^{\beta\nu}_{\gamma Q} \right), \\
                 \mathsf{G}_\pm^{\mu\alpha,\beta\nu}(\gamma,b) &:= \tfrac{1}{2}\left(\varepsilon^{\mu\alpha}_{Q'\gamma} \, t^{\beta\nu}_{bQ} \pm t^{\mu\alpha}_{Q'b'} \, \varepsilon^{\beta\nu}_{ \gamma Q} \right),
             \end{split}
             \end{equation}
             where again $b^\mu \in \{p^\mu, Q^\mu, {Q'}^\mu\}$.
             While these combinations do not yield any new basis elements, they provide a relatively compact representation of the basis in Ref.~\cite{Drechsel:1997xv}
             which is given in Table~\ref{drechsel-tarrach}.

             The authors of Ref.~\cite{Drechsel:1997xv} note that in virtual Compton scattering (${Q'}^2=0$)
             the elements $T_{3,6,19}$ decouple from the cross section when contracted with the photon polarization vectors, and $T_{8,9}$, $T_{12,13}$, $T_{20,21}$
             become pairwise identical, thus leaving 12 independent elements.
             Expressed in terms of our Lorentz tensors with their inherent ${Q'}^2$ dependence, the origin of this behavior is apparent in Table~\ref{drechsel-tarrach}.
             In our basis of Eq.~\eqref{qcv-onshell-basis}, the same argument entails
             that in virtual Compton scattering $\mathsf{F}_+(Q,Q)$ decouples and, upon contraction with the photon polarization vectors, $\mathsf{F}_+(p,Q)$ and $\mathsf{F}_-(p,Q)$
             on the one hand and $\mathsf{G}_+(p,Q)$ and $\mathsf{G}_-(p,Q)$ on the other hand become identical, which
             leaves again 12 independent tensor structures.
             We note that an alternative representation of Table~\ref{drechsel-tarrach} in the VCS limit in terms of the electromagnetic field-strength tensor has been given in Ref.~\cite{Gorchtein:2009wz}.

             \bigskip
             \textbf{Forward VVCS.}
             Finally, another interesting case is the onshell basis~\eqref{qcv-onshell-basis} in the case of forward VVCS, where $Q=Q'$:
             since $\Delta=Q-Q'=0$, the incoming and outgoing nucleon momenta are identical, $p_f = p_i = p$,
             and one has $\Lambda_+^f = \Lambda_+^i = \Lambda_+$. This implies
             \begin{equation}
                 \Lambda_+ \gamma_5 \,\Lambda_+ = 0\,, \qquad \Lambda_+ \gamma_5 \, \rho_4^{\alpha\beta}(p)\, \Lambda_+ = 0\,,
             \end{equation}
             so that the tensor structures with $\mathsf{G}_\pm$ vanish as they contain one instance of $\gamma_5$.
             In general, only five of the 18 elements remain linearly independent, and we call them for the moment
             \begin{equation}\label{FWD-basis}
             \begin{split}
                 \mathsf{T}_1^{\mu\nu} &= \mathsf{E}_+^{\mu\nu}(p,p)\,\Lambda_+\,, \\[0.8mm]
                 \mathsf{T}_2^{\mu\nu} &= \mathsf{F}_+^{\mu\nu}(p,p)\,\Lambda_+\,, \\
                 \mathsf{T}_3^{\mu\nu} &= \mathsf{E}_+^{\mu\alpha,\beta\nu}(p,p)\,\Lambda_+ \,\rho_4^{\alpha\beta}  \Lambda_+\,, \\
                 \mathsf{T}_4^{\mu\nu} &= \mathsf{F}_+^{\mu\alpha,\beta\nu}(p,p)\,\Lambda_+ \,\rho_4^{\alpha\beta} \Lambda_+\,, \\
                 \mathsf{T}_5^{\mu\nu} &= \mathsf{F}_-^{\mu\alpha,\beta\nu}(p,Q)\,\Lambda_+ \,\rho_4^{\alpha\beta}  \Lambda_+\,.
             \end{split}
             \end{equation}
             The coefficients of $\mathsf{T}_{1\dots 4}^{\mu\nu}$ are related to the structure functions $f_T$, $f_L$, $g_{TT}$ and $g_{LT}$ in the forward limit.
             In the literature (see, for example, Eq.~(114) in Ref.~\cite{Drechsel:2002ar}),
             the two spin-independent tensors are typically chosen as $T^{\mu\nu}_Q$ and $p_T^\mu \,p_T^\nu$,
             where $T^{\mu\nu}_Q = t^{\mu\nu}_{QQ}/Q^2$ is the usual projector defined in Eq.~\eqref{transverse-projector} and $p_T^\mu = T^{\mu\nu}_Q p^\nu$.
             One can work out the relations
             \begin{equation}
             \begin{split}
                 T^{\mu\nu}_Q \,\Lambda_+ &= \frac{(\mathsf{T}_1+\mathsf{T}_2)^{\mu\nu}}{p^2 \,Q^2}\,, \\
                 p_T^\mu \,p_T^\nu\,\Lambda_+ &= \frac{1}{Q^2} \left[ \mathsf{T}_2^{\mu\nu} - \frac{(p\cdot Q)^2}{p^2 \,Q^2} (\mathsf{T}_1+\mathsf{T}_2)^{\mu\nu}\right]
             \end{split}
             \end{equation}
             with our basis elements above, 
             and the spin-dependent structures can be written as
             \begin{equation}
             \begin{split}
                  \mathsf{T}_3^{\mu\nu} &= -M^2\,\Lambda_+ \, \varepsilon^{\mu\nu}_{Qp} \,\slashed{Q} \, \Lambda_+ \,, \\
                 \mathsf{T}^{\mu\nu}_4 &= M^2\,p\cdot Q\,\,  \Lambda_+ \,\varepsilon^{\mu\nu}_{Q\gamma}\, \Lambda_+ \,, \\
                 \mathsf{T}^{\mu\nu}_5 &= Q^2\,  \Lambda_+  \left( p^\mu_T\,\varepsilon^{\nu\alpha}_{Q\gamma}  + p^\nu_T\,\varepsilon^{\mu\alpha}_{Q\gamma} \right) \frac{p^\alpha}{2} \,\Lambda_+
                       \,,
             \end{split}
             \end{equation}
             where we exploited the definitions~\eqref{new-transverse-projectors-2} and~\eqref{new-transverse-projectors-3}.
             $\mathsf{T}^{\mu\nu}_4$ and $(\mathsf{T}_3+\mathsf{T}_4)^{\mu\nu}$ are proportional to the two spin-dependent tensors in Ref.~\cite{Drechsel:2002ar}
             if we define a covariant 'spin vector' via
             \begin{equation}
             \begin{split}
                 S^i &= \frac{\Sigma^i}{2} = \frac{1}{4}\,\varepsilon_{ijk}\,\sigma^{jk} \quad \longrightarrow  \\
                 S^\mu &= \frac{1}{4}\,\varepsilon^{\mu\alpha\beta\lambda}\,\hat{p}^\lambda\,\sigma^{\alpha\beta} = \frac{i}{2}\,\gamma^\mu_T\,\gamma_5 \,\hat{\slashed{p}}\,,
             \end{split}
             \end{equation}
             where $\sigma^{\alpha\beta} = -\frac{i}{2}\,[\gamma^\alpha,\gamma^\beta]$ and $\gamma^\mu_T = T^{\mu\nu}_p\,\gamma^\nu$.
             We used the properties of the positive-energy projectors: $\hat{\slashed{p}} \,\Lambda_+  =\Lambda_+$ and $\Lambda_+ \gamma_5 \gamma^\mu_T \,\Lambda_+ = \Lambda_+ \gamma_5 \gamma^\mu  \Lambda_+$.


  \section{Nucleon handbag diagrams} \label{sec:nucleon-compton-amplitude}

            We have now gathered sufficient knowledge to proceed with the calculation of the nucleon Compton scattering amplitude.
            We focus exclusively on the handbag contribution from Eq.~\eqref{scattering-handbag} which is depicted in Fig.~\ref{fig:current-kinematics}.
            Within the rainbow-ladder truncation, the scattering amplitude depends on
            the DSE solution for the dressed quark propagator and the solution for the nucleon's bound-state amplitude from its
            covariant Faddeev equation. For these ingredients we refer the reader to the literature~\cite{Eichmann:2009qa,Eichmann:2011vu}.
            We use the quark-gluon interaction of Ref.~\cite{Maris:1999nt} with the input parameters from Ref.~\cite{Eichmann:2011pv}, and we work with a
            quark mass  $m_u=m_d=3.7$ MeV at the renormalization point $\mu=19$ GeV so that the resulting pion mass is $m_\pi=138$ MeV.

            The remaining component is the quark Compton vertex that is calculated from its inhomogeneous BSE~\eqref{ibse-qcv},
            illustrated in Fig.~\ref{fig:qcv-born-small}.
             We compute the full vertex including its 128 components in general spacelike kinematics, i.e.
             for
             \begin{equation}
                 p^2, \, t, \, \sigma > 0  \quad \text{and} \quad z, \, y, \, Z \in (-1,1)\,,
             \end{equation}
             with the Lorentz-invariant variables defined in Eq.~\eqref{qcv-variables}.
             It turns out that the structure of the basis in Table~\ref{qcv-basis-1} simplifies the analysis enormously;
             the details of the calculation are provided in App.~\ref{sec:qcv-bse}.

             Combining all ingredients yields the handbag contribution to the nucleon's Compton amplitude $\widetilde{J}^{\mu\nu}$.
             Its explicit Lorentz-Dirac, color-flavor and momentum structure is worked out in App.~\ref{sec:nucleon:handbag}.
             The full Compton amplitude can be spanned onto a transverse basis according to Eqs.~\eqref{qcv-transverse-basis-Y} and~\eqref{qcv-nucleon-transverse}
             or, alternatively, using the regular basis elements from Eq.~\eqref{qcv-onshell-basis}. The corresponding dressing functions
             are Lorentz-invariant and can be expressed in terms of the four variables $\{t,X,Y,Z\}$ defined in Eqs.~(\ref{euclidean-variables-1}--\ref{XYZ-def}),
             or the variables $\{t,\nu,\tau,\tau'\}$ from Sec.~\ref{sec:phasespace}, or any other suitable combination thereof.

  \subsection{Born terms and gauge invariance}\label{sec:bornterms}

             The extraction of nucleon polarizabilities is complicated by the fact that the handbag part
             does not respect electromagnetic gauge invariance on its own. As a consequence, it is not transverse with respect to the photon momenta
             and the basis elements in Eqs.~\eqref{qcv-transverse-basis-Y} or~\eqref{qcv-onshell-basis} are not sufficient to fully describe its structure.
             Unfortunately the lack of transversality will also affect the transverse projection of the handbag amplitude as it does
             not satisfy the analyticity constraints that are manifest in Eq.~\eqref{qcv-onshell-basis}.

             According to their definition, the polarizabilities are related to the Lorentz-invariant dressing functions of the residual part $J_\text{R}^{\mu\nu}$ of the nucleon amplitude,
             \begin{equation}
                 J^{\mu\nu}_\text{R} = \widetilde{J}^{\mu\nu} - J^{\mu\nu}_\text{B}\,,
             \end{equation}
             where the following specific form of the resonant nucleon Born part is subtracted from the full amplitude:
             \begin{equation}\label{nca-born}
             \begin{split}
                 J^{\mu\nu}_\text{B}  = & -\Lambda_+^f\,\Gamma_N^\mu(-Q')\,S_N(P+\Sigma)\,\Gamma_N^\nu(Q)\,\Lambda_+^i \\
                                        &-\Lambda_+^f\,\Gamma_N^\nu(Q)\,S_N(P-\Sigma)\,\Gamma_N^\mu(-Q')\,\Lambda_+^i \,.
             \end{split}
             \end{equation}
             Here, $S_N(p) = (-i\slashed{p}+M)/(p^2+M^2)$ is the nucleon propagator with nucleon mass $M$, and
             \begin{equation}
                -i \Gamma^\mu_N(Q) =  F_1(Q^2)\,\gamma^\mu + \frac{iF_2(Q^2)}{4M}\,[\gamma^\mu,\slashed{Q}]
             \end{equation}
             is the onshell nucleon-photon vertex from Eq.~\eqref{current-onshell-standard} with the Dirac and Pauli form factors $F_1$ and $F_2$.
             If the intermediate nucleon momenta in Eq.~\eqref{nca-born} are onshell, we have
             \begin{equation}
                 S_N(p) \stackrel{p^2 \rightarrow -M^2}{\longlonglongrightarrow} \frac{2M \Lambda_+(p)}{p^2+M^2}  \,, \quad
                 \Lambda_+(p)=\frac{\mathds{1}+\hat{\Slash{p}}}{2}\,,
             \end{equation}
             and the positive-energy projectors ensure that the terms in Eq.~\eqref{nca-born} are transverse for an arbitrary nucleon-photon vertex,
             so that their sum is transverse in the Born limit $s=u=0$.
             However, with the ansatz for $\Gamma^\mu_N$ from above,
             the sum of both terms for $J^{\mu\nu}_\text{B}$ in~\eqref{nca-born} is also transverse in general kinematics
             so that the remainder $J^{\mu\nu}_\text{R}$ is gauge invariant as well.
             This is in general not true for other onshell-equivalent formulations of $\Gamma^\mu_N$, see~\cite{Scherer:1996ux} for a detailed discussion.

     \begin{figure*}[t]
     \center{
     \includegraphics[scale=0.104]{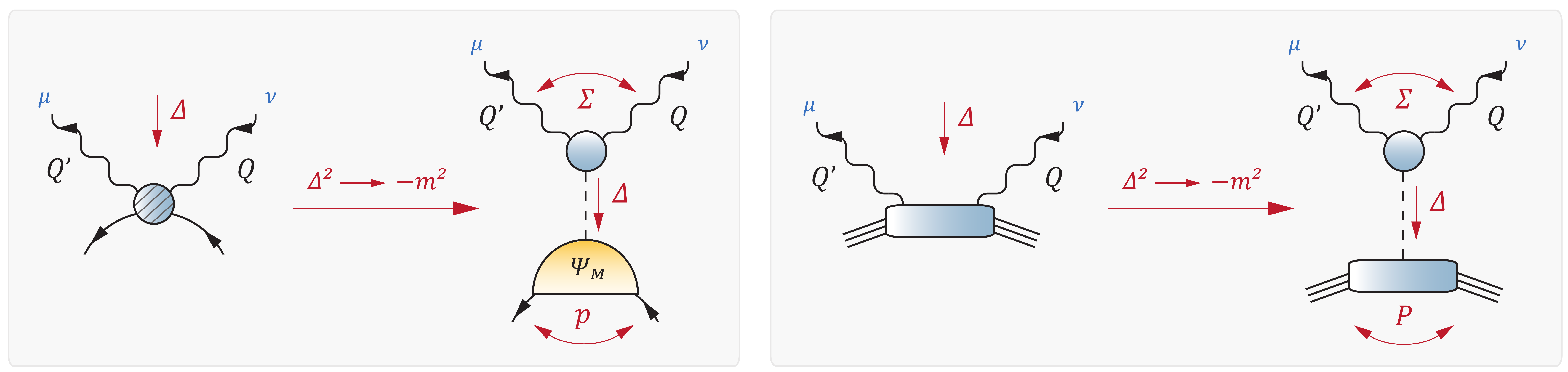}}
        \caption{$t-$channel meson poles in the quark Compton vertex (\textit{left panel}) and the nucleon Compton scattering amplitude (\textit{right panel}).
                 At the pion pole, the four-point functions separate into the pion's bound-state amplitude/the nucleon-pion current (i.e., the pole residue of the nucleon's pseudoscalar current)
                 and the $\pi\gamma\gamma$ transition current.}
        \label{fig:t-channel}
     \end{figure*}

             In order to obtain $J^{\mu\nu}_\text{R}$ from a quark-level calculation
             one would need to implement the full set of diagrams from Eq.~\eqref{scattering-reduced}:
             \begin{equation}\label{nca-full-set}
                \widetilde{J}^{\mu\nu} = J^{\mu\nu}_\text{Handbag} + J^{\mu\nu}_\text{Cat's ears} + J^{\mu\nu}_{G}\,.
             \end{equation}
             The third term involves the quark six-point function $G$ minus the quark Born parts that we shuffled from the third to the first term via Eq.~\eqref{scattering-handbag}.
             The sum of diagrams in Eq.~\eqref{nca-full-set} satisfies electromagnetic gauge invariance by construction and is therefore transverse.
             The residual part is then obtained by subtracting Eq.~(\ref{nca-born}),
             with nucleon electromagnetic form factors calculated consistently.
             Thereby $J^{\mu\nu}_\text{B}$ cancels the nucleon pole encapsulated in $J^{\mu\nu}_{G}$, and the transverse remainder encodes the polarizabilities.

             While the direct calculation of $J^{\mu\nu}_{G}$ might be feasible in the long term, it is certainly beyond our present capacities.
             Under the assumption of handbag dominance, a sensible alternative would be the construction of a 'gauge-invariant completion' of the calculated handbag diagrams.
             If the dynamics of $J^{\mu\nu}_\text{R}$ are dominated by the handbag contribution, one can attempt to add a kinematic term
             \begin{equation}\label{nca-residual}
                 J^{\mu\nu}_\text{R} = J^{\mu\nu}_\text{Handbag} + J^{\mu\nu}_\text{Kinematic}
             \end{equation}
             whose sole purpose is to restore transversality (while maintaining analyticity), so that all non-transverse contributions in the sum cancel exactly.
             Such constructions are possible and we will explore them in future work.
             For the remaining part we will focus on a feature of the Compton amplitude that is unhampered by such problems, namely, the $t-$channel resonance structure.

  \subsection{Meson poles in the $t$ channel}

             We recall from Eq.~\eqref{qcv-T-matrix} that the quark Compton vertex can be written as the contraction of the $t-$channel $q\bar{q}$ Green function $G$
             with the inhomogeneous term $\Gamma_0^{\mu\nu}$. If a rainbow-ladder truncation is employed, the last quantity simplifies to the quark Born diagrams $\Gamma_\text{B}^{\mu\nu}$.
             The Green function includes all meson-exchange poles at the respective location
             \begin{equation}
                 t = \frac{\Delta^2}{4M^2} \longrightarrow -\frac{m_\text{M}^2}{4M^2}
             \end{equation}
             where it factorizes into the structure
             \begin{equation}
                 G \stackrel{\Delta^2\rightarrow-m_\text{M}^2}{\longlonglongrightarrow} G_0\,\frac{ \Psi_\text{M} \conjg{\Psi}_\text{M} }{\Delta^2+m_\text{M}^2}\,G_0\,.
             \end{equation}
             Here, $\Psi_\text{M}$ defines the meson's homogeneous bound-state amplitude and $m_\text{M}$ the meson mass,
             and $G_0$ is the disconnected product of a dressed quark and antiquark propagator.
             Via Eq.~\eqref{qcv-T-matrix}, these poles also appear in the quark Compton vertex, cf.~Fig.~\ref{fig:t-channel},
            \begin{equation}\label{G-Pole2}
                \widetilde{\Gamma}^{\mu\nu} \stackrel{\Delta^2 \rightarrow -m_\text{M}^2}{\longlonglongrightarrow} \frac{J^{\mu\nu}_{\text{M}\gamma\gamma}}{\Delta^2+m_\text{M}^2}\,\Psi_\text{M}\,,
            \end{equation}
            as long as the bound-state wave function $\chi_\text{M}=G_0 \Psi_\text{M}$ has non-vanishing overlap with the Born part $\Gamma_\text{B}$, i.e.,
             \begin{equation}\label{r-def}
                J^{\mu\nu}_{\text{M}\gamma\gamma} := \text{Tr} \int\limits_{p} \conjg{\chi}_\text{M}(p,\Delta)\,\Gamma_\text{B}^{\mu\nu}(p,\Sigma,\Delta)\,\Big|_{\Delta^2\rightarrow -m_\text{M}^2}\,  \neq 0\,.
             \end{equation}
             The quantity $J^{\mu\nu}_{\text{M}\gamma\gamma}(\Sigma,\Delta)$ defines the two-photon transition current for a meson M with mass $m_\text{M}$ and given $J^{PC}$ quantum numbers.
             If it is non-zero, the respective pole will appear in the quark Compton vertex and consequently also in the nucleon Compton amplitude.
             Since these contributions are resonant and the onshell transition currents are conserved, they are transverse by themselves
             and we can extract them without concern for analyticity constraints.

             It is interesting that the pole structure of the quark Compton vertex can already be read off from the basis decomposition in Table~\ref{qcv-basis-1}.
             Consider, for example, a scalar two-photon current $J^{\mu\nu}_{\text{sc},\gamma\gamma}(\Sigma,\Delta)$. Since the relative momentum $p$ is integrated out,
             we have only the two unit vectors $s^\mu$ and $d^\mu$ from Eq.~\eqref{orthonormal-momenta} for its construction at our disposal, and a complete tensor decomposition consists of the elements
             \begin{equation}
                 \delta^{\mu\nu}, \quad s^\mu s^\nu \pm d^\mu d^\nu, \quad s^\mu d^\nu \pm d^\mu s^\nu\,.
             \end{equation}
             With Eq.~\eqref{vrsd-delta}, they correspond to the elements $\mathsf{X}_1^{\mu\nu}$, $\mathsf{X}_5^{\mu\nu}$, $\mathsf{X}_6^{\mu\nu}$, $\mathsf{X}_7^{\mu\nu}$ and $\mathsf{X}_8^{\mu\nu}$.
             Implementing the transversality constraint  with respect to the external photon legs leaves only
             two possible transverse Lorentz structures, namely $\mathsf{Y}_1^{\mu\nu}$ and $\mathsf{Y}_3^{\mu\nu}$
             from Eq.~\eqref{qcv-transverse-basis-Y}, for the onshell current.\footnote{Expressed in
             terms of the regular basis from Eq.~\eqref{transverse-scalar}, they are linear combinations of $t^{\mu\nu}_{Q'Q}$ and $t_{Q'Q'}^{\mu\alpha}\,t_{QQ}^{\alpha\nu}$.}
             Since the $\mathsf{Y}_i$ are orthogonal, scalar poles can only appear in the dressing functions associated with these two basis elements in the quark Compton vertex,
             combined with those Dirac structures in Table~\ref{qcv-basis-1} that also contribute to the bound-state amplitude of a scalar meson.
             The same observation holds also for the nucleon Compton scattering amplitude when expressed in the basis of Eq.~\eqref{qcv-nucleon-transverse}.

             Similarly, the only possible structure for a pseudoscalar two-photon current is $\mathsf{Y}_4^{\mu\nu} \sim \varepsilon^{\mu\nu\alpha\beta} s^\alpha d^\beta$, cf.~Eq.~\eqref{vrsd-epsilon},
             so that pseudoscalar poles in the Compton vertex can only appear in the dressing functions associated with $\mathsf{Y}_4^{\mu\nu}$.
             The same analysis for the axialvector case, whose onshell current is transverse to all three external legs, entails that the tensor structures $\mathsf{Y}_4^{\mu\nu}$, $\mathsf{Y}_6^{\mu\nu}$, $\mathsf{Y}_7^{\mu\nu}$, $\mathsf{Y}_8^{\mu\nu}$ and $\mathsf{Y}_9^{\mu\nu}$
             contribute axialvector-meson poles.

\begin{figure}[t]
     \center{
     \includegraphics[scale=0.35]{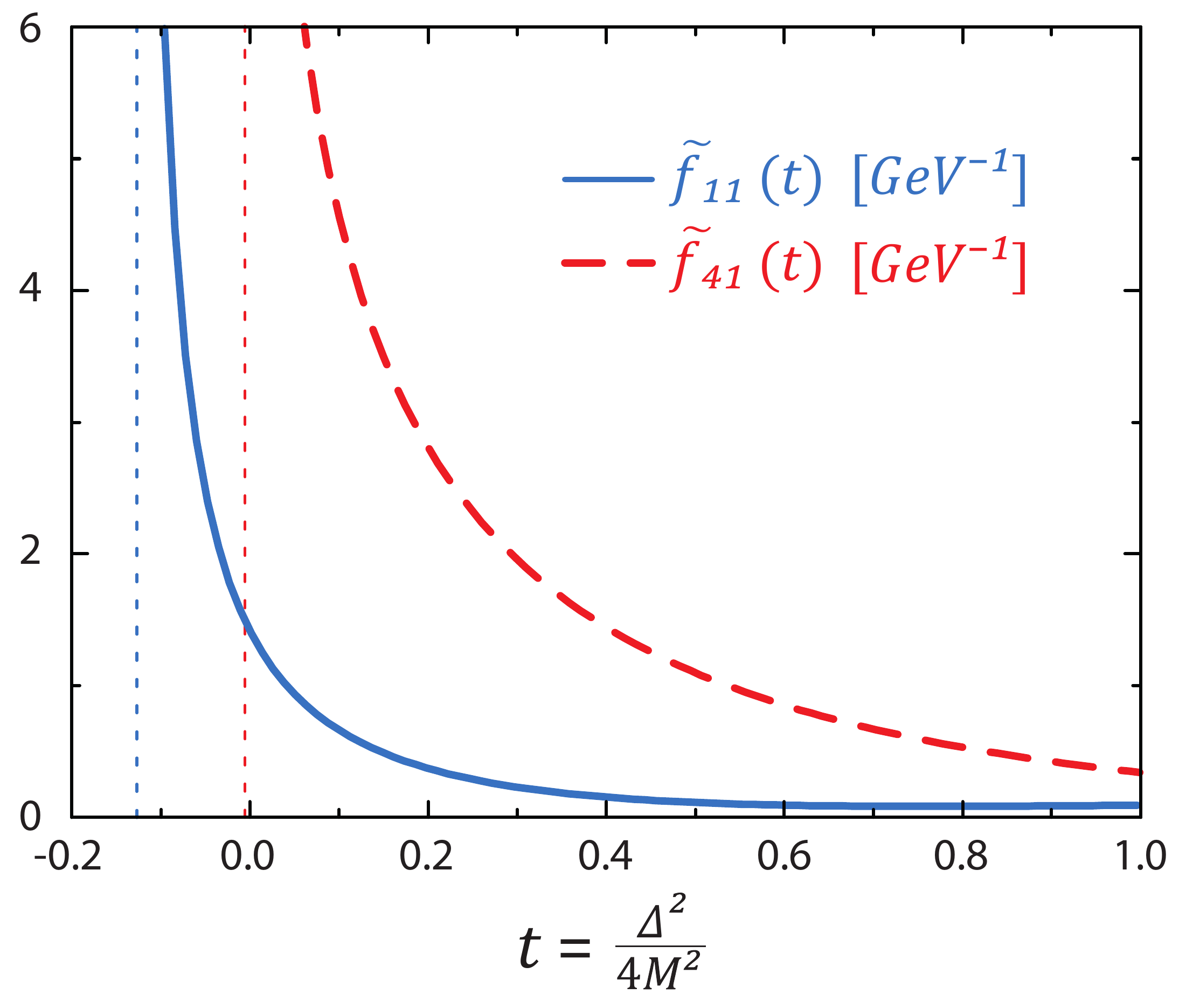}}
        \caption{Dressing functions $\tilde{f}_{1,1}(t)$ and $\tilde{f}_{4,1}(t)$ of the quark Compton vertex in the basis~\eqref{qcv-basis-decomposition-Y} at $p^2=0$, $Z=0$ and $\sigma=1$. We plot the $0^\text{th}$
                 Chebyshev moments in $z$ and $y$. The scalar and pion pole locations are shown by dotted vertical lines.}
        \label{fig:results}
     \end{figure}

             The dressing functions $\tilde{f}_{1,1}(t)$ and $\tilde{f}_{4,1}(t)$ of the quark Compton vertex in the basis decomposition of Eq.~\eqref{qcv-transverse-basis-Y} are shown in Fig.~\ref{fig:results}.
             In principle, these functions depend on the six momentum variables $p^2$, $z$, $y$, $t$, $\sigma$ and $Z$. We performed a Chebyshev expansion in the angular
             variables $z$ and $y$ and show only the zeroth Chebyshev moments. Furthermore, we set $p^2=Z=0$ and $\sigma=1$,
             so that the only remaining variable is $t=\Delta^2/(4M^2)$.
             In contrast to the nucleon Compton amplitude, we can solve the inhomogeneous BSE for the vertex at timelike (i.e., negative) values of $t$ directly
             since the only needed ingredients are the calculated quark propagator and the quark-photon vertex in the complex plane.
             Both dressing functions $\tilde{f}_{1,1}(t)$ and $\tilde{f}_{4,1}(t)$ diverge, as they should, at the respective scalar and pion pole locations which are denoted by dotted lines. With our rainbow-ladder input,
             the scalar meson mass is $m_\text{sc}=0.67$ GeV and the pion mass is $m_\pi=0.14$ GeV.

\begin{figure*}[t]
     \center{
     \includegraphics[scale=0.35]{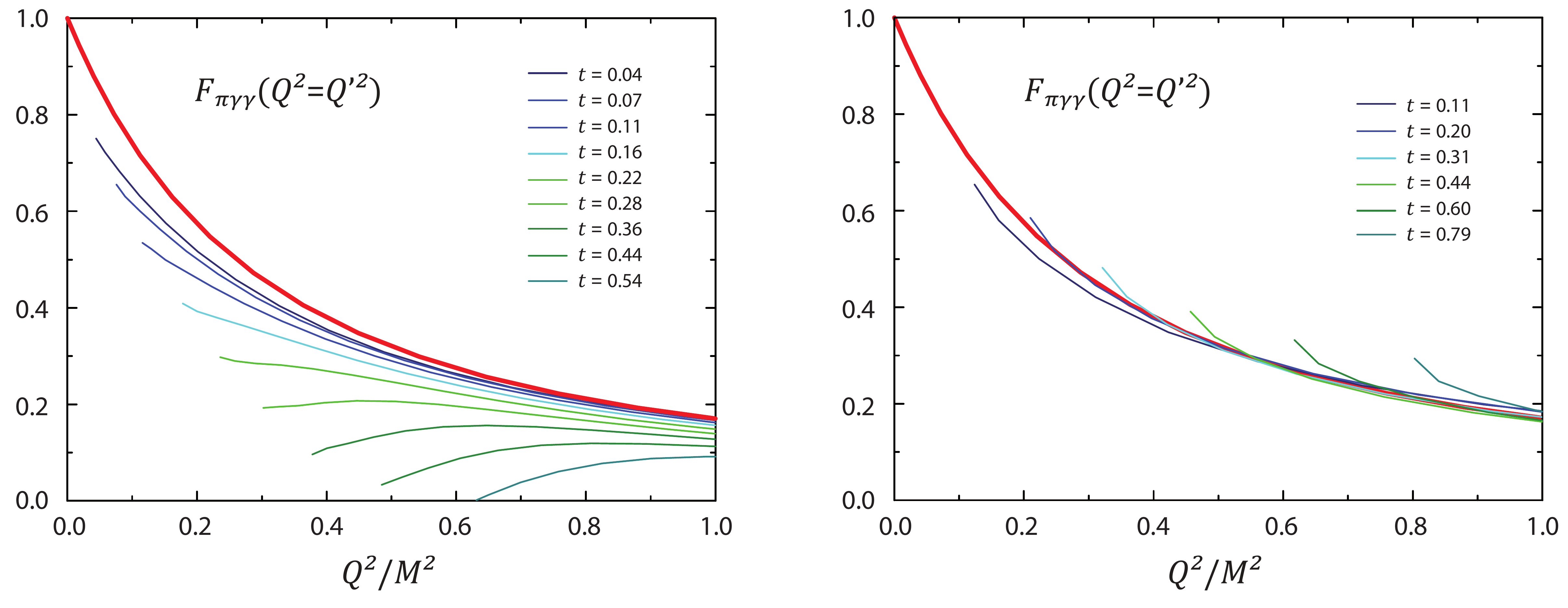}}
        \caption{Reconstruction of $F_{\pi\gamma\gamma}$ from the quark Compton vertex (\textit{left panel}) and the nucleon Compton amplitude (\textit{right panel}) via
                 Eq.~\eqref{pigammagamma-extraction}. The thick red line in each panel is the result of the direct calculation for the $\pi\gamma\gamma$ transition current;
                 the thin lines are the extrapolated values at various values of $t$.}
        \label{fig:results-2}
     \end{figure*}

             In order to check our numerical accuracy we can further extract the residues at these poles and compare them with known results from the literature.
             As discussed before, the basis elements involving $\mathsf{Y}_4^{\mu\nu}$ contain the pion $t-$channel pole via the $\pi^0\gamma\gamma$ form factor.
             The respective current reads
             \begin{equation}
                 J^{\mu\nu}_{\pi\gamma\gamma}(\Sigma,\Delta) = \frac{i g_{\pi\gamma\gamma}}{2\pi^2 f_\pi}\,F_{\pi\gamma\gamma}(\tau,\tau')\,\varepsilon^{\mu\nu\alpha\beta} \Sigma^\alpha \Delta^\beta\,,
             \end{equation}
             where $F_{\pi\gamma\gamma}(\tau,\tau')$ is normalized to unity at $\tau=\tau'=0$.
             In the chiral limit one has the exact result $g_{\pi\gamma\gamma} = \nicefrac{1}{2}$.
             For non-zero current-quark masses, the transition form factor $F_{\pi\gamma\gamma}$ in rainbow-ladder has been calculated in Ref.~\cite{Maris:2002mz}.
             In order to compare with our Compton dressing functions, we repeat that calculation without giving details.
             The same phase-space considerations from earlier apply here, and the accessible spacelike region in $\tau$ and $\tau'$ is shown in Fig.~\ref{fig:phasespace}.
             The result for the transition form factor in the symmetric case $\tau=\tau'$ is plotted by the thick (red) lines in Fig.~\ref{fig:results-2}.

             Next, we want to reconstruct the same form factor from the quark Compton vertex and the nucleon Compton amplitude.
             Although we calculate the latter only for spacelike $t$, the pion mass is sufficiently small so that the extraction of the pole residue is feasible.
             For $\Delta^2 \rightarrow -m_\pi^2$, the Compton vertex reduces to
             \begin{equation}\label{pigammagamma-qcv-1}
                 \widetilde{\Gamma}^{\mu\nu}(p,\Sigma,\Delta) \longrightarrow \frac{J^{\mu\nu}_{\pi\gamma\gamma}(\Sigma,\Delta)}{\Delta^2+m_\pi^2}\,\Psi_\pi(p,\Delta)\,,
             \end{equation}
             where $\Psi_\pi$ is the pion's bound-state amplitude
             \begin{equation}\label{pion-WF}
                \Psi_\pi(p,\Delta) = \sum_i h_i^{(\pi)}(p^2,z)\,i\gamma_5 \,\tau_i(p,\Delta)\left[\sigma_k\right]_\mathsf{ab}\,\delta_{AB}\,,
             \end{equation}
             with four Dirac structures that can be chosen as $\tau_i \in \{ \mathds{1},\, \Slash{p},\, \Slash{\Delta},\, [\Slash{p},\Slash{\Delta}] \}$.
             We have stated the color and flavor factors explicitly (the $\sigma_k$ are the Pauli matrices); with our color-flavor convention
             one obtains in the chiral limit: $h_1^{(\pi)}(p^2,z) = B(p^2)/f_\pi$, where $B(p^2)$ is the scalar dressing function of the inverse quark propagator~\cite{Maris:1997hd}.
             Similarly, the nucleon's Compton amplitude at the pion pole becomes
             \begin{equation}\label{pigammagamma-ncv-1}
                 \widetilde{J}^{\mu\nu}(P,\Sigma,\Delta) \longrightarrow \frac{J^{\mu\nu}_{\pi\gamma\gamma}(\Sigma,\Delta)}{\Delta^2+m_\pi^2}\,J_{\pi NN}(P,\Delta)\,,
             \end{equation}
             where $J_{\pi NN}$ is the residue of the nucleon's pseudoscalar current
             at the pion pole:
             \begin{equation}
                 J_{\pi NN}(P,\Delta) = G_{\pi NN}(\Delta^2)\,\Lambda_+^f i\gamma_5 \,\Lambda_+^i \left[\frac{\sigma_k}{2}\right]_\mathsf{ab}\,.
             \end{equation}
             The rainbow-ladder calculation of the pion-nucleon form factor $G_{\pi NN}$
             has been performed by us recently~\cite{Eichmann:2011pv}.
             The relations~\eqref{pigammagamma-qcv-1} and~\eqref{pigammagamma-ncv-1} are illustrated in Fig.~\ref{fig:t-channel}.

             With the bases of Eq.~\eqref{qcv-basis-decomposition-Y} for the quark Compton vertex and~\eqref{qcv-nucleon-transverse} for the nucleon Compton amplitude,
             the tensor structure of the pion transition form factor is proportional to $\mathsf{Y}_4$:
             \begin{equation}\label{Y4-pionpole}
                \mathsf{Y}_4^{\mu\nu} = \tfrac{1}{\sqrt{2}}\, \varepsilon^{\mu\nu\alpha\beta} s^\alpha d^\beta = \frac{\varepsilon^{\mu\nu\alpha\beta} \Sigma^\alpha \Delta^\beta}{2  M^2 \sqrt{2\sigma t\,(1-Z^2)}}\,,
             \end{equation}
             and the pion pole will appear in the respective dressing functions $\tilde{f}_{4,a}$ and $F_4$.
             If we work out Eqs.~(\ref{pigammagamma-qcv-1}--\ref{Y4-pionpole}) and define
             \begin{equation}
                  \mathcal{H}(t,\sigma,Z)  = \frac{4\pi^2 f_\pi}{ \sqrt{2\sigma t \,(1-Z^2)}}\left( t+\frac{m_\pi^2}{4M^2}\right),
             \end{equation}
             we can extract the form factor at the symmetric point $\tau=\tau'$ from the relations
             \begin{equation}\label{pigammagamma-extraction}
             \begin{split}
                  g_{\pi\gamma\gamma}\,F_{\pi\gamma\gamma}(\tau,\tau) &=  \mathcal{H}(t,\sigma,Z)\,\frac{\tilde{f}_{4,1}^v(p^2,z,y,t,\sigma,Z)}{2h_1^{(\pi)}(p^2,z)}\,, \\
                  g_{\pi\gamma\gamma}\,F_{\pi\gamma\gamma}(\tau,\tau) &= \mathcal{H}(t,\sigma,Z)\,\frac{F_4^v(t,\sigma,Y,Z)}{G_{\pi NN}(\Delta^2)}
             \end{split}
             \end{equation}
             which are valid in the limit $\tau=\tau' \Rightarrow Z=0$, $\sigma=4\tau-t$ and $t\rightarrow -m_\pi^2/(4M^2)$.
             We also set $p^2=z=y=0$ in the quark Compton vertex and $Y=0$ in the nucleon Compton amplitude.
             The superscript $v$ denotes the flavor projection to the isovector part.
             For the quark Compton vertex, flavor is just an overall factor, namely the square of the quark charge matrix in Eq.~\eqref{flavor-Q2},
             so that the isovector projection is $f_4^v = f_4/3$.
             For the nucleon Compton amplitude, the Dirac-Lorentz and flavor parts are entangled, cf. Eq.~\eqref{flavor-iso},
             and the isovector projection is
             \begin{equation}
                 F_4^v = \frac{3F_4^{11} - F_4^{22}}{3} = F_4^p-F_4^n\,.
             \end{equation}

             The extracted transition form factor is shown in Fig.~\ref{fig:results-2} for various values of $t$.
             For $t\rightarrow -m_\pi^2/(4M^2) \approx -0.005$, the curves should converge toward the result from the direct onshell calculation of the current.
             In the case of the quark Compton vertex, shown in the left panel, this behavior is nicely realized as long is $t$ is not too small.
             For small values of $t$, the iteration of the vertex BSE becomes
             unstable close to the pion pole.

             The extraction of $F_{\pi\gamma\gamma}$ from the scattering amplitude in the right panel of Fig.~\ref{fig:results-2} turned out to be more difficult.
             In order to implement the quark Compton vertex dressing functions $\tilde{f}_{4,1}$ in the nucleon Compton amplitude,
             we are so far restricted to zeroth Chebyshev moments in the variables $z$ and $y$.
             The reason is that the real relative momentum $p$ in the vertex is shifted to $p_3=p+P/3$ in the Compton amplitude, cf.~App.~\ref{sec:handbag-dirac},
             where $P=(P_i+P_f)/2$ is the average nucleon momentum. The vertex depends now on $\Gamma^{\mu\nu}(p_3,\Sigma,\Delta)$, and $p_3$ is complex because $P^2$ is negative.
             Therefore, we must calculate the vertex for complex $p_3^2$ and complex
             values of the new variables $z$ and $y$ that are constructed from $p_3$, $\Sigma$ and $\Delta$.
             The first task is straightforward, but a Chebyshev expansion in $z$ and $y$ is impractical because $|z|$ and $|y|$
             are no longer restricted to the interior of a unit circle.
             If the angular dependencies were small, keeping only the zeroth moments would be sufficient,
             but this is usually only true in certain basis decompositions of the vertex.

             For the purpose of generating the right panel of Fig.~\ref{fig:results-2}, we transformed the eight Dirac basis elements $\tau_a$ attached to $\mathsf{Y}_4^{\mu\nu}$
             to the basis
             \begin{equation*}
                 \mathds{1}\,,  \;
                 \Slash{p}\,,\;
                 \Slash{\Delta}\,, \;
                 [ \Slash{p},\,\Slash{\Delta} ]\,,\;
                 \Slash{\Sigma_t}\,,\;
                 [ \Slash{p},\,\Slash{\Sigma}_t ]\,,\;
                 [ \Slash{\Sigma_t},\,\Slash{\Delta} ]\,,\;
                 [ \Slash{p},\,\Slash{\Sigma_t},\,\Slash{\Delta} ]\,,
             \end{equation*}
             equipped with appropriate prefactors $p\cdot\Delta$, $p\cdot\Sigma$ and $\Sigma\cdot\Delta$ to ensure
             positive charge-conjugation and Bose symmetry. $\Sigma_t$ is chosen to be transverse to $\Delta$ and $p$
             so that only the first four elements contribute to the pion pole,
             as they are the only ones that appear in the pion's bound-state amplitude. In this basis the angular dependencies
             close to the pion pole are weak and a zeroth Chebyshev approximation is justified; however, the same approximation
             in a different basis can lead to large inaccuracies.

             We note that the problem could be cured by a moving-frame calculation of the quark Compton vertex,
             where its inhomogeneous BSE is solved directly in the kinematics that appear in the scattering amplitude.
             In that case all variables are real and no complex continuation is necessary, however at the price of a dependence on two further angular variables.
             Such a course might turn out indispensable for the extraction of polarizabilities where the full 128-dimensional basis decomposition of the vertex is integrated over.
             We will investigate this option in the~future.

     \section{Conclusions and Outlook} \label{sec:conclusions}

            We have provided a theoretical analysis as well as first numerical results
            for the nucleon's Compton scattering amplitude in the Dyson-Schwinger approach.
            We have previously derived a decomposition of the scattering amplitude at the quark-gluon level that satisfies electromagnetic gauge invariance by construction.
            However, its full numerical calculation is complicated by the presence of the six-quark Green function
            which includes the $s-$ and $u-$channel nucleon resonance terms.

            In this paper we have calculated the nonperturbative handbag contribution to the scattering amplitude.
            It is microscopically represented by the quark Compton vertex, which
            includes the quark Born terms and the quark-antiquark Green function in the $t$ channel that comprises all $t-$channel meson resonances.
            Based on a well-established rainbow-ladder ansatz for the quark-gluon interaction,
            we implemented the Dyson-Schwinger solution for the dressed quark propagator,
            the covariant Faddeev-equation solution for the nucleon's bound-state amplitude,
            and we solved the quark Compton vertex from its inhomogeneous Bethe-Salpeter equation.

            For the purpose of our calculations we have constructed a 128-dimensional orthogonal tensor basis for the quark Compton vertex
            and studied its symmetry properties.
            The same basis can be used for the nucleon Compton scattering amplitude,
            where the number of independent elements reduces to 32 and further down to 18 if the transversality conditions are implemented.
            Moreover, we have constructed a fully transverse tensor basis
            that is free of kinematic singularities in the relevant kinematic limits.
            We have established its relation with bases used in the literature,
            and we studied its simplification in the limits of virtual and forward Compton scattering.

            We have also generalized the Ball-Chiu construction to the two-photon case
            and established the general form of the fermion (and also scalar) Compton vertex
            that is compatible with electromagnetic gauge invariance.
            This construction can be used for devising models for the scattering amplitude at the quark level
            based on the dressed quark propagator, the nucleon wave function, and meson pole terms in the one- and two-photon vertices.
            On the other hand, it can also be useful for model building on the nucleon level, as it
            provides a gauge-invariant completion of the nucleon Born terms in general kinematics with an offshell nucleon propagator and nucleon-photon vertex.

            We have extracted the $\pi\gamma\gamma$ transition form factor at the $t-$channel pion pole,
            both from the calculated quark Compton vertex and the nucleon scattering amplitude.
            We obtained reasonable agreement with the result from a direct calculation.
            We found that the angular dependencies in the quark Compton vertex are not negligible, and
            in view of obtaining observables in other kinematic regions it might be necessary
            to solve the vertex in a moving frame.
            We note that an analogous analysis for the pion electroproduction amplitude, which is possible with the methods devised here,
            can in principle yield input for the experimental extraction of the pion's electromagnetic form factor from offshell momenta.

            We plan to continue our studies by extracting the experimentally measured polarizabilities in virtual and forward Compton scattering.
            This task has not been possible so far since the handbag contribution is not gauge invariant by itself, and therefore not transverse,
            so that the extracted polarizabilities would not satisfy the analyticity constraints.
            In the long term, electromagnetic gauge invariance can be restored by taking into account the full set of diagrams in the decomposition of the scattering amplitude.
            As an alternative one can attempt to construct ans\"atze for the scattering amplitude
            that include the dynamics of the calculated handbag diagrams while restoring transversality and analyticity.
            With that in mind, we also plan to investigate two-photon contributions to electromagnetic form factors
            that are determined from the nucleon Compton amplitude at spacelike momenta.

    \section{Acknowledgements}

     We are grateful to R. Alkofer, A. Bashir, M. Lutz and R. Williams for useful discussions.
     This work was supported by the Austrian Science Fund FWF under
     Erwin-Schr\"odinger-Stipendium No.~J3039,
            the Helmholtz International Center for FAIR
            within the LOEWE program of the State of Hesse,
            the Helmholtz Young Investigator Group No.~VH-NG-332, and by DFG TR16.

\begin{appendix}


  \section{Euclidean conventions}\label{sec:euclidean}

            We work in Euclidean momentum space with the conventions
            \begin{equation}
                p\cdot q = \sum_{k=1}^4 p_k \, q_k,\quad
                p^2 = p\cdot p,\quad
                \Slash{p} = p\cdot\gamma\,.
            \end{equation}
            A vector $p$ is spacelike if $p^2 > 0$ and timelike if $p^2<0$.
            The hermitian $\gamma-$matrices $\gamma^\mu = (\gamma^\mu)^\dag$ satisfy the anticommutation relations
            $\left\{ \gamma^\mu, \gamma^\nu \right\} = 2\,\delta^{\,\mu\nu}$, and we define
            \begin{equation}
                \sigma^{\mu\nu} = -\frac{i}{2} \left[ \gamma^\mu, \gamma^\nu \right]\,, \quad
                \gamma_5 = -\gamma^1 \gamma^2 \gamma^3 \gamma^4\,.
            \end{equation}
            In the standard representation one has:
            \begin{equation*}
                \gamma^k  =  \left( \begin{array}{cc} 0 & -i \sigma_k \\ i \sigma_k & 0 \end{array} \right), \;
                \gamma^4  =  \left( \begin{array}{c@{\quad}c} \mathds{1} & 0 \\ 0 & \!\!-\mathds{1} \end{array} \right), \;
                \gamma_5  =  \left( \begin{array}{c@{\quad}c} 0 & \mathds{1} \\ \mathds{1} & 0 \end{array} \right),
            \end{equation*}
            where $\sigma_k$ are the three Pauli matrices.
            The charge conjugation matrix is given by
            \begin{equation}
                C = \gamma^4 \gamma^2, \quad C^T = C^\dag = C^{-1} = -C\,,
            \end{equation}
            where $T$ denotes a Dirac transpose.
            Four-momenta are conveniently expressed through hyperspherical coordinates:
            \begin{equation}\label{APP:momentum-coordinates}
                p^\mu = \sqrt{p^2} \left( \begin{array}{l} \sqrt{1-z^2}\,\sqrt{1-y^2}\,\sin{\phi} \\
                                                           \sqrt{1-z^2}\,\sqrt{1-y^2}\,\cos{\phi} \\
                                                           \sqrt{1-z^2}\;\;y \\
                                                           \;\; z
                                         \end{array}\right),
            \end{equation}
            and a four-momentum integration reads:
            \begin{equation*} \label{hypersphericalintegral}
                 \int\limits_p := \frac{1}{(2\pi)^4}\,\frac{1}{2}\int\limits_0^{\infty} dp^2 \,p^2 \int\limits_{-1}^1 dz\,\sqrt{1-z^2}  \int\limits_{-1}^1 dy \int\limits_0^{2\pi} d\phi \,.
            \end{equation*}

     \begin{figure*}[t]
     \center{
     \includegraphics[scale=0.11]{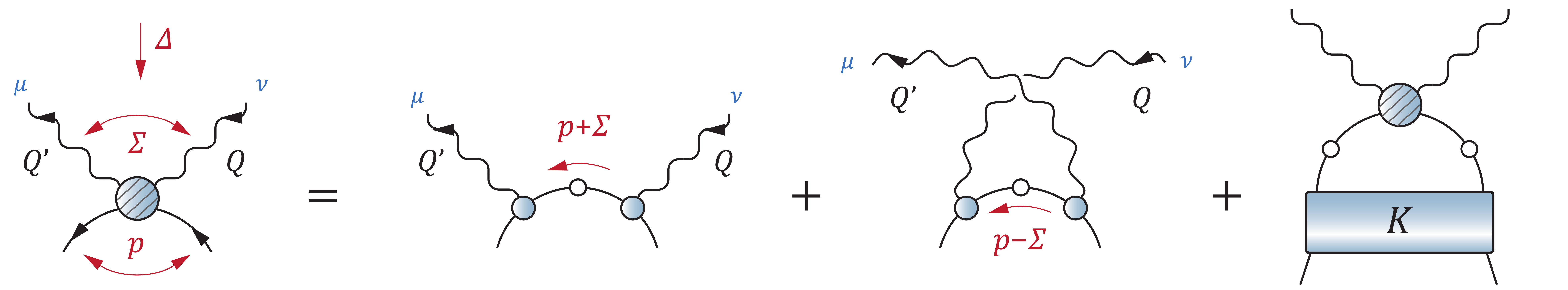}}
        \caption{Inhomogeneous BSE for the quark Compton vertex, Eqs.~\eqref{ibse-qcv} and~\eqref{qcv-bse}.}
        \label{fig:qcv-ibse}
     \end{figure*}

  \section{Inhomogeneous BSE for the quark Compton vertex}\label{sec:qcv-bse}

             \subsection{Inhomogeneous BSE}

             In this appendix we provide details on solving the inhomogeneous Bethe-Salpeter equation~\eqref{ibse-qcv} for the quark Compton vertex.
             Reattaching indices and momentum labels, it has the explicit form
             \begin{equation}\label{qcv-bse}
             \begin{split}
                 \big[\widetilde{\Gamma}-\Gamma_\text{B}\big]^{\mu\nu}_{\alpha\beta}(p,\Sigma,\Delta) &=
                                                                   \! \int\limits_{p'} \mathcal{K}_{\alpha\alpha'\beta'\beta}\, \widetilde{\chi}_{\alpha'\beta'}^{\mu\nu}(p',\Sigma,\Delta)\,,\\
                 \widetilde{\chi}^{\mu\nu}(p,\Sigma,\Delta) &= S(p_+)\,\widetilde{\Gamma}^{\mu\nu}(p,\Sigma,\Delta)\,S(p_-)
             \end{split}
             \end{equation}
             which is illustrated in Fig.~\ref{fig:qcv-ibse}. $S(p_\pm)$ with $p_\pm = p \pm \tfrac{\Delta}{2}$ denotes the dressed quark propagator from Eq.~\eqref{quark-propagator}, and
             the rainbow-ladder kernel is given by
             \begin{equation}
                 \mathcal{K}_{\alpha\alpha'\beta'\beta}= \frac{4}{3}\,Z_2^2\,\frac{4\pi \alpha(k^2)}{k^2}\,T^{\rho\sigma}_k i\gamma^\rho_{\alpha\alpha'}\,i\gamma^\sigma_{\beta'\beta}\,,
             \end{equation}
             Here, $Z_2$ is the quark renormalization constant, $4/3$ is the color trace, $T_k^{\rho\sigma}=\delta^{\rho\sigma}-\hat{k}^\rho \hat{k}^\sigma$
             is the transverse projector with respect to the gluon momentum $k=p-p'$,
             and $\alpha(k^2)$ is the rainbow-ladder quark-gluon interaction from Ref.~\cite{Maris:1999nt};
             see also Eqs.~(13)--(14) in Ref.~\cite{Eichmann:2011ec}.

             The rainbow-ladder BSE respects the WTI for the Compton vertex: contracting the BSE with ${Q'}^\mu$
             and inserting the WTIs \eqref{qcv-wti-r} and~\eqref{qcv-wti-wf} for the Compton vertices $\widetilde{\Gamma}^{\mu\nu}$ and $\widetilde{\chi}^{\mu\nu}$ yields the condition
             \begin{equation}
             \begin{split}
                 & \Gamma^\nu(p_f^+,Q) - \Gamma^\nu(p_f^-,Q) \\
                 & = \! \int\limits_{p'} \mathcal{K}_{\alpha\alpha'\beta'\beta} \left( \chi^\nu({p^+_f}',Q) - \chi^\nu({p^-_f}',Q) \right)
             \end{split}
             \end{equation}
             which is satisfied since the quark-photon vertex is calculated from its own BSE,
             \begin{equation}
                 \Gamma^\nu(p,Q) = Z_2 \,i\gamma^\nu + \! \int\limits_{p'} \mathcal{K}_{\alpha\alpha'\beta'\beta} \,\chi^\nu(p',Q)\,,
             \end{equation}
             and the rainbow-ladder kernel only depends on the exchange momentum $p-p'$.

             \subsection{Explicit form of the BSE}

             We work in the frame defined by Eq.~\eqref{simple-frame} for the external momenta $\Sigma$ and $\Delta$, Eq.~\eqref{simple-frame-p}
             for the external relative momentum $p$, and
             \begin{equation}
                p'=\sqrt{{p'}^2}\left(\begin{array}{l} \sqrt{1-{z'}^2}\,\sqrt{1-{y'}^2}\,\sin\phi \\ \sqrt{1-{z'}^2}\,\sqrt{1-{y'}^2}\,\cos\phi \\ \sqrt{1-{z'}^2}\,y' \\ z' \end{array}\right)
             \end{equation}
             for the internal relative momentum. The unit vectors that characterize the external momenta are then given by Eq.~\eqref{vrsd}.
             The unit vectors constructed from the internal relative momentum $p'$ are defined in analogy to Eqs.~\eqref{orthonormal-momenta} and~\eqref{def-v}:
             $d$ and $s$ remain unchanged, but we have in addition
             \begin{equation}
                 {r'}^\mu = \widehat{p'_t}^\mu\,, \quad {v'}^\mu = \epsilon^{\alpha\beta\gamma\delta} {r'}^\alpha s^\beta d^\gamma\,,
             \end{equation}
             which yields
             \begin{equation}
                 r'=\left(\begin{array}{c} \sin\phi \\ \cos\phi \\ 0 \\ 0 \end{array}\right), \quad
                 v'=\left(\begin{array}{c} \cos\phi \\ -\sin\phi \\ 0 \\ 0 \end{array}\right).
             \end{equation}
             Lorentz-invariant relations that combine external and internal
             basis elements, but otherwise involve no further momentum, can therefore only depend on the variable $r\cdot r' = v\cdot v' = \cos\phi$.
             Furthermore, in analogy to the set $\omega=\{ p^2,\,z,\,y\}$ of Eq.~\eqref{qcv-variables},
             we collect the loop variables via
             \begin{equation}
                 \omega' := \left\{ \; {p'}^2\,,  \;
                                   z'=\hat{p'}\cdot\hat{\Delta}\,, \;
                                   y'=\widehat{p'_T}\cdot\widehat{\Sigma_T} \; \right\}.
             \end{equation}

             To start with, we express the vertices $\Gamma^{\mu\nu}$ and $\chi^{\mu\nu}$ by their basis decomposition~\eqref{qcv-basis-decomposition}, where
             $f_{i,a}$ and $\tilde{f}_{i,a}$ denote the Lorentz-invariant dressing functions in either case.
             Second, we exploit the orthogonality relation~\eqref{qcv-orthonormality} for the basis elements by multiplying the equations
             with $\conjg{\tau}_{i,a}^{\nu\mu}$ from the left and taking traces. This yields:
             \begin{equation}\label{qcv-bse-detail}
             \begin{split}
                 f_{i,a}(\omega,\lambda) &= f^0_{i,a}(\omega,\lambda) +\int_{p'} g(k^2) \,K_{ij,ab}\,\tilde{f}_{j,b}(\omega',\lambda)\,, \\
                 \tilde{f}_{i,a}(\omega,\lambda) &= G_{ij,ab}(\omega,t)\,f_{j,b}(\omega,\lambda)\,,
             \end{split}
             \end{equation}
             where
             \begin{equation}\label{KGg}
             \begin{split}
                 K_{ij,ab} &= \tfrac{1}{4} \text{Tr} \left\{ \conjg{\tau}_{i,a}^{\nu\mu}(r,s,d)\,\gamma^\rho\, \tau_{j,b}^{\mu\nu}(r',s,d)\,\gamma^\sigma\right\} T_k^{\rho\sigma}\,,\\
                 G_{ij,ab} &= \tfrac{1}{4} \text{Tr} \left\{ \conjg{\tau}_{i,a}^{\nu\mu}(r,s,d)\,S(p_+)\, \tau_{j,b}^{\mu\nu}(r,s,d)\,S(p_-)\right\},\\
                 g(k^2) &= -\frac{4}{3}\,Z_2^2\,\frac{4\pi \alpha(k^2)}{k^2}\,.
             \end{split}
             \end{equation}

             The simplicity of the rainbow-ladder kernel, which depends only on the gluon momentum $k=p-p'$ and enters through the projector
             $T_k^{\rho\sigma}=\delta^{\rho\sigma}-\hat{k}^\rho \hat{k}^\sigma$,
             entails that the kernel matrix $K_{ij,ab}$ only depends on the external and internal relative momenta $p$ and $p'$.
             The propagator matrix $G_{ij,ab}$ does not involve any integration and thus depends only on $p$, but now also on the external momentum $\Delta$,
             i.e., on the Lorentz-invariant variables $\omega=\{p^2,z,y\}$ and $t$. The remaining invariants $\sigma$ and $Z$
             enter the equation merely as external parameters in the driving term $f^0_{i,a}$.

             The typical strategy in an iterative solution of Eq.~\eqref{qcv-bse-detail} would be to compute and store the kernel and propagator matrices
             for given values of the external parameters $t$, $\sigma$ and $Z$ in advance, i.e., before the actual iteration is performed.
             Upon integrating the combination $g(k^2)\,K_{ij,ab}$ in Eq.~\eqref{qcv-bse-detail} over the angle $\phi$,
             which does not appear in the dressing functions, we obtain a kernel of the form
             \begin{equation}\label{kernel-cheby}
                 \tilde{K}_{ij,ab}(p^2,{p'}^2, z, z', y, y') \; \rightarrow \; \tilde{K}_{ij,ab}^{mm',nn'}(p^2,{p'}^2)\,.
             \end{equation}
             The additional indices in the second step shall indicate that we have employed further Chebyshev expansions in the angular variables for efficiency,
             which is justified since the angular dependencies are typically small.
             Unfortunately, even with economical numerical accuracy, the above expression requires several hundred GB of memory for storage,
             and so we must look for better strategies.

             \subsection{Breaking down the kernel} \label{sec:ibse-kernel}

        \renewcommand{\arraystretch}{1.2}

\begin{table*}
             \begin{equation*}
                K_{ij,ab}=\left( \begin{array}{c|cc|c|c|c|c|c|cc|cc|cc|cc}
                                               K   &               &                 &     &   &   &   &   &    &    &    &    &    &    &    &     \\ \hline
                                                   & (2c^2-1) \,K  &  -2c\, K'       &     &   &   &   &   &    &    &    &    &    &    &    &     \\
                                                   & -2c \,K'      & (1-2c^2)\, K    &     &   &   &   &   &    &    &    &    &    &    &    &     \\ \hline  
                                                   &               &                 & -K  &   &   &   &   &    &    &    &    &    &    &    &     \\ \hline
                                                   &               &                 &     & K &   &   &   &    &    &    &    &    &    &    &     \\ \hline
                                                   &               &                 &     &   & K &   &   &    &    &    &    &    &    &    &     \\ \hline
                                                   &               &                 &     &   &   & K &   &    &    &    &    &    &    &    &     \\ \hline
                                                   &               &                 &     &   &   &   & K &    &    &    &    &    &    &    &     \\ \hline                                                                                                     &               &                 &     &   &   &   &   & cK & K' &    &    &    &    &    &     \\
                                                   &               &                 &     &   &   &   &   & K' &-cK &    &    &    &    &    &     \\ \hline
                                                   &               &                 &     &   &   &   &   &    &    & cK & K' &    &    &    &     \\
                                                   &               &                 &     &   &   &   &   &    &    & K' &-cK &    &    &    &     \\ \hline
                                                   &               &                 &     &   &   &   &   &    &    &    &    & cK & K' &    &     \\
                                                   &               &                 &     &   &   &   &   &    &    &    &    & K' &-cK &    &     \\ \hline
                                                   &               &                 &     &   &   &   &   &    &    &    &    &    &    & cK & K'       \\
                                                   &               &                 &     &   &   &   &   &    &    &    &    &    &    & K' &-cK       \\
                                          \end{array}\right).
             \end{equation*}
             \caption{Full kernel of the Bethe-Salpeter equation for the quark Compton vertex. The kernels $K$ and $K'$ are given in Eqs.~\eqref{K_ab} and \eqref{K'_ab}, respectively.}
             \label{table:kernel-lorentz}
\end{table*}

         \renewcommand{\arraystretch}{1.0}

             Let us analyze the kernel in Eq.~\eqref{KGg} in more detail. We can exploit the basis decomposition~\eqref{basis-element}
             for the quark Compton vertex and write
             \begin{equation}\label{qcv-bse-kernel-explicit}
                 K_{ij,ab}  = \mathsf{X}_i^{\mu\nu} \,{\mathsf{X}'_j}^{\mu\nu}  \,\tfrac{1}{4} \,\text{Tr} \left\{ \conjg{\tau}_a \left[\begin{array}{c} \mathds{1} \\ \gamma_5 \end{array}\right]_i
                     \gamma^\rho_T \left[\begin{array}{c} \mathds{1} \\ \gamma_5 \end{array}\right]_j \tau'_b \,\gamma^\rho_T\right\}  \,,
             \end{equation}
             where we have abbreviated
             \begin{equation}
                 \mathsf{X}_i^{\mu\nu} = \mathsf{X}_i^{\mu\nu}(r,s,d)\,, \quad
                 {\mathsf{X}'_j}^{\mu\nu} = \mathsf{X}_j^{\mu\nu}(r',s,d)
             \end{equation}
             and similarly for the $\tau$'s.
             We have also contracted the $\gamma$-matrices with the gluon projector, so that $\gamma^\rho_T = \gamma^\rho - \hat{\Slash{k}}\,\hat{k}^\rho$
             is transverse to the gluon.
             The simple structure of the basis elements, where Lorentz and Dirac parts are completely factorized,
             should have consequences for the kernel as well.
             $\mathsf{X}_i^{\mu\nu} \mathsf{X}_j^{\mu\nu}=\delta_{ij}$ is diagonal in $i,j$, and the only complication in
             $\mathsf{X}_i^{\mu\nu} \,{\mathsf{X}'_j}^{\mu\nu}$ is the additional dependence on $r'$.
             Thus, the only scalar product that can enter the Lorentz trace is $r\cdot r'$, which suggests near orthogonality as well.

             The following Dirac traces can occur in Eq.~\eqref{qcv-bse-kernel-explicit}:
             \begin{equation}\label{qcv-bse-kernel-dirac}
             \begin{split}
                 \mathds{1} \times \mathds{1} \rightarrow\quad &  \tfrac{1}{4} \,\text{Tr} \left\{ \conjg{\tau}_a  \gamma^\rho_T \,\tau'_b \,\gamma^\rho_T\right\}=:K_{ab} \,,\\
                 \gamma_5 \times \mathds{1}\rightarrow\quad   &  \tfrac{1}{4} \,\text{Tr} \left\{ \conjg{\tau}_a  \gamma_5\,\gamma^\rho_T \,\tau'_b \,\gamma^\rho_T\right\} =: \frac{K'_{ab}}{\sqrt{1-(r\cdot r')^2}}\,,  \\
                 \mathds{1} \times \gamma_5\rightarrow\quad   &  \tfrac{1}{4} \,\text{Tr} \left\{ \conjg{\tau}_a \gamma^\rho_T \, \gamma_5\,\tau'_b \,\gamma^\rho_T\right\} = -\frac{K'_{ab}}{\sqrt{1-(r\cdot r')^2}}\,,\\
                 \gamma_5 \times \gamma_5\rightarrow\quad     &  \tfrac{1}{4} \,\text{Tr} \left\{ \conjg{\tau}_a \gamma_5\, \gamma^\rho_T \,\gamma_5\,\tau'_b \,\gamma^\rho_T\right\}=-K_{ab} \,,
             \end{split}
             \end{equation}
             where we have pulled out the prefactor in $K'_{ab}$ for later convenience.

             Combining this with the Lorentz traces, the result for the kernel $K_{ij,ab}$ is given in Table~\ref{table:kernel-lorentz}.
             It is a $16\times 16$ matrix in the indices $i,j$, where
             each entry is a further $8\times 8$ matrix proportional to $K_{ab}$ or $K'_{ab}$. We abbreviated $c=r\cdot r'=\cos\phi$.
             With the basis of Table~\ref{qcv-basis-1}, the kernel is indeed almost diagonal in $i,j$ (entries not shown are zero).
             We observe that so far we have reduced the $128\times 128$ kernel elements to $5\times 64$ independent entries
             \begin{equation}\label{5-entries}
                 K_{ab}\,, \;\; c\,K_{ab}\,, \;\; c^2\,K_{ab}\,, \;\; K'_{ab}\,, \;\; c\,K'_{ab}
             \end{equation}
             that appear in Table~\ref{table:kernel-lorentz} and have to be integrated over separately.
             It is interesting to note that this structure is quite general since it is merely the consequence of our choice of basis.
             The actual ansatz for the Bethe-Salpeter kernel only enters via the Dirac traces $K$ and $K'$ to which we will turn now.

             The $8\times 8$ matrices $K_{ab}$ and $K'_{ab}$ defined in Eq.~\eqref{qcv-bse-kernel-dirac}
             depend on the momenta $\hat{k}$, $r$, $r'$, $s$ and $d$, and therefore only on the Lorentz-invariants
             \begin{equation}
                   r\cdot r'\,,\quad \hat{k}\cdot r\,, \quad \hat{k}\cdot r'\,, \quad  \hat{k}\cdot s\,, \quad  \hat{k}\cdot d\,.
             \end{equation}
             Since the normalized gluon momentum enters through the projector, $\hat{k}$ can only appear quadratically.
             The explicit calculation yields
        \renewcommand{\arraystretch}{1.2}
             \begin{equation}\label{K_ab}
                K_{ab}=\left( \begin{array}{c|ccc|crr|c}
                                          a_1  &      &     &     &      &       &       &   \\ \hline
                                               & a_2  & b_1 & b_2 &      &       &       &   \\
                                               & b'_1 & a_3 & b_3 &      &       &       &   \\
                                               & b'_2 & b_3 & a_4 &      &       &       &   \\ \hline
                                               &      &     &     & a_5  &   b_4 &  b_2  &   \\
                                               &      &     &     & b_4  &  a_6  & -b_1  &   \\
                                               &      &     &     & b'_2 & -b'_1 & a_7   &   \\ \hline
                                               &      &     &     &      &       &       & a_8
                                          \end{array}\right),
             \end{equation}
             where an empty slot again means that the respective entry is zero.
             $K_{ab}$ is characterized by 12 independent entries:
        \renewcommand{\arraystretch}{1.2}
             \begin{equation}
                \begin{array}{rl}
                   a_1 &= 3 \,, \\
                   a_2 &= -(r\cdot r') - 2\,(r\cdot \hat{k})(\hat{k}\cdot r') \,, \\
                   a_3 &= -1-2\, (\hat{k}\cdot s)^2  \,,  \\
                   a_4 &= -1-2\, (\hat{k}\cdot d)^2  \,, \\
                   a_5 &=  -a_2-(a_1+a_3) (r\cdot r')  \,, \\
                   a_6 &=  -a_2-(a_1+a_4) (r\cdot r') \,, \\[5mm]
                \end{array}
             \end{equation}
             and
             \begin{equation}
                \begin{array}{rl}
                   b_1 &=   -2\,(s\cdot \hat{k})\,(\hat{k}\cdot r) \,, \\
                   b'_1 &=  -2\,(s\cdot \hat{k})\,(\hat{k}\cdot r') \,, \\
                   b_2 &=   -2\,(d\cdot \hat{k})\,(\hat{k}\cdot r) \,,  \\
                   b'_2 &=  -2\,(d\cdot \hat{k})\,(\hat{k}\cdot r')  \,,  \\
                   b_3 &=   -2\,(d\cdot \hat{k})\,(\hat{k}\cdot s)  \,,  \\
                   b_4 &= -b_3\,(r\cdot r')\,,
                \end{array}
             \end{equation}
             with $a_7 =  -(a_1+a_3+a_4)$, $a_8 =  -(a_2+a_5+a_6)$.
             The opposite-parity kernel $K'_{ab}$ becomes:
         \renewcommand{\arraystretch}{1.2}
             \begin{equation}\label{K'_ab}
                K'_{ab}=\left( \begin{array}{c|ccc|ccc|r}
                                               &      &     &     &      &       &        &    \\ \hline
                                               &      &     &     &      &       &        & -c'_0  \\
                                               &      &     &     &      &       &        & -c'_1  \\
                                               &      &     &     &      &       &        & -c'_2  \\ \hline
                                               &      &     &     & c_4  &   c_3 &  c_1   &   \\
                                               &      &     &     & c'_3 &  -c_4 &  c_2   &   \\
                                               &      &     &     & c'_1 &  c'_2 &        &   \\ \hline
                             \protect{\;\;}   & -c_0 &-c_1 &-c_2 &      &       &        &
                                          \end{array}\right),
             \end{equation}
             with
             \begin{equation}
                \begin{array}{rl}
                   c_0  &= 1+a_2\, r\cdot r' + 2\,(\hat{k}\cdot r')^2\,,  \\
                   c'_0 &= 1+a_2\, r\cdot r' + 2\,(\hat{k}\cdot r)^2\,,  \\
                   c_1  &= b_1\, r\cdot r' - b'_1\,,   \\
                   c'_1 &= b'_1\, r\cdot r' - b_1 \,,  \\
                   c_2  &= b_2\, r\cdot r' - b'_2 \,,  \\
                   c'_2 &= b'_2\, r\cdot r' - b_2 \,,  \\
                   c_3  &= c_0 - (a_1+a_4)- (a_2+a_6)\,r\cdot r' \,, \\
                   c'_3 &= c'_0 - (a_1+a_4)- (a_2+a_6)\,r\cdot r' \,, \\
                   c_4  &= -(b_3+b_4\, r\cdot r')\,.
                \end{array}
             \end{equation}
             Since $c_3-c'_3=c_0-c'_0$, there are only $8$ independent entries here, which yields $20$ in total
             for $K_{ab}$ and $K'_{ab}$.
             In combination with the 5 independent elements from Eq.~\eqref{5-entries}, we arrive at $12\times 3 + 8\times 2=52$ entries
             for the kernel $K_{ij,ab}$ and therefore
             have reduced memory requirements by a factor $\sim 300$.

             The structure of Table~\ref{table:kernel-lorentz} owes its simplicity to our chosen basis,
             in particular, to the definite properties with respect to the columns $\mathsf{S}$ and $\mathsf{T}$
             in Table~\ref{qcv-basis-1}. Operations that can be performed on the Lorentz indices alone do not affect
             the inhomogeneous BSE~\eqref{qcv-bse}, so that each $\{\mathsf{S}, \mathsf{T}\}$ subspace
             entails a decoupled equation. The same is true for operations that can be performed on the external momenta
             $s$ and $d$ that enter the equation. (The remaining momenta $v$ and $r$ can appear both outside and
             inside  the loop integral, via $v'$ and $r'$, and therefore do not decouple.) This leads to 11 decoupled blocks in
             Table~\ref{table:kernel-lorentz} and 11 decoupled inhomogeneous BSEs which can be solved independently.
             Their only difference is
             the projection of the driving term onto each basis element, i.e., the $f_{i,a}^0$ in Eq.~\eqref{qcv-bse-detail}.

             Analogous observations hold for simpler BSEs, for example those for the vector and axialvector vertices whose twelve basis elements
             decouple into transverse and longitudinal BSEs with respect to the external total momentum~\cite{Eichmann:2011vu,Eichmann:2011pv}.
             Another closely related example
             is the tensor vertex where our analysis can be immediately applied and yields six decoupled equations.

    \renewcommand{\arraystretch}{0.8}

             We can perform a similar dissection for the propagator matrix $G_{ij,ab}$ in Eq.~\eqref{KGg}. Evaluating Eq.~\eqref{basis-element}, with $p_\pm = p \pm \tfrac{\Delta}{2}$, yields
             \begin{equation}
             \begin{split}
                  G_{ij,ab} &= \delta_{ij}\,\tfrac{1}{4} \,\text{Tr} \left\{ \conjg{\tau}_a \left[\begin{array}{c} \mathds{1} \\ \gamma_5 \end{array}\right]_i
                     S(p_+) \left[\begin{array}{c} \mathds{1} \\ \gamma_5 \end{array}\right]_j \tau_b \,S(p_-)\right\} \\[2mm]
                    & = \delta_{ij}\,\tfrac{1}{4} \,\text{Tr} \left\{ \conjg{\tau}_a \, S(\pm p_+)  \,\tau_b \,S(p_-)\right\}.
             \end{split}
             \end{equation}
             Since no integration is involved here, all basis elements depend only on the external momenta $r$, $s$ and $d$, and the Lorentz trace is diagonal in $i,j$.
             This also means that only the diagonal components $\mathds{1}\times \mathds{1}$ and $\gamma_5\times \gamma_5$ survive.
             The resulting two $8\times 8$ matrices in $a,b$ are, however, more complicated
             than their kernel counterparts because of the momentum dependence of the quark propagators,
             and it is more convenient to perform these traces numerically.
             $G_{ij,ab}$ carries an explicit dependence on the external variable $t$, and a Chebyshev expansion analogous to Eq.~\eqref{kernel-cheby} yields the form
             \begin{equation}
                 G_{ij,ab}(p^2,z,y,t) \; \longrightarrow \; G_{ij,ab}^{mm',nn'}(p^2,t).
             \end{equation}

  \section{Nucleon handbag diagrams} \label{sec:nucleon:handbag}

         \renewcommand{\arraystretch}{1.0}

\begin{figure*}[t]
            \begin{center}
            \includegraphics[scale=0.17]{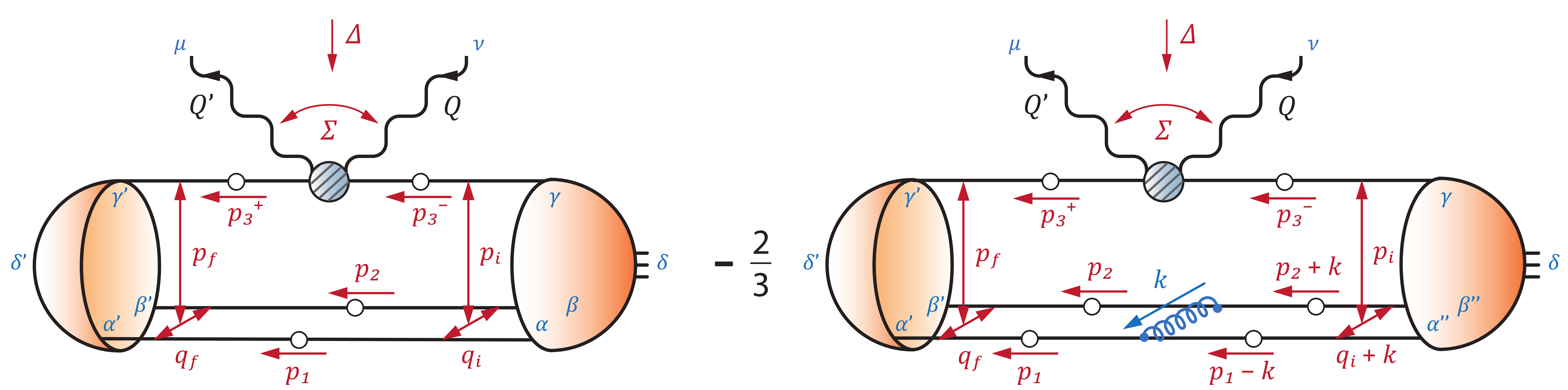}
            \caption{Notation and kinematics for the handbag diagrams of Eq.~\eqref{scattering-handbag} in the nucleon's rainbow-ladder truncated Compton scattering amplitude.
                     } \label{fig:current-kinematics}
            \end{center}
            \end{figure*}

            In the following we detail the calculation of the handbag contribution to the nucleon's Compton scattering amplitude,
            illustrated in Fig.~\ref{fig:current-kinematics}.
            In rainbow-ladder truncation, it is the sum of the impulse-approximation and spectator-kernel diagrams.
            In principle one must add up all three permutations $a=1,2,3$, where $a$ is the label of the quark that couples to the current.
            However, the symmetry properties of the nucleon amplitude relate the permuted diagrams among each other so that
            it is sufficient to consider only one of the three permutations explicitly. In the following we choose $a=3$,
            which corresponds to a coupling to the upper quark line as shown in Fig.~\ref{fig:current-kinematics}.

             \subsection{Nucleon bound-state amplitude}

            We first collect the properties of the nucleon's covariant bound-state amplitude that appears in the diagrams.
            The nucleon amplitude including its full Dirac, flavor and color dependence reads
              \begin{equation}\label{FE:nucleon_amplitude_full}
                  \mathbf{\Psi}(p,q,P) = \left( \sum_{n=1}^2  \Psi_n \, \mathsf{F}_n\right) \frac{\varepsilon_{ABC}}{\sqrt{6}}\,,
              \end{equation}
            where $[\Psi_n]_{\alpha\beta\gamma\delta}(p,q,P)$ is the spin-momentum amplitude and $[\mathsf{F}_n]_{abcd}$
            are the two flavor tensors. They transform as doublets under
            the permutation group $\mathds{S}^3$ so that the bracket in Eq.~\eqref{FE:nucleon_amplitude_full} is a permutation-group singlet.
            The Dirac amplitudes $\Psi_n$ carry three spinor indices $\alpha,\beta,\gamma$ for the quark legs and one
            spinor index $\delta$ for the nucleon, and they are mixed-antisymmetric $(\Psi_1)$ or mixed-symmetric $(\Psi_2)$
            under exchange of the indices $\alpha$, $\beta$ and related quark momenta.
            They can be decomposed in a basis of 64 orthonormal Dirac structures $\mathsf{X}^i_{\alpha\beta\gamma\delta}$
            which correspond to $s-$, $p-$ and $d-$waves in the nucleon's rest frame; see Refs.~\cite{Eichmann:2009qa,Eichmann:2009en,Eichmann:2011vu} for details:
            \begin{equation}\label{amplitude:reconstruction1}
                [\Psi_n]_{\alpha\beta\gamma\delta}(p,q,P) = \sum_{i=1}^{64} f_i^n \,\mathsf{X}^i_{\alpha\beta\gamma\delta}(p,q,P)\,.
            \end{equation}
            The momentum-dependent dressing functions $f_i^n$ are the Lorentz-invariant solutions of the covariant Faddeev equation.
            They depend on the five Lorentz invariants that can be constructed from the momenta $p$, $q$ and $P$ ($P^2=-M^2$ is fixed).
            The dressing functions for $n=1,2$ are not independent but related via permutation-group symmetry.

            Similarly, the two isospin-$\nicefrac{1}{2}$ flavor tensors $\mathsf{F}_n$ carry three isospin indices
            $\mathsf{a,b,c}$ for the quarks and one for the nucleon. They are mixed-antisymmetric ($\mathsf{F}_1$) or mixed-symmetric ($\mathsf{F}_2$)
            with respect to $\mathsf{a}$, $\mathsf{b}$:
              \begin{equation} \label{FAD:flavor}
              \begin{split}
                  [\mathsf{F}_1]_\mathsf{abcd} &= \textstyle\frac{1}{\sqrt{2}}\,[i \sigma_2]_\mathsf{ab} \,\delta_\mathsf{cd} \,,\\
                  [\mathsf{F}_2]_\mathsf{abcd} &= -\textstyle\frac{1}{\sqrt{6}}\, [\vect{\sigma}\,i\sigma_2]_\mathsf{ab} \,\vect{\sigma}_\mathsf{cd} \,,
              \end{split}
              \end{equation}
              where the $\sigma_k$ are the Pauli matrices,
             and they are normalized to unity: $[\mathsf{F}^\dag_{n'}]_{\mathsf{bad'c}}[\mathsf{F}_n]_{\mathsf{abcd}} = \delta_{n'n} \,\delta_{\mathsf{d'd}}$.
             To project onto proton or neutron flavor states, the nucleon index $\mathsf{d}$ must be contracted with either of the two isospin vectors $(1,0)$ or $(0,1)$, respectively.
             Finally, the antisymmetric tensor $\varepsilon_{ABC}$ in Eq.~\eqref{FE:nucleon_amplitude_full} denotes the color part which is also normalized to unity via the factor $1/\sqrt{6}$.

 \renewcommand{\arraystretch}{1.1}

             \subsection{Flavor}

            We proceed by combining the ingredients to the Compton scattering amplitude.
            Starting with the flavor traces, the quark Compton vertex
             comes with a flavor factor
              \begin{equation}\label{flavor-Q2}
                 \mathsf{Q}^2 = \left(\begin{array}{cc} q_u^2 & 0 \\ 0 & q_d^2 \end{array}\right) = \frac{5}{9}\,\mathsf{\sigma_s}+ \frac{1}{3}\,\mathsf{\sigma_v}\,,
             \end{equation}
             i.e., the squared quark charge matrix.
             We have also given the isoscalar-isovector decomposition
             with isospin matrices $\mathsf{\sigma_s}=\mathds{1}/2$ and $\mathsf{\sigma_v}=\sigma_3/2$,
             where $\mathsf{\sigma_3}=\text{diag}(1,-1)$ is the Pauli matrix.
             The rainbow-ladder kernel is flavor-independent; hence,
             the flavor trace in both $a=3$ diagrams is given by
             \begin{equation} \label{flavor-current}
                 \left[\mathsf{F}^{(3)}_{n'n}\right]_\mathsf{d'd} := [\mathsf{F}^\dag_{n'}]_{\mathsf{bad'c'}}\, \mathsf{Q}^2_{\mathsf{c'c}} \,[\mathsf{F}_n]_{\mathsf{abcd}}\,,
             \end{equation}
             with $\mathsf{d'}=\mathsf{d}=1$ for the proton and $\mathsf{d'}=\mathsf{d}=2$ for the neutron.
             We keep the doublet indices $n'$ and $n$ for the outgoing and incoming nucleon amplitude general
             so we can treat the Dirac and flavor parts in the scattering amplitude separately.
             The above expression vanishes if $n\neq n'$, and for $n=n'$ it yields:
             \begin{equation}\label{current-flavor-3}
             \begin{split}
                 \mathsf{F}^{(3)}_{11} &= \left(\begin{array}{cc} q_u^2 & 0 \\ 0 & q_d^2 \end{array}\right),\\ 
                 \mathsf{F}^{(3)}_{22}  &= \frac{1}{3}  \left(\begin{array}{cc} q_u^2 + 2 q_d^2 & 0 \\ 0 & q_d^2 + 2 q_u^2 \end{array}\right),
             \end{split}
             \end{equation}
             where the matrices are again defined in isospin space.
            The total scattering amplitude
            is then the sum over the three permutations and the final and initial doublet configurations:
             \begin{equation}\label{current-total}
                 \left[\widetilde{J}^{\mu\nu}_{\delta'\delta}(P,\Sigma,\Delta)\right]_{\mathsf{d'd}} =
                 \sum_{a=1}^3 \sum_{n'n} \left[J_{n'n}^{(a)}\right]^{\mu\nu}_{\delta'\delta}\,\left[\mathsf{F}^{(a)}_{n'n}\right]_\mathsf{d'd}\,,
             \end{equation}
            where the first bracket includes the Dirac-Lorentz and color parts to which we will return below.
             The permuted diagrams for $a=1,2$ can be inferred from the $a=3$ diagram by applying the two-dimensional matrix representations $\mathcal{M}'$, $\mathcal{M}''$
             of the permutation group $\mathds{S}^3$,
             \begin{equation*}
                   \mathcal{M}' = \frac{1}{2}\left( \begin{array}{cc} -1 & -\sqrt{3} \\ \sqrt{3} & -1 \end{array}\right), \quad
                   \mathcal{M}'' = \frac{1}{2}\left( \begin{array}{cc} -1 & \sqrt{3} \\ -\sqrt{3} & -1 \end{array}\right)
             \end{equation*}
             which act upon the doublet indices $n',n$ (here without summation) via~\cite{Eichmann:2011vu}
             \begin{equation*}\label{current:permutations}
             \begin{split}
                 J^{(1)}_{n'n}\,\mathsf{F}^{(1)}_{n'n} &= \left[ \mathcal{M}' J^{(3)} {\mathcal{M}'}^T \right]_{n'n} \left[ \mathcal{M}' \mathsf{F}^{(3)} {\mathcal{M}'}^T \right]_{n'n} \,, \\
                 J^{(2)}_{n'n}\,\mathsf{F}^{(2)}_{n'n} &= \left[ \mathcal{M}'' J^{(3)} {\mathcal{M}''}^T \right]_{n'n} \left[ \mathcal{M}'' \mathsf{F}^{(3)} {\mathcal{M}''}^T \right]_{n'n} \,.
             \end{split}
             \end{equation*}
             Inserting this together with\eqref{current-flavor-3} into Eq.~\eqref{current-total} yields the result for the total scattering amplitude:
             \begin{equation}\label{current-finalfinal}
             \begin{split}
                 \left[\widetilde{J}^{\mu\nu}_{\delta'\delta}(P,\Sigma,\Delta)\right]_{\mathsf{11}} &= \frac{2}{3}\left[ 2 J_{11}^{(3)} + J_{22}^{(3)} \right]^{\mu\nu}_{\delta'\delta} \,, \\
                 \left[\widetilde{J}^{\mu\nu}_{\delta'\delta}(P,\Sigma,\Delta)\right]_{\mathsf{22}} &= \frac{1}{3}\left[  J_{11}^{(3)} + 3 J_{22}^{(3)} \right]^{\mu\nu}_{\delta'\delta} \,,
             \end{split}
             \end{equation}
             where $J_{nn}^{(3)}$ are the Dirac contributions to the $a=3$ diagrams.
             The first line in~\eqref{current-finalfinal} is the result for the proton and the second corresponds to the neutron.
             Correspondingly, the isoscalar-isovector decomposition reads
             \begin{equation}\label{flavor-iso}
             \begin{split}
                 \widetilde{J}^{\mu\nu}_{\delta'\delta}(P,\Sigma,\Delta) &=  \frac{5}{3}\left[ J_{11}^{(3)} + J_{22}^{(3)} \right]^{\mu\nu}_{\delta'\delta} \,\mathsf{\sigma_s} \\
                                             &+ \frac{1}{3}\left[  3J_{11}^{(3)} - J_{22}^{(3)} \right]^{\mu\nu}_{\delta'\delta}\, \mathsf{\sigma_v}\,,
             \end{split}
             \end{equation}
             with the isospin matrices from Eq.~\eqref{flavor-Q2}.

             \subsection{Dirac structure} \label{sec:handbag-dirac}

             The remaining task is to determine the Dirac-Lorentz and color structure of the $a=3$ diagram.
             The color trace in the impulse-approximation diagram equals $1$ while the second diagram involving the rainbow-ladder kernel picks up
             a color factor $-\nicefrac{2}{3}$. We can combine both diagrams and write:
             \begin{equation}\label{current:diagram-3}
             \begin{split}
                 &\left[ J^{(3)}_{n'n} \right]^{\mu\nu}_{\delta'\delta} =  \int\limits_p \!\!\!\int\limits_q [\conjg{\Psi}_{n'}]_{\beta'\alpha'\delta'\gamma'}(p_f,q_f,P_f)   \\
                 & \times S_{\alpha'\alpha}(p_1)\,S_{\beta'\beta}(p_2)\,\left[ S(p_3^+)\,\widetilde{\Gamma}^{\mu\nu}(p_3,\Sigma,\Delta)\,S(p_3^-)\right]_{\gamma'\gamma}  \\
                 & \times \big[\Psi_n- \textstyle\frac{2}{3}\,\Psi_n^{(3)}\big]_{\alpha\beta\gamma\delta}(p_i,q_i,P_i)  \,.
             \end{split}
             \end{equation}
             The momentum routing is illustrated in Fig.~\ref{fig:current-kinematics}.
             $P$, $\Sigma$ and $\Delta$ are the external momenta according to their definition in Section~\ref{sec:kinematics-variables},
             with $P_i$ and $P_f$ the incoming and outgoing nucleon momenta.
             $p_i$, $q_i$ and $p_f$, $q_f$ are the incoming and outgoing relative momenta;
             $p_1$, $p_2$, $p_3$ and $p_3^\pm=p_3 \pm \Delta/2$ are the quark momenta;
             and $p$ and $q$ are the (real) loop momenta.
             For a symmetric momentum-partitioning parameter $\nicefrac{1}{3}$ the relative momenta are explicitly given by
             \begin{equation}\label{ff-momenta-1}
                 p_f = p + \frac{\Delta}{3}\,, \quad
                 p_i = p - \frac{\Delta}{3}\,, \quad
                 q_f = q_i = q
             \end{equation}
             and the quark momenta by
             \begin{equation}\label{ff-momenta-2}
                 p_1 = -q - \frac{p}{2} + \frac{P}{3}\,, \quad
                 p_2 =  q - \frac{p}{2} + \frac{P}{3}
             \end{equation}
             and $p_3 = p+P/3$.
             $\widetilde{\Gamma}^{\mu\nu}$ is the quark Compton vertex and the $S$ are the dressed quark propagators.
             Here we also see that it is sufficient to know the nucleon bound-state amplitudes
             for real values of $p_i^2$ and $p_f^2$ as long as the four-vector $\Delta$ is real,
             which is no longer the case for timelike values $t<0$.

             The second diagram with the kernel appears in Eq.~\eqref{current:diagram-3} via
             \begin{equation}\label{current-psi-3}
             \begin{split}
                 &[\Psi_n^{(3)}]_{\alpha\beta\gamma\delta}(p_i,q_i,P_i) =  \int_k K_{\alpha\alpha'\beta\beta'}(k)  \\
                 & \qquad\qquad \times S_{\alpha'\alpha''}(p_1-k)\, S_{\beta'\beta''}(p_2+k)  \\
                 & \qquad\qquad \times [\Psi_n]_{\alpha''\beta''\gamma\delta}(p_i,q_i+k,P_i) \,.
             \end{split}
             \end{equation}
             Since $\Psi_n^{(3)}$ is one of the three diagrams in the Faddeev equation, it is not necessary to compute
             it anew in the form-factor calculation.
             If we attach dressed quark propagators to the third line of~\eqref{current:diagram-3} and denote the result by $\Phi_n$,
             we can define a nucleon-quark four-point function
             \begin{equation}\label{current-M-def}
             \begin{split}
                 & \mathcal{W}^{n'n}_{\delta'\delta\gamma\gamma'}(p,P,\Delta) = \\
                  &= \int_q [\conjg{\Psi}_{n'}]_{\beta\alpha\delta'\gamma'}(p_f,q_f,P_f) \,  [\Phi_n]_{\alpha\beta\gamma\delta}(p_i,q_i,P_i)
             \end{split}
             \end{equation}
             and write Eq.~\eqref{current:diagram-3} in a more compact form:
             \begin{equation}\label{current:alternate-def}
                 \left[ J^{(3)}_{n'n} \right]^{\mu\nu}_{\delta'\delta} =  \int\limits_p \mathcal{W}^{n'n}_{\delta'\delta\gamma\gamma'} \left[ S(p_3^+)\,\widetilde{\Gamma}^{\mu\nu}(p_3,\Sigma,\Delta) \right]_{\gamma'\gamma}\,.
             \end{equation}
             Note that we could have removed the appearance of $S(p_3^+)$ as well by absorbing the quark leg  into the nucleon amplitude $\conjg{\Psi}_{n'}$.
             The quantity $\mathcal{W}$ is 'universal' in the sense that it only depends on the nucleon's bound-state amplitude,
             and the quark propagator and quark-gluon interaction that enter its calculation in the first place.
             If the quark Compton vertex is replaced by a vector, pseudoscalar or axialvector $q\bar{q}$ vertex,
             Eq.~\eqref{current:alternate-def} yields the nucleon's electromagnetic, pseudoscalar or axial current,
             cf. Appendix D in Ref.~\cite{Eichmann:2011vu} and App. B in~\cite{Eichmann:2011pv}.
             Nevertheless, we recall that Compton scattering requires further diagrams beyond the handbag structure
             for a consistent and electromagnetically gauge-invariant description.

 \renewcommand{\arraystretch}{1.1}

              \subsection{Calculation of $\mathcal{W}$}

             While a brute-force calculation of the integral in Eq.~\eqref{current-M-def} is relatively straightforward,
             it represents also the bottleneck in the numerical determination of the scattering amplitude.
             We can break up the integral into smaller pieces by choosing a suitable
             orthogonal basis decomposition for the incoming and outgoing nucleon amplitudes.
             Let us write Eq.~\eqref{amplitude:reconstruction1} in the form
             \begin{equation}\label{faddeev-alternate-basis}
                 [\Psi_n]_{\alpha\beta\gamma\delta}(p,q,P) = \sum_{kl\omega \lambda} f_{kl\omega \lambda}^n  \, \mathsf{X}^{kl\omega \lambda}_{\alpha\beta\gamma\delta}(R,S,\hat{P}) \,,
             \end{equation}
             with vectors $V$, $R$ and $S$ defined similarly as in Eq.~\eqref{orthonormal-momenta}:
             \begin{equation}\label{W-SRV}
                 S^\mu = \widehat{p_T}^\mu, \quad
                 R^\mu = \widehat{q_t}^\mu, \quad
                 V^\mu = \varepsilon^{\mu\alpha\beta\gamma} R^\alpha S^\beta \hat{P}^\gamma\,.
             \end{equation}
             Orthogonal projection is here understood with respect to the nucleon momentum $P$,
             so that in the nucleon's rest frame the vectors $V$, $R$, $S$ and $\hat{P}$
             would reduce to the Euclidean unit vectors in analogy to Eq.~\eqref{vrsd}.
             The coefficients $f_{kl\omega r}^n$ depend on the five Lorentz-invariant momentum variables that can
             be formed from the three momenta $p$, $q$ and $P$, with $P^2=-M^2$ fixed.

             The simplest 64-dimensional orthonormal basis for the nucleon's bound-state amplitude is defined by the elements
             \begin{equation}\label{faddeev-alternate-basis-2}
             \begin{split}
                & \mathsf{X}^{kl\omega \lambda}_{\alpha\beta\gamma\delta}(R,S,\hat{P}) =
                  \left[ \tau_\lambda\,\Gamma_k\, \Lambda_\omega \gamma_5 C \right]_{\alpha\beta}
                  \left[ \tau_\lambda\,\Gamma_l \,\Lambda_+\right]_{\gamma\delta}\,, \\
                & \qquad\quad k,l=1\dots 4, \qquad \lambda=\pm, \qquad \omega=\pm\,.
             \end{split}
             \end{equation}
             $\Lambda_\omega = (\mathds{1} + \omega\, \hat{\slashed{P}})/2$ are the positive- and negative-energy projectors
             that depend on the normalized total momentum; $\tau_+= \mathds{1}$ and $\tau_- = \gamma_5$; and the four Dirac structures are given by
             \begin{equation}
                   \Gamma_0 = \mathds{1} \,, \quad
                   \Gamma_1 = \gamma_5 \slashed{V} \,, \quad
                   \Gamma_2 = \slashed{R}\,, \quad
                   \Gamma_3 = \slashed{S}  \,.
             \end{equation}
             The basis elements satisfy the orthonormality relation
             \begin{equation}
                 \tfrac{1}{4}\,\conjg{\mathsf{X}}^{k'l'\omega'\lambda'}_{\beta\alpha\delta\gamma} \,\mathsf{X}^{kl\omega \lambda}_{\alpha\beta\gamma\delta}
                 = \delta_{kk'}\,\delta_{ll'}\,\delta_{\omega\omega'}\,\delta_{\lambda\lambda'}\,,
             \end{equation}
             where the charge-conjugated elements are given by
             \begin{equation}
                 \conjg{\mathsf{X}}^{kl\omega \lambda}_{\beta\alpha\delta\gamma}
                 = \left[ C^T \gamma_5\,\Lambda_\omega\,\conjg{\Gamma}_k\,\tau_\lambda \right]_{\beta\alpha} \left[ \Lambda_+ \conjg{\Gamma}_l \,\tau_\lambda\right]_{\delta\gamma}\,,
             \end{equation}
             with $\conjg{\Gamma}_1=-\Gamma_1$ and all other $\conjg{\Gamma}_i = \Gamma_i$. The charge-conjugated nucleon amplitude is
             \begin{equation}
                 [\conjg{\Psi}_n]_{\beta\alpha\delta\gamma}(p,q,P) = \sum_{kl\omega \lambda}f_{kl\omega\lambda}^n  \, \conjg{\mathsf{X}}^{kl\omega \lambda}_{\beta\alpha\delta\gamma}\,.
             \end{equation}
             The decomposition for the amplitude $\Phi_n$ in Eq.~\eqref{current-M-def} that includes the spectator kernel is analogous.

             When implemented in the form-factor diagrams, the momentum dependence on $\{ p,q,P \}$ becomes a dependence on
             the incoming and outgoing momenta $\{p_i,q_i,P_i\}$ or $\{p_f,q_f,P_f\}$.
             The orthogonal unit vectors must be adapted accordingly so that
             one has $\{ V_i,R_i,S_i\}$ and $\{V_f,R_f,S_f\}$.
             Inserting~\eqref{faddeev-alternate-basis} and~\eqref{faddeev-alternate-basis-2} into~\eqref{current-M-def}, the function $\mathcal{W}$ becomes
             \begin{equation}\label{current:W-calc2}
             \begin{split}
                 & \mathcal{W}^{n'n}_{\delta'\delta\gamma\gamma'} =   \int\limits_q  \sum_{ll'  \lambda\lambda'}
                     \big[ \Lambda_+^f \,\conjg{\Gamma}_{l'}^f \,\tau_{\lambda'}\big]_{\delta'\gamma'} \big[\tau_\lambda \,\Gamma_l^i \,\Lambda_+^i\big]_{\gamma\delta}  \\
                 &  \qquad \qquad \times\sum_{kk'\omega\omega'}  K_{\lambda\lambda' kk' \omega\omega'} \,f^{n'}_{k'l'\omega'\lambda'}\, g^n_{kl\omega \lambda}    \,,
             \end{split}
             \end{equation}
             where the $f^{n'}_{k'l'\omega'\lambda'}$ denote the dressing functions of $\conjg{\Psi}_{n'}$ and the $g^n_{kl\omega \lambda}$ those of $\Phi_n$, and the kernel
             is given by
             \begin{equation}
                  K_{\lambda\lambda' kk' \omega\omega'} = \text{Tr}\left\{ \Lambda_{\omega'}^f\,\conjg{\Gamma}_{k'}^f\,\tau_{\lambda'}\,\tau_\lambda\,\Gamma_k^i\,\Lambda_\omega^i\right\}.
             \end{equation}
             Since $\tau_\lambda=\mathds{1}$ or $\gamma_5$, it will have the form
             \begin{equation}\label{K-lambda-plusminus}
                  K_{\lambda\lambda' kk' \omega\omega'} = \delta_{\lambda\lambda'}\,K^{(+)}_{kk' \omega\omega'} + \delta_{\lambda,-\lambda'}\,K^{(-)}_{kk' \omega\omega'}\,,
             \end{equation}
             where the positive and negative-parity contributions can be calculated explicitly:
        \renewcommand{\arraystretch}{1.2}
             \begin{equation} \label{M-Kernel-1}
                K^{(+)}_{kk' \omega\omega'}=\left( \begin{array}{c ccc}
                                          c_0  & 0                  & c_R'                &  c_S'      \\
                                          0    & c_{VV}             & \widetilde{c}_{VR}  &  \widetilde{c}_{VR}    \\
                                          c_R  & \widetilde{c}_{RV} & c_{RR}              & c_{RS}    \\
                                          c_S  & \widetilde{c}_{SV} & c_{SR}              & c_{SS}   \\
                                          \end{array}\right),
             \end{equation}
             \begin{equation} \label{M-Kernel-2}
                K^{(-)}_{kk' \omega\omega'}=\left( \begin{array}{c ccc}
                                          0    & c_V'               & 0                   &  0         \\
                                          c_V  & \widetilde{c}_{VV} & c_{VR}              &  c_{VS}                \\
                                          0    & c_{RV}             & \widetilde{c}_{RR}  &  \widetilde{c}_{RS}    \\
                                          0    & c_{SV}             & \widetilde{c}_{SR}  &  \widetilde{c}_{SS}   \\
                                          \end{array}\right),
             \end{equation}
             with
             \begin{equation}\label{M-Kernel-3}
             \begin{split}
                 c_0 &= 2\left((1+t)\,\delta_{\omega\omega'} - t\,\delta_{\omega,-\omega'}\right)\,,\\
                 c_a &= \omega' \,a_i\cdot \widehat{P}_f\,, \\
                 c_b' &= \omega\,b_f\cdot \widehat{P}_i\,,\\
                 c_{ab} &= c_0\,(a_i\cdot b_f) - \omega\,\omega'\,(a_i\cdot \widehat{P}_f)(b_f\cdot \widehat{P}_i)\,,\\
                 \widetilde{c}_{ab} &= -\omega\,\omega'\,\varepsilon^{\mu\nu\alpha\beta} \,a_i^\mu \,b_f^\nu \,\widehat{P}_i^\alpha \,\widehat{P}_f^\beta\,,
             \end{split}
             \end{equation}
             and $a,b \in \{ V, R, S \}$.

             Going further and exploiting Eq.~\eqref{K-lambda-plusminus}, we write the second line of Eq.~\eqref{current:W-calc2} as
             \begin{equation}\label{F-Kfg}
                 \mathcal{F}_{n'n,\,ll',\lambda}^\pm  = \sum_{kk'\omega\omega'}  K^{(\pm)}_{kk' \omega\omega'} \,f^{n'}_{k'l'\omega',\pm \lambda}\, g^n_{kl\omega \lambda}\,.
             \end{equation}
             The $q-$dependence in the first line of Eq.~\eqref{current:W-calc2} appears only in the Dirac structures $\Gamma_l^i$ and $\conjg{\Gamma}_{l'}^f$ .
             We can separate the Dirac from the Lorentz parts by shuffling the momentum dependencies from the first into the second line.
             If we perform the sum over $ll'$ and the $q-$integration,
             the resulting Lorentz structures depend on the three momenta $p$, $P$ and $\Delta$ and have positive or negative parity. They
             carry zero, one or two Lorentz indices, and as a consequence of Eq.~\eqref{W-SRV} they
             are transverse to the incoming or outgoing nucleon momenta:
             \begin{equation*}
                 A \sim \int_q   \dots  , \quad
                 A^\mu_a \sim \int_q a_f^\mu \dots  , \quad
                 A^{\mu\nu}_{ab} \sim \int_q a_f^\mu \,b_i^\nu  \dots
             \end{equation*}
             and so on, with $a,b \in \{ V, R, S \}$.
             We can now apply our findings
             from the Compton-vertex analysis in Sec.~\ref{sec:vertex:tensorbasis} to construct orthonormal tensor bases for these quantities.
             In the former case, we had to find transverse basis elements with respect to the incoming and outgoing photon momenta;
             here we consider transversality with respect to incoming and outgoing nucleon momenta.
             Let us define another normalized set of four-vectors, constructed from $p$, $P$ and $\Delta$:
             \begin{equation} \label{df-di-r-v}
             \begin{split}
                 d_f^\mu &= \sqrt{1+t}\,\widehat{\Delta}^\mu + i\sqrt{t}\,\widehat{P}^\mu, \\
                 d_i^\mu &= \sqrt{1+t}\,\widehat{\Delta}^\mu - i\sqrt{t}\,\widehat{P}^\mu, \\
                 r^\mu &= \widehat{p_t}^\mu \quad \text{with} \quad p_t^\mu = T^{\mu\alpha}_P\,T^{\alpha\beta}_\Delta\,p^\beta, \\
                 v^\mu &= \varepsilon^{\mu\alpha\beta\gamma} r^\alpha \widehat{P}^\beta \widehat{\Delta}^\gamma\,.
             \end{split}
             \end{equation}
             We will use the symbols $v$ and $r$ in this context only in the remainder of this subsection;
             they should not be confused with the quantities defined in Section~\ref{sec:vertex:tensorbasis}.
             The unit vectors $\{v,r,d_i\}$ are by construction transverse to $P_i=P-\tfrac{\Delta}{2}$, and $\{v,r,d_f\}$ are transverse to $P_f=P+\tfrac{\Delta}{2}$.
             (Since $P_i$ and $P_f$ are linear combinations of $P$ and $\Delta$, $v$ and $r$ are transverse to both.)
             Thus, one can immediately write down the complete set of basis elements for these integrals:
             \begin{equation}\label{Amunu-1}
             \begin{split}
                 A & \; \rightarrow \; 1 \\
                 A^\mu_a & \; \rightarrow \; r^\mu, \;\; d_f^\mu \\
                 A^\nu_b & \; \rightarrow \; r^\nu, \;\; d_i^\nu \\
                 A^{\mu\nu}_{ab} & \; \rightarrow \;  v^\mu v^\nu, \;\; r^\mu r^\nu, \;\; d_f^\mu \,r^\nu , \;\; r^\mu d_i^\nu , \;\; d_f^\mu \,d_i^\nu
             \end{split}
             \end{equation}
             with coefficients that depend on the Lorentz-invariants $p^2$, $p\cdot P$, $p\cdot \Delta$ and $\Delta^2$.
             Eq.~\eqref{Amunu-1} applies if the resulting tensor has positive parity; for negative parity
             (induced by an odd power of $a_i=V_i$ or $b_f=V_f$) one has instead
             the general form
             \begin{equation}
             \begin{split}
                 A & \; \rightarrow \; 0 \\
                 A^\mu_a & \; \rightarrow \; v^\mu,   \\
                 A^\nu_b & \; \rightarrow \; v^\nu,   \\
                 A^{\mu\nu}_{ab} & \; \rightarrow \;  v^\mu r^\nu, \;\; r^\mu v^\nu, \;\; d_f^\mu \,v^\nu , \;\; v^\mu d_i^\nu .
             \end{split}
             \end{equation}
             Since, with the exception of $d_f\cdot d_i \neq 0$, the vectors in Eq.~\eqref{df-di-r-v} are normalized and orthogonal,
             the coefficients of these basis decompositions are  obtained simply by contracting $A^\mu_a$,
             $A^\nu_b$ and $A^{\mu\nu}_{ab}$ with each basis element.

             If we work out these relations  we arrive at the final result for Eq.~\eqref{current-M-def}:
             \begin{equation} \label{current:W-result}
             \begin{split}
                 & \mathcal{W}^{n'n}_{\delta'\delta\gamma\gamma'} =   \sum_{kk'  \lambda} c^{n'n}_{kk'\lambda}
                     \big[ \Lambda_+^f \conjg{\Omega}_{k'}^f \,\tau_\lambda\big]_{\delta'\gamma'} \big[\tau_\lambda \,\Omega_k^i \,\Lambda_+^i\big]_{\gamma\delta}  \,,
             \end{split}
             \end{equation}
             where the new Dirac structures are given by
             \begin{equation}
                \begin{array}{rl}
                   \conjg{\Omega}_0^f &= \mathds{1} \,,\\
                   \conjg{\Omega}_1^f &= \slashed{v}\,\gamma_5\,,\\
                   \conjg{\Omega}_2^f &= \slashed{r}\,,\\
                   \conjg{\Omega}_3^f &= \slashed{d}_f \,.
                \end{array}
                \qquad
                \begin{array}{rl}
                   \Omega_0^i &= \mathds{1} \,,  \\
                   \Omega_1^i &= \gamma_5\,\slashed{v}\,,   \\
                   \Omega_2^i &= \slashed{r}\,,  \\
                   \Omega_3^i &= \slashed{d}_i \,.
                \end{array}
             \end{equation}
             Note that one obtains $4\times 4\times 2=32$ independent basis elements for the nucleon-quark four-point function $\mathcal{W}^{n'n}_{\delta'\delta\gamma\gamma'}$,
             as it should be for a spinor four-point function with two onshell legs.
             The coefficients are obtained from the integral
             \begin{equation}\label{current-cn'n}
                 c^{n'n}_{kk'\lambda} =  \int\limits_q \sum_{ll'}  C^i_{kl}\,C^{f}_{k'l'}\,\mathcal{F}_{n'n,\,ll',\,\mathcal{P}(k)\mathcal{P}(l)\lambda}^{\mathcal{P}(k)\mathcal{P}(k')\mathcal{P}(l)\mathcal{P}(l')}\,,
             \end{equation}
             where $C^i_{0l} = C^f_{0l} = \delta_{0l}$, and for $k,l=1,2,3$ one has:
             \begin{equation}
                 C^i_{kl}  = (u_k)_i \cdot (a_l)_i \,, \qquad
                 C^f_{kl}  = (u_k)_f \cdot (a_l)_f \,,
             \end{equation}
             with $(u_k)_i \in \{ v, r, d_i \}$, $(u_k)_f \in \{ v, r, d_f \}$ and $a_l  \in \{ V, R, S \}$.
             Positive parity for all coefficients leads to the signs $\mathcal{P}$ which we define as
             \begin{equation}
                  \mathcal{P}(k) = \left\{ \begin{array}{rl}
                                          +1 &\quad k=0,2,3, \\
                                          -1 & \quad k=1.
                                          \end{array}  \right.
             \end{equation}

   \renewcommand{\arraystretch}{1.0}

              In precomputing the quantity $\mathcal{W}(p,P,\Delta)$ we have seen that it does not depend on the average external momentum $\Sigma$ in the form-factor
              diagrams, so that its Lorentz-invariant dressing functions $c^{n'n}_{kk'\lambda}$ are only functions of $p^2$, $p\cdot P$, $p\cdot\Delta$ and $\Delta^2$.
              This also means that the Lorentz frame defined by Eqs.~\eqref{simple-frame}--\eqref{simple-frame-nucleon} is not the most economical choice since the
              dependence on the external variable $Y$ becomes redundant. We compute these functions instead in the following frame:
             \begin{align}
                 \frac{\Delta}{M} &= 2\sqrt{t}\left(\begin{array}{c} 0 \\ 0 \\ 0 \\ 1 \end{array}\right), \quad
                  \frac{P}{M}=i \sqrt{1+t}\left(\begin{array}{c} 0 \\ 0 \\ 1 \\ 0 \end{array}\right),   \label{Delta-P-Compton} \\
                 & \quad  p=\sqrt{p^2}\left(\begin{array}{l} 0 \\ \sqrt{1-z^2}\,\sqrt{1-y^2} \\ \sqrt{1-z^2}\,y \\ z \end{array}\right),  \label{p-frame-1}
             \end{align}
             with the integration momentum
             \begin{equation}
                 q=\sqrt{q^2}\left(\begin{array}{l} \sqrt{1-{z'}^2}\,\sqrt{1-{y'}^2}\,\sin\phi' \\ \sqrt{1-{z'}^2}\,\sqrt{1-{y'}^2}\,\cos\phi' \\ \sqrt{1-{z'}^2}\,y' \\ z' \end{array}\right).
             \end{equation}
             The vectors $v^\mu$ and $r^\mu$ from Eq.~\eqref{df-di-r-v} are then the Euclidean unit vectors in the 1-- and 2--directions whereas
             $V_i$, $R_i$, $S_i$ and $V_f$, $R_f$, $S_f$ are more complicated.

             When reinserting the result \eqref{current:W-result}  for $\mathcal{W}(p,P,\Delta)$ into the scattering amplitude~\eqref{current:alternate-def}, we keep $\Delta$ and $P$ from above and introduce
             the average photon momentum $\Sigma$ via
             \begin{equation}\label{Sigma-Compton}
                 \frac{\Sigma}{M}=\sqrt{\sigma}\left(\begin{array}{l} 0 \\ \sqrt{1-Z^2}\,\sqrt{1-Y^2} \\ \sqrt{1-Z^2}\,Y \\ Z \end{array}\right).
             \end{equation}
             The Lorentz-invariant definitions in Eqs.~(\ref{euclidean-variables-1}--\ref{XYZ-def}) remain of course unaltered by this choice.
             In that way the scattering amplitude depends on the external variables $\sigma$, $Y$ and $Z$ only through the quark Compton vertex.
             $p$ is now the loop momentum and we have an additional dependence on the angle $\phi$:
             \begin{equation} \label{p-frame-2}
                 p=\sqrt{p^2}\left(\begin{array}{l} \sqrt{1-z^2}\,\sqrt{1-y^2}\,\sin\phi \\ \sqrt{1-z^2}\,\sqrt{1-y^2}\,\cos\phi \\ \sqrt{1-z^2}\,y \\ z \end{array}\right).
             \end{equation}
             Note that no interpolation for the coefficients of $\mathcal{W}$ is required when changing the frame from \eqref{p-frame-1} to \eqref{p-frame-2} since the
             explicit form of all variables $p^2$, $p\cdot P$, $p\cdot\Delta$ and $\Delta^2$ remains the same.
             With the methods described in this appendix we have reduced the original two- and three-loop integrals
             from Fig.~\ref{fig:current-kinematics}, which carry a dependence on four external variables,
             to a successive calculation of intermediate one-loop diagrams which depend only on a subset of these variables.

              \subsection{Extraction of Compton dressing functions}

        \renewcommand{\arraystretch}{1.0}

             In the final step we want to extract the Compton form factors, i.e., the Lorentz-invariant dressing functions
             of the nucleon Compton amplitude $\widetilde{J}^{\mu\nu}(P,\Sigma,\Delta)$, from Eq.~\eqref{current-finalfinal}.
             For simplicity we use the basis decomposition of Eq.~\eqref{qcv-nucleon-transverse} with the dressing functions  $F_i$ and $G_i$;
             dressing functions in other bases can be obtained by appropriate basis transformations in the end.
             If we work in the frame defined by Eqs.~\eqref{Delta-P-Compton} and~\eqref{Sigma-Compton} instead of Eqs.~\eqref{simple-frame}--\eqref{simple-frame-nucleon},
             the orthogonal unit vectors $v$, $r$, $s$ and $d$ for the external momenta
             assume the form
             \begin{equation*}\label{vrsd-nucleon}
                 v=\left(\begin{array}{c} -1 \\ 0 \\ 0 \\ 0 \end{array}\right), \;
                 r=\left(\begin{array}{c} 0 \\ -Y \\ \sqrt{1-Y^2} \\ 0 \end{array}\right), \;
                 s=\left(\begin{array}{c} 0 \\ \sqrt{1-Y^2} \\ Y \\ 0 \end{array}\right),
             \end{equation*}
             and $d=(0,0,0,1)$ as before.
             We exploit the orthonormality relations for the basis elements $\mathsf{Y}_i^{\mu\nu}$ to write
             \begin{equation}
                 J_i := \mathsf{Y}_i^{\mu\nu} \widetilde{J}^{\mu\nu} = \Lambda_+^f \left[\begin{array}{c} \mathds{1} \\ \gamma_5 \end{array}\right]_i
                 \left( F_i + G_i \, \Slash{s}\right)\Lambda_+^i\,,
             \end{equation}
             where one has to pick $\gamma_5$ instead of $\mathds{1}$ for the elements $i=4,5,8,9$.
             This yields
             \begin{equation}\label{current-extraction-1}
             \begin{split}
                 F_i &= \frac{1}{2t\,(1-Y^2)}\,\text{Tr}\left[ \left(\frac{t}{1+t}-Y^2 + \frac{Y}{\sqrt{1+t}}\,\Slash{s}\right) J_i\right], \\
                 G_i &= \frac{1}{2t\,(1-Y^2)}\,\text{Tr}\left[ \left(\frac{Y}{\sqrt{1+t}}-\Slash{s}\right) J_i\right]
             \end{split}
             \end{equation}
             for the elements with $i=1,2,3,6,7$ and
             \begin{equation}\label{current-extraction-2}
             \begin{split}
                 F_i &= -\frac{1}{2t}\,\text{Tr}\left[\gamma_5\, J_i\right], \\
                 G_i &= \frac{1}{2(1+t) (1-Y^2)}\,\text{Tr}\left[ \Slash{s} \gamma_5 \, J_i\right]
             \end{split}
             \end{equation}
             for the remaining ones. The functions $F_i$ and $G_i$ depend on the four variables $t$, $\sigma$, $Y$ and $Z$.

             To conclude, let us summarize the steps to arrive at the Lorentz-invariant dressing functions that constitute the
             handbag part of the nucleon's Compton amplitude:
             \begin{itemize}
             \item Write the Faddeev-equation solution for the nucleon bound-state amplitude $\Psi_n$, and for the combination $\Phi_n$ that
                   absorbs the spectator kernel,
                   in the basis~\eqref{faddeev-alternate-basis} and extract their invariant dressing functions $f_{kl\omega \lambda}^n$ and $g_{kl\omega \lambda}^n$.
             \item Combine them in the quantity $\mathcal{F}$ defined in Eq.~\eqref{F-Kfg}, using the kernels~\eqref{M-Kernel-1}--\eqref{M-Kernel-3}.
             \item Integrate over $q$ to obtain the Lorentz-invariant coefficients of the four-point function $\mathcal{W}$ in Eq.~\eqref{current-cn'n}.
             \item Contract $\mathcal{W}$ of Eq.~\eqref{current:W-result} with the Dirac structures $\mathcal{O}_{\delta\delta'}$
                   that are needed to extract the Compton form factors. From Eqs.~\eqref{current-extraction-1}--\eqref{current-extraction-2} we have
                      $\mathcal{O} \in \{ \mathds{1}, \, \Slash{s}, \,  \gamma_5, \,  \Slash{s} \gamma_5 \}$. Contract the result with the
                      precalculated quark Compton vertex according to~\eqref{current:alternate-def}.
             \item Add up the permutation-group doublet contributions according to Eq.~\eqref{current-finalfinal} and extract the Compton form factors $F_i$ and $G_i$ for proton and neutron.
             \item Apply appropriate transformation matrices to obtain the dressing functions in other bases, for example those defined in Eq.~\eqref{qcv-onshell-basis} or Table~\ref{drechsel-tarrach}.
             \end{itemize}

             \bigskip

  \section{Basis decomposition into transverse and longitudinal parts}\label{qcv-basis-LT}

            In Section~\ref{sec:wti} we have decomposed the fermion Compton vertex into a part that satisfies its Ward-Takahashi identity and a fully transverse part.
            Another possible decomposition is the one into \textit{fully} transverse, \textit{fully} longitudinal and \textit{mixed} transverse-longitudinal parts.
            Both decompositions are not congruent since the WTI-preserving contribution has components in all three subspaces.
            The second decomposition is useful for the numerical calculation of the quark Compton vertex: when implemented in the nucleon Compton amplitude,
            only its fully transverse part can contribute to physics since
            the remainders must be cancelled by analogous pieces in the non-handbag diagrams of the scattering amplitude.

             We use the usual transverse projectors with respect to the incoming and outgoing photon momentum:
             \begin{equation}\label{transverse-projectors}
                 T_Q^{\mu\nu} = \delta^{\mu\nu} - \frac{Q^\mu  Q^\nu}{Q^2}\,, \qquad
                 T_{Q'}^{\mu\nu} = \delta^{\mu\nu} - \frac{{Q'}^\mu  {Q'}^\nu}{{Q'}^2}\,.
             \end{equation}
             With their help, the three contributions to the Compton vertex, which are independently Bose- and charge-conjugation invariant, can be written as
             \begin{equation}\label{LT-decomposition}
             \begin{split}
                 \Gamma^{\mu\nu} &= \left[T_{Q'}^{\mu\alpha}+\frac{{Q'}^\mu  {Q'}^\alpha}{{Q'}^2}\right] \Gamma^{\alpha\beta} \left[T_Q^{\beta\nu}+\frac{Q^\beta  Q^\nu}{Q^2}\right] \\
                                 &= \Gamma^{\mu\nu}_{TT}  + \frac{{Q'}^\mu  Q^\nu}{{Q'}^2  Q^2}\,\Gamma_{LL} + \left[\frac{{Q'}^\mu}{{Q'}^2}\,\Gamma_{LT}^\nu + \Gamma_{TL}^\mu\,\frac{Q^\nu}{Q^2}\right]\,,
             \end{split}
             \end{equation}
             where $\Gamma^{\mu\nu}_{TT}$, $\Gamma_{LT}^\nu$, $\Gamma_{TL}^\mu$ and $\Gamma_{LL}$ have been defined in an obvious way.
             Bose symmetry entails
             \begin{equation}
                 \Gamma_{LT}^\mu(\Sigma) = -\Gamma_{TL}^\mu(-\Sigma)\,, \quad  \Gamma_{LL}(\Sigma)=\Gamma_{LL}(-\Sigma).
             \end{equation}
             We can now evaluate the WTI, either Eq.~\eqref{qcv-wti-r} for the 1PI vertex or Eq.~\eqref{qcv-wti-full} for the full vertex, to obtain:
             \begin{equation}
             \begin{split}
                 {Q'}^\mu  \Gamma^{\mu\nu} &= \Gamma^\nu_\text{LT} + \frac{Q^\nu}{Q^2}\,\Gamma_\text{LL} = W_1^\nu \\
                 \Gamma^{\mu\nu} Q^\nu    &= \Gamma^\mu_\text{TL} + \frac{{Q'}^\mu}{{Q'}^2}\,\Gamma_\text{LL} = W_2^\mu\,,
             \end{split}
             \end{equation}
             where $W_1^\mu$ and $W_2^\mu = -W_1^\mu(-\Sigma)$ generically denote the right-hand side of the respective WTI.
             This yields
             \begin{equation}\label{WTI-constraints-for-long-parts}
             \begin{split}
                 \Gamma_\text{LL} &= W_1 \cdot Q = W_2 \cdot {Q'}\,, \\
                 \Gamma_\text{LT}^\nu &= W_1^\mu \,T^{\mu\nu}_Q\,, \\
                 \Gamma_\text{TL}^\mu &= T^{\mu\nu}_{Q'} \,W_2^\nu\,,
             \end{split}
             \end{equation}
             and therefore all three parts are completely determined from the fermion-photon vertices.
             Note that the $W_i^\mu$ vanish for the onshell Compton amplitude, cf.~Eq.~\eqref{ncv-wti}, so that there only the fully transverse piece survives.

             The basis elements in Table~\ref{qcv-basis-1} can be rearranged to correspond to either one of the three
             terms in Eq.~\eqref{LT-decomposition}. Since the basis factorizes the Dirac from the Lorentz structure,
             it is sufficient to analyze the transversality relations at the level of the 16 Lorentz structures.
             To single out the transverse part $\Gamma^{\mu\nu}_{TT}$, we evaluate the condition
             \begin{equation}
                 {Q'}^\mu\,\Gamma^{\mu\nu} = 0 \quad \cap \quad
                 Q^\nu\,\Gamma^{\mu\nu} = 0\,.
             \end{equation}
             With the auxiliary variables
             \begin{equation}
                 a = \sqrt{X}\,Z\,, \quad b = \sqrt{X}\,\sqrt{1-Z^2}\,,
             \end{equation}
             where $X$ was defined in Eq.~\eqref{XYZ-def},
             and the orthonormal momenta from Eq.~\eqref{orthonormal-momenta} we can write
             \begin{equation}
                 \left.\begin{array}{c} {Q'}^\mu \\[1mm] Q^\mu \end{array}\right\} = \sqrt{t}\left(b\,s^\mu + (a \mp 1)\,d^\mu\right),
             \end{equation}
             so that the transversality conditions become
             \begin{equation}
             \begin{split}
                 \left( b\,s^\mu + (a-1)\,d^\mu\right) \Gamma^{\mu\nu} = 0\,, \\
                 \left( b\,s^\nu + (a+1)\,d^\nu\right) \Gamma^{\mu\nu} = 0\,.
             \end{split}
             \end{equation}
             The resulting relations between the dressing functions
             allow to construct a fully transverse orthonormal Lorentz basis which consists of 9 elements and is given in Eq.~\eqref{qcv-transverse-basis-Y}.
             We note that the $9\times 2=18$ transverse Lorentz-Dirac basis elements for the onshell amplitude, cf.~Table~\ref{qcv-basis-2},
             are linear combinations of those in Eq.~\eqref{qcv-onshell-basis} since both cover the same subspace.

             The remaining longitudinal Lorentz structures are constructed analogously.
             One basis element corresponds to the fully longitudinal part $\Gamma_{LL}$ and is obtained from
             \begin{equation}
                 T_{Q'}^{\mu\alpha}\, \Gamma^{\alpha\nu}=0\quad \cap \quad
                 \Gamma^{\mu\alpha} T_Q^{\alpha\nu}=0\,,
             \end{equation}
             namely:
             \begin{equation}
                \mathsf{Y}_{10} = \frac{(1-X)(\mathsf{X}_5-\mathsf{X}_6) -2b\,(\mathsf{X}_8 +a \,\mathsf{X}_7+ b\,\mathsf{X}_6)}{\sqrt{2\,n_1 n_2}} \,,
             \end{equation}
             where we used again
             \[ n_1 = 1 + X\,, \qquad n_2 = n_1 - \frac{4a^2 }{n_1}\,.\]
             The remaining seven linear combinations satisfy
             \begin{equation}
                 T_{Q'}^{\mu\alpha}\,\Gamma^{\alpha\beta}\,T_Q^{\alpha\nu}=0 \quad \cap \quad
                 {Q'}^\alpha\,\Gamma^{\alpha\beta}\,Q^\beta=0
             \end{equation}
             and are given by
             \begin{align*}
                \mathsf{Y}_{11} &= \frac{(1-X)\,\mathsf{X}_7+2b\,(b\,\mathsf{X}_7-a\,\mathsf{X}_6)}{\sqrt{n_1 n_2}}  \,, \\
                \mathsf{Y}_{12} &= \frac{(1-X)\,\mathsf{X}_8+2b\,\mathsf{X}_5}{\sqrt{n_1 n_2}}\,, \\[2mm]
                \mathsf{Y}_{13} &= \frac{1}{\sqrt{n_1}} \left(\mathsf{X}_{15} +a\, \mathsf{X}_{13}+b\, X_{9}\right)  \\
                \mathsf{Y}_{14} &= \frac{1}{\sqrt{n_2}} \left(\mathsf{X}_{13} +a\, \mathsf{X}_{15}+b\, X_{11}\right) - \frac{2a\,\mathsf{Y}_{13}}{\sqrt{n_1 n_2}}\, \,,\\[1mm]
                \mathsf{Y}_{15} &= \frac{1}{\sqrt{n_1}} \left(\mathsf{X}_{16} +a\, \mathsf{X}_{14}+b\, X_{10}\right),\\
                \mathsf{Y}_{16} &= \frac{1}{\sqrt{n_2}} \left(\mathsf{X}_{14} +a\, \mathsf{X}_{16}+b\, X_{12}\right)- \frac{2a\,\mathsf{Y}_{15}}{\sqrt{n_1 n_2}}\,.
             \end{align*}
             The complete 128-dimensional basis in Eq.~\eqref{qcv-basis-decomposition-Y}, constructed from the 16 $\mathsf{Y}_i$ and the eight $\tau_a$,
             is again orthonormal in the sense of Eq.~\eqref{qcv-orthonormality}.

\end{appendix}

\bigskip

\bibliographystyle{apsrev4-1-mod}

\bibliography{lit-compton}

\end{document}